\documentclass[manuscript]{aastex}

\usepackage{natbib}
\usepackage{graphicx}
\usepackage{amsmath}
\usepackage{lscape}
\usepackage{longtable}
\usepackage[version=3]{mhchem}

\slugcomment{}

\shorttitle{Constraining the abundances of complex organics in the inner regions of solar-type protostars}
\shortauthors{Taquet et al.}

\begin{document}


\title{Constraining the abundances of complex organics in the inner regions of solar-type protostars}

\author{Vianney Taquet \altaffilmark{1,2,3}, Ana L\'opez-Sepulcre
  \altaffilmark{4,5,6}, Cecilia Ceccarelli \altaffilmark{4,5}, Roberto Neri
  \altaffilmark{7}, Claudine Kahane \altaffilmark{4,5} and Steven B. Charnley \altaffilmark{1}}

\altaffiltext{1}{Astrochemistry Laboratory and The
  Goddard Center for Astrobiology, Mailstop 691, NASA Goddard Space
  Flight Center, 8800 Greenbelt Road, Greenbelt, MD 20770, USA}
\altaffiltext{2}{NASA Postdoctoral Program Fellow}
\altaffiltext{3}{Current address: Leiden Observatory, Leiden University, P.O. Box 9513,
2300-RA Leiden, The Netherlands}
\altaffiltext{4}{Univ. Grenoble Alpes, IPAG, F-38000 Grenoble, France}
\altaffiltext{5}{CNRS, IPAG, F-38000 Grenoble, France}
\altaffiltext{6}{Department of Physics, The University of Tokyo,
  Bunkyo-ku, Tokyo 113-0033, Japan} 
\altaffiltext{7}{Institut de Radioastronomie Millim{\'e}trique,
  Grenoble, France}
\email{taquet@strw.leidenuniv.nl}

\begin{abstract}

The high abundances of Complex Organic Molecules (COMs) with respect 
to methanol, the most abundant COM, detected towards low-mass
protostars, tend to be underpredicted by astrochemical models. 
This discrepancy might come from the large beam of  the single-dish
telescopes, encompassing several components of the studied protostar,
commonly used to detect COMs. 
To address this issue, we have carried out multi-line observations of
methanol and several COMs towards the two low-mass protostars
NGC1333-IRAS2A and -IRAS4A  with the Plateau de Bure interferometer
at an angular resolution of 2 arcsec, resulting in the first
multi-line detection of the O-bearing species glycolaldehyde and
ethanol and of the N-bearing species ethyl cyanide towards low-mass
protostars other than IRAS 16293. 
The high number of detected transitions from COMs (more than 40 methanol 
transitions for instance) allowed us to accurately derive the source size of 
their emission and the COMs column densities.
The COMs abundances with respect to methanol derived towards IRAS2A and 
IRAS4A are slightly, but not substantitally, lower than those derived
from previous single-dish observations.  
The COMs abundance ratios do not vary significantly with the
protostellar luminosity, over five orders of magnitude, implying
that low-mass hot corinos are quite chemically rich as high-mass hot
cores. 
Astrochemical models still underpredict the abundances of key COMs, 
such as methyl formate or di-methyl ether, suggesting that our
understanding of their formation remains incomplete. 

\end{abstract}

\keywords{astrochemistry - ISM: abundances -ISM: individual objects
  (NGC 1333-IRAS2A, NGC 1333-IRAS4A) - ISM: molecules - stars: formation}

\section{Introduction}

The early stages of low-mass star formation are known to be
accompanied by the increase of the molecular complexity. 
Most of the lines detected in the sub-millimetric spectra of Class 0
protostars are attributed to Complex Organic Molecules 
\citep[COMs, i.e. molecules based on carbon chemistry with 6 or more
atoms;][]{Herbst2009}, as shown by unbiased spectral surveys of
low-mass protostars \citep[see][for the spectral survey of
the low-mass protostar IRAS 16293-2422 for instance]{Caux2011}. 

The bright protostars IRAS 16293-2422 and NGC1333-IRAS4A (hereafter
IRAS 16293 and IRAS4A, respectively) have been the first two
protostars where COMs, such as methyl formate, di-methyl 
ether, formic acid, methyl cyanide, or ethyl cyanide, have been
detected with single-dish telescopes \citep{Cazaux2003, Bottinelli2004a}.  
The subsequent detection of a few COMs towards IRAS 16293 with
interferometers, providing better angular resolutions of $\sim$ 2
\arcsec, by \citet{Bottinelli2004b} and \citet{Kuan2004} confirmed
that most of the COMs emission likely comes from the warm inner region of
protostellar envelopes, called ``hot corinos'': the low-mass
counterparts of high-mass hot cores. 
Since then, the number of low-mass protostars showing COMs has increased 
with the detection of COMs towards $\sim 10$ other low-mass protostars
by \citet{Bottinelli2007} and \citet{Oberg2011, Oberg2014}.
\citet{Palau2011} and \citet{Fuente2014} also reported the
detection of several COMs towards four low/intermediate-mass protostars. 

Thanks to a larger number of transitions detected in the broad bands
of their receivers, single-dish telescopes have been first used to
derive the column densities of COMs, allowing them to constrain their
abundances averaged over a relatively large beam of 10-30 \arcsec. 
The abundances of COMs, usually compared to that of their probable mother
molecules (formaldehyde and methanol, see next paragraph), are found
to be relatively high: $\gtrsim$ 10 \% for methyl formate, di-methyl
ether, and formic acid, and $\sim$ 1 \% for methyl cyanide and
ethyl cyanide \citep[see][]{Bottinelli2007, Oberg2011, Oberg2014} although
the number of constrained abundance ratios remains relatively low and
the abundances show some scattering between the sources and molecules. 

Warm gas phase chemistry triggered by the sublimation of the main
ice components, and of methanol in particular, has been first
invoked by \citet{Millar1991} and \citet{Charnley1992} to explain the
presence of COMs observed towards the high-mass hot core Orion KL by
\citet{Blake1987}.  
However, more recent laboratory experiments and theoretical calculations have
contradicted several key assumptions made in the gas phase models: 
dissociative recombination of large ions do not lead predominantly to
the formation of COMs but rather to their fragmentation into small pieces
\citep{Geppert2006, Hamberg2010} whilst ion-molecule reactions have
been found to be not sufficiently efficient to produce the observed
amount of methyl formate \citep{Horn2004}.   
The current scenario of COMs formation is now based on the
recombination of radicals at the surface of interstellar grains
during the warm-up phase (30 K $<T<$ 100 K) occurring in the envelopes
surrounding Class 0 protostars \citep{Garrod2006, Garrod2008}. 
In these models, the radicals are
generated by the UV photodissocation of the main ice components or
they have survived to the incomplete hydrogenation process of CO
leading to CH$_3$OH during the ice formation \citep{Garrod2006, Taquet2012}.
Since the 90s, laboratory experiments have shown that UV irradiation
of interstellar ice analogs containing methanol, formaldehyde, and
ammonia can lead to the formation of a plethora of
complex molecules and even amino-acids \citep{Allamandola1988,
  Gerakines1996, Hudson2000, MunozCaro2002, Oberg2009}.
However, the quantitative efficiency of the COMs formation in ices and
their actual chemical pathways are still highly uncertain.

Although the current models produce a large set of COMs in significant
quantities, they are not able to explain the very high abundance ratio
($> 10$ \%) with respect to methanol seen for a few COMs, such as methyl
formate or di-methyl ether \citep[see][for a discussion of this
problem]{Taquet2012}.  
The discrepancy between observations and models could be due to the
large beams of single-dish telescopes used to derive the abundance
ratios of COMs. Typical single-dish beams of $\sim$ 10 $\arcsec$ are
much larger than the size of hot corinos \citep[$\sim$ 0.5
\arcsec;][]{Maret2005,   Maury2014} and encompass the cold envelopes
and possible outflows  driven by the central protostars, where COMs
have also been detected \citep{Arce2008, Oberg2010, Jaber2014}.  
In addition, COMs have also been found in cold and quiescent cores by
\citet{Bacmann2012} and \citet{Cernicharo2012} who claimed that
quiescent cold gas phase chemistry can produce COMs but in lower
quantities. 
However, new observations by \citet{Vastel2014} rather suggest that
the emission from COMs observed in another pre-stellar core originates
in an outer ring, so that the previous conclusions may need some
cautions. 

To better constrain the abundances of COMs originating from the hot
corinos surrounding the low-mass protostars, a large number of
transitions of methanol and COMs need to be observed with
interferometers, providing angular resolutions of $\sim$ 1-2
$\arcsec$. The emission originating from hot corinos can be
distinguished from other components of the envelope, allowing us to
directly derive the abundance of COMs in the hot corinos.
Moreover, CH$_3$OH emission is likely optically thick towards the
continuum peak of protostars \citep{Zapata2013}, observations of its optically
thin isotopologue $^{13}$CH$_3$OH are therefore required to derive an
accurate estimate of the methanol column density. 
Although several publications have reported the
interferometric detection of COMs in hot corinos
\citep{Bottinelli2004b, Kuan2004, Jorgensen2005, Bisschop2008,
  Jorgensen2011, Persson2012, Jorgensen2012, Maury2014}, to our
knowledge, none of them led to an accurate and simultaneous
estimation of the column densities of methanol and COMs. 

In this work, we present multi-line observations of methanol
($^{12}$CH$_3$OH, and $^{13}$CH$_3$OH) as well as several COMs (methyl
formate, di-methyl ether, ethanol, glycolaldehyde methyl cyanide, and
ethyl cyanide) performed with the Plateau de Bure interferometer
towards the two low-mass protostars IRAS2A and IRAS4A located in the
NGC1333 star-forming region. 
Although the angular resolution does not allow us to spatially resolve the
emission of COMs, it is sufficiently high to distinguish the emission
from the hot corinos from other components of protostellar envelopes.
The paper is structured as follows: section 2 describes the
observational strategy, section 3 presents the continuum maps as well
as the spectra and the maps of molecular transitions, section 4
explains the adopted methodology to derive the abundances
of COMs, section 5 discusses the results; and section 6 summarizes this
work with the conclusions.

\section{Observations}

The two low-mass Class 0 protostars IRAS2A and IRAS4A
were observed with the IRAM Plateau de Bure Interferometer (PdBi)
on 2010 July 20, July 21, August 1, August 3, November 24, and 2011
March 10 in the C and D configurations of the array. Due to the
proximity to each other, the two sources were observed in the same
track. 
Phase and amplitude were calibrated by performing regular observations
of the nearby point sources 3C454.3, 3C84, and 0333+321. The amplitude
calibration uncertainty is estimated to be $\sim$20\%. 
The WIDEX backends have been used at 143.4 and 165.2 GHz, providing a
bandwidth of 3.6 GHz each with a spectral resolution
of 1.95 MHz ($\sim 3.5 - 4$ km s$^{-1}$). 
High-resolution narrow band backends focused on two CH$_3$OH lines and
12 HCOOCH$_3$ lines have also been used. They provide a bandwidth of
80 MHz with a spectral resolution of 0.04 MHz (0.08 km s$^{-1}$). Due
to the low signal-to-noise ratio (S/N) obtained for the methyl formate lines
at high spectral resolution, we decreased the spectral resolution to
0.4 MHz (0.8 km s$^{-1}$) to obtain a S/N ratio higher than 3.
The data calibration and imaging were performed using the CLIC and
MAPPING packages of the GILDAS software
\footnote{http://www.iram.fr/IRAMFR/GILDAS}. Continuum images were 
produced by averaging line-free channels in the WIDEX correlator
before the Fourier transformation of the data. The coordinates of the
source and the size of the synthesized beams are reported in Table
1. 

\begin{table}[htp]
\centering
\caption{Properties of NGC 1333 IRAS2A and IRAS4A.}
\begin{footnotesize}
\begin{tabular}{l c c}
\hline
\hline
Source & IRAS2A & IRAS4A \\
\hline
R. A. (J2000) & 03:28:55:57 & SE: 03:29:10.52 \\
 & & NW: 03:29:10.42 \\
Decl. (J2000) & 31:14:37:22 & SE: 31:13:31.06 \\
 & & NW: 31:13:32.04 \\
d (pc) $^a$ & 235 & 235 \\
$V_{\textrm{LSR}}$ (km/s) & +7.7 & +7.2 \\
$L_{\textrm{bol}}$ ($L_{\odot}$) $^b$ & 36 & 9.1 \\
$M_{\textrm{env}}$ ($M_{\odot}$) $^c$& 5.1 & 5.6 \\
\hline
\multicolumn{3}{c}{Frequency = 143 GHz} \\
\hline
Beam size (\arcsec) & 2.2$\times$1.7 & 2.1$\times$1.7 \\
Beam PA ($^{\circ}$) & 25 & 25 \\
rms(WideX) $^d$ & 2.57 & 3.34 \\
rms(Cont.) $^e$ & 1.56 & 10.8 \\
Flux (Jy) $^f$ & 0.13 & 1.1 \\
Size (\arcsec) $^f$ & 1.7$\times$1.7 & 2.1$\times$1.7 \\
PA ($^{\circ}$) $^f$ & +51 & +25 \\
$M$ ($M_{\odot}$) $^g$ & 0.4 & 3.8 \\
$N$(H$_2$) (cm$^{-2}$) $^g$ & $5.0 \times 10^{24}$ & $3.7 \times 10^{25}$ \\
\hline
\multicolumn{3}{c}{Frequency = 165 GHz} \\
\hline
Beam size (\arcsec) & 2.3$\times$1.7 & 2.4$\times$1.8 \\
Beam PA ($^{\circ}$) & 110 & 115 \\
rms(WideX) $^d$ & 3.50 & 4.02 \\
rms(Cont.) $^e$ & 1.84 & 10.8 \\
Flux (Jy) $^f$& 0.19 & 1.6 \\
Size (\arcsec) $^f$ & 1.9$\times$1.7 & 2.5$\times$1.6 \\
PA ($^{\circ}$) $^f$ & -67 & -44 \\
$M$ ($M_{\odot}$) $^g$ & 0.4 & 3.5 \\
$N$(H$_2$) (cm$^{-2}$) $^g$ & $4.5 \times 10^{24}$ & $3.1 \times 10^{25}$ \\
\hline
\end{tabular}
\end{footnotesize}
\tablecomments{
$a$: \citet{Hirota2008}; $b$: \citet{Karska2013}; $c:$
\citet{Kristensen2012};
$d$: Units of mJy/beam/channel for a channel width of 1.95 MHz. 
$e$: Units of mJy/beam.
$f$: Continuum integrated fluxes and sizes were obtained from elliptical 
Gaussian fits in the ($u,v$) plane (i.e., deconvolved FWHM size);
$g$: The envelope mass and averaged column density were derived from
the continuum fluxes obtained within the deconvolved FWHM size (see
text for more details).}
\label{prop_cont}
\end{table}

\section{Results}

\subsection{Continuum Maps}

\begin{figure}[htp]
\centering
\includegraphics[width=80mm]{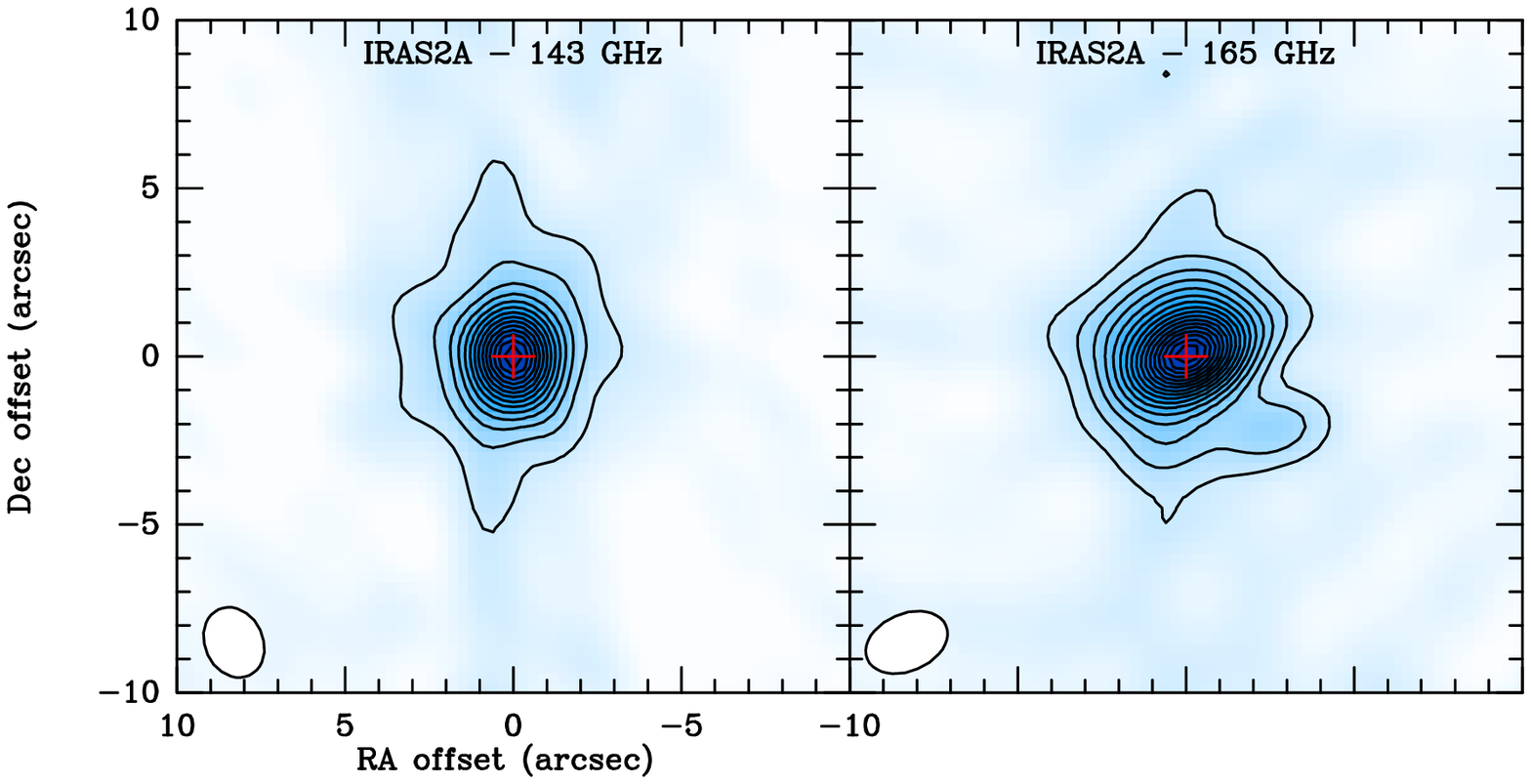}
\includegraphics[width=80mm]{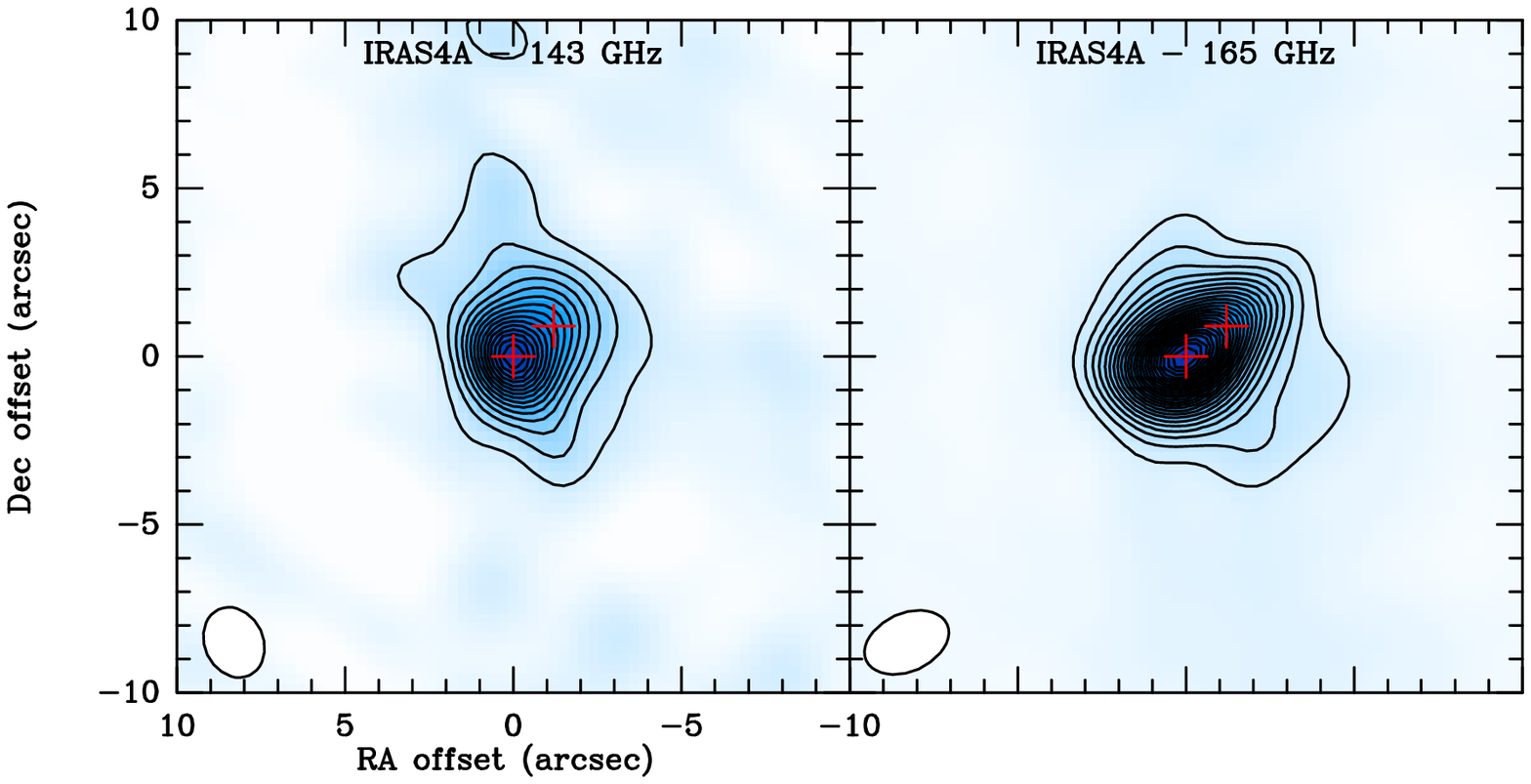}
\caption{Continuum maps at 145 and 165 GHz of IRAS2A (rms of $1.56$
  and $1.84$ mJy/beam, respectively) and IRAS4A (rms of $10.8$ and $10.8$
  mJy/beam, respectively) . The positions of the sources IRAS2A,
  IRAS4A-NW, and IRAS4A-SE are marked by a red plus sign. Contour
  levels are in steps of 3$\sigma$. 
The white ellipse represents the half-power beamwidth of the
  synthesized beam. } 
\label{maps_cont}
\end{figure}

Figure \ref{maps_cont} shows the maps of the
continuum emission of IRAS2A and IRAS4A at 143 and 165 GHz obtained
after natural weighted cleaning.  
Parameters of the continuum emission (integrated flux and 
deconvolved FWHM size), obtained from elliptical Gaussian fits in the
($u,v$) plane, are given in Table \ref{prop_cont}.  
For the two settings, the FWHM size of the continuum emission is
slightly smaller than the size of the synthesized 
beam, the continuum emission is consequently not resolved.
In particular, IRAS4A is known to be a binary system with a 1.8 
$\arcsec$ separation \citep{Looney2000}, as depicted by the two red crosses in
Fig. \ref{maps_cont} that indicate the positions of IRAS4A-SE and
-NW. Although the continuum emission of IRAS4A is peaked at the 
southeast (SE) position rather than at the northwest (NW) position for
the two settings, we cannot resolve the two sources.

We estimated the mass of the envelope and the H$_2$ column density
from the continuum fluxes and the sizes derived from elliptical
Gaussian fits listed in Table \ref{prop_cont}. Assuming an optically
thin dust emission, the mass $M$ of the envelope is given by
\begin{equation}
M = \frac{S_{\nu} d^2}{K_{\nu} B_{\nu}(T_d) R_d}
\end{equation}
where $S_{\nu}$ is the continuum flux integrated over the gaussian
ellipse and listed in Table \ref{prop_cont}, $d$ is the distance to 
the two low-mass protostars \citep[235pc;][]{Hirota2008}, $B_{\nu}(T_d)$ 
is the Planck black body function for a temperature $T_d$ assumed to be 
30 K, $R_d$ is the dust-to-gas mass ratio equal to 0.01. 
$K_{\nu}$ is the opacity per dust mass taken from column 5 in Table 5 of
\citet{Ossenkopf1994} (corresponding to grains showing a MRN size
distribution covered by thin ice mantles at $n_{\textrm{H}} = 10^6$
cm$^{-3}$) and extrapolated to 145 GHz ($K_{\nu} = 0.38$ g/cm$^2$) and
165 GHz ($K_{\nu} = 0.48$ g/cm$^2$).
The column density of H$_2$ averaged over the gaussian ellipse can be deduced from $M$ following this formula
\begin{equation}
N(\textrm{H}_2) = \frac{M}{\mu m_{\textrm{H}} \Omega d^2}
\end{equation}
where $M$ is the envelope mass, $\mu$ = 2.38 is the mean molecular mass 
in units of hydrogen atom masses, $m_{\textrm{H}}$ is the hydrogen atom 
mass, and $\Omega$ the solid angle subtended by the gaussian ellipse.
The envelope mass $M$ and H$_2$ column density $N$(H$_2$) derived for the
two frequencies are listed in Table \ref{prop_cont}. 

\subsection{Spectra}

\begin{figure*}[htp]
\centering
\includegraphics[width=\columnwidth]{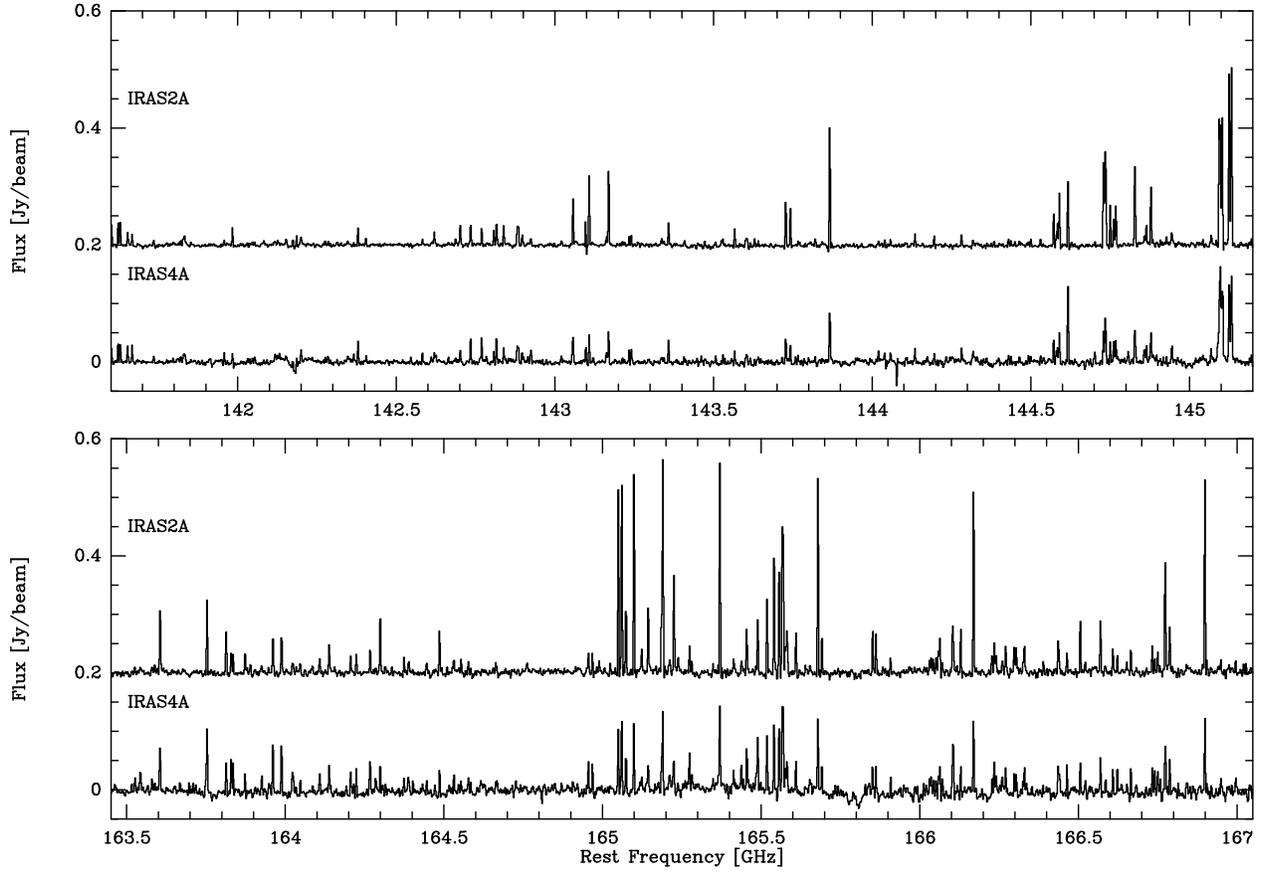}
\caption{PdBi continuum-subtracted spectra of the WideX backends
  around 143 and 165 GHz towards the peak position of IRAS2A and
    IRAS4A-NW. The rms noise levels
  are 2.57 and 3.50 mJy/beam/channel at 143 GHz and 165 GHz towards IRAS2A and
  3.34 and 4.02 mJy/beam/channel at 143.5 GHz and 165 GHz towards
  IRAS4A, respectively.} 
\label{widex_spectra}
\end{figure*}

\begin{figure}[htp]
\centering
\includegraphics[height=45mm]{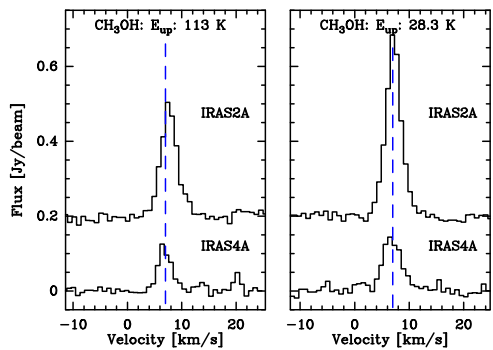}
\caption{PdBi spectra of methanol obtained with the narrow band
  correlators towards the peak position of IRAS2A and
    IRAS4A-NW The rms noise levels are 7.80 and 7.86
  mJy/beam/channel towards IRAS2A and IRAS4A, respectively.}  
\label{narrow_meth}
\end{figure}

\begin{figure*}[htp]
\centering
\includegraphics[height=45mm]{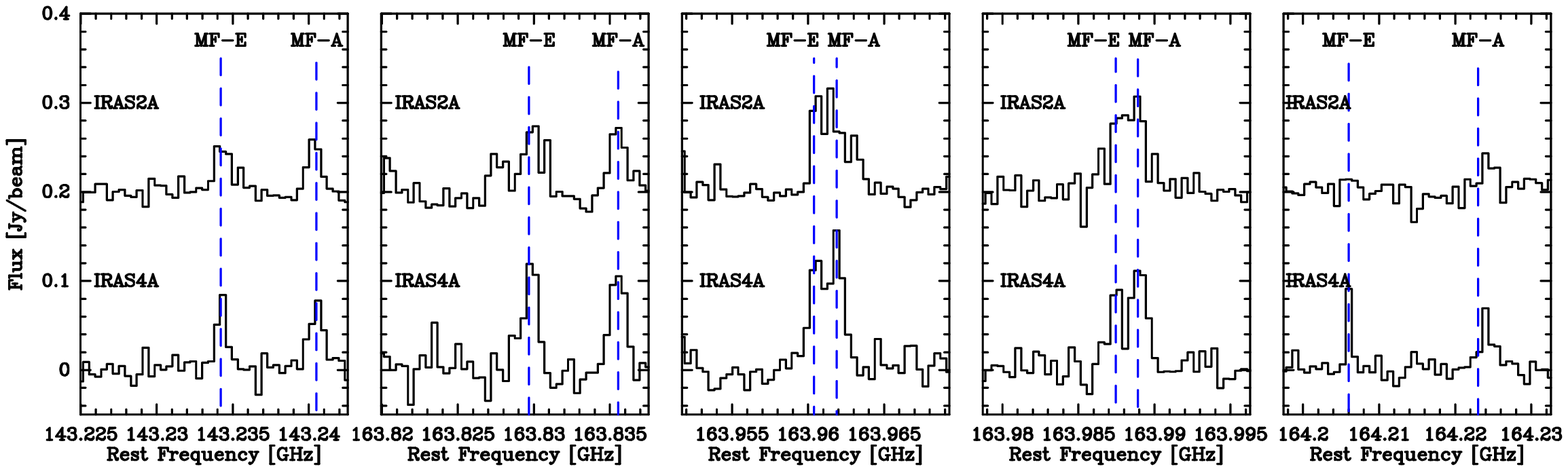} 
\caption{PdBi spectra of methyl formate 
  obtained with the narrow band correlators towards the peak position
  of IRAS2A and IRAS4A-NW. The rms noises   are about 7-10
  mJy/beam/channel towards IRAS2A and IRAS4A.}  
\label{narrow_mf}
\end{figure*}

For the two sources and for all the settings, we obtained the spectral
cubes by subtracting the continuum visibilities to the whole 
(line+continuum) datacube. For IRAS4A, the baseline has been flattened
by importing the data cubes into CLASS and subtracting a polynomial
function to each individual spectrum. 
The 3.6 GHz wide spectra obtained at the coordinates of IRAS2A and
IRAS4A-NW with the two WideX backends are presented in Figure 
\ref{widex_spectra}. We also present the narrow band spectra of
CH$_3$OH and HCOOCH$_3$ in Figures \ref{narrow_meth}
and \ref{narrow_mf}. The 1$\sigma$ rms noise in the line-free channels
of all spectra are given in the caption of Fig. \ref{widex_spectra}
and \ref{narrow_meth} and \ref{narrow_mf}.
The two sources are chemically rich since the WideX spectra towards IRAS2A and
IRAS4A-NW display $\sim 200$ and $\sim 170$ lines detected with a
signal-to-noise ratio higher than 3, resulting in a line
density of 28 and 23 detected lines per GHz, respectively. 
For comparison, \citet{Maury2014} detected 86 lines above the 3-sigma
level between 216.9 and 220.5 GHz towards IRAS2A with the PdBi,
resulting in a similar line density of 24 detected lines per GHz. 

The line identification was carried out using the JPL
\citep{Pickett1998} and the CDMS \citep{Muller2005} spectroscopic
catalogues. 
Line identifications were performed by eye, by taking into account the
upper energy level, the line strength, and the velocity of each
transition. 
The detected molecules are listed in Table 2 with
the number of detected transitions, the energy range of their upper
energy levels, and the spectroscopic reference. 
We detected about 35 transitions for the main isotopologue of methanol
with upper energy levels up to $\sim 1020$ K and about 13 lines for its
isotopologue $^{13}$CH$_3$OH. 
In addition to methanol, we report the first multi-line detection of glycol
aldehyde HCOCH$_2$OH, ethyl cyanide C$_2$H$_5$CN, and ethanol
C$_2$H$_5$OH towards low-mass protostars other than IRAS 16293
\footnote{ The detection of glycolaldehyde was reported in IRAS2A by
  Coutens et al. (2015) while this article was in the review process.}. We
also detected several transitions originating from methyl formate
HCOOCH$_3$, di-methyl ether CH$_3$OCH$_3$, and methyl cyanide CH$_3$CN
towards the two sources.  
Due to the low spectral resolution, we carefully checked that the
detected lines do not suffer from any blending with other transitions
from similar or other molecules by using the JPL and CDMS databases
but also with the Splatalogue database \footnote{http://www.cv.nrao.edu/php/splat/}.    
In total, about 70 \% of the lines detected at a 3 sigma level in the
two 3.6 GHz WideX correlators are attributed to the complex organics
listed in Table 2. Other lines are attributed to
the deuterated methanol isotopologues CH$_2$DOH, CH$_3$OD, CHD$_2$OH, 
other deuterated molecules such as HDO \citep[studied in a previous
work;][]{Taquet2013b}, DCN, NH$_2$D, and DCO$^+$ (in absorption) and
the sulphur-bearing species SO$_2$, and C$^{34}$S. The analysis of the
deuterated methanol transitions will be published in a separate article. 
$\sim$20 lines are unidentified.

\begin{table}[htp]
\begin{footnotesize}
\centering
\caption{List of molecules detected towards IRAS2A and IRAS4A.}
\begin{tabular}{l c c c c c}
\hline
\hline
	&	\multicolumn{2}{c}{IRAS2A}			&	\multicolumn{2}{c}{IRAS4A}			&	Ref.	\\
\cline{2-5}											
Molecule	&	$N_{\textrm{lines}}$	&	$E_{\textrm{up}}$	&	$N_{\textrm{lines}}$	&	$E_{\textrm{up}}$	&		\\
	&		&	(K)	&		&	(K)	&		\\
\hline											
CH$_3$OH	&	34	&	14 - 1022	&	35	&	14 - 1022	&	1	\\
HCOOCH$_3$	&	20	&	43 - 248	&	20	&	43 - 237	&	2	\\
CH$_2$DOH	&	13	&	33 - 230	&	13	&	33 - 230	&	3	\\
$^{13}$CH$_3$OH	&	13	&	14 - 222	&	12	&	14 - 222	&	4	\\
CH$_3$OCH$_3$	&	8	&	11 - 314	&	7	&	11 - 314	&	5	\\
C$_2$H$_5$OH	&	8	&	37 - 216	&	7	&	37 - 216	&	6	\\
CH$_3$CN	&	7	&	40 - 390	&	7	&	40 - 390	&	7	\\
CHD$_2$OH	&	6	&	20 - 67	&	6	&	20 - 67	&	8	\\
HCOCH$_2$OH	&	4	&	53 - 68	&	7	&	53 - 177	&	9	\\
SO$_2$	&	4	&	24 - 102	&	4	&	24 - 102	&	10	\\
C$_2$H$_5$CN	&	4	&	63 - 130	&	3	&	63 - 130	&	11	\\
CH$_3$OD	&	1	&	40	&	1	&	40	&	12	\\
H$_2$$^{13}$CO	&	1	&	10	&	1	&	10	&	13	\\
H$_2$C$^{18}$O	&	1	&	22	&	1	&	22	&	14	\\
HC$_3$N	&	1	&	75	&	1	&	75	&	15	\\
NH$_2$CHO	&	1	&	30	&	1	&	30	&	16	\\
CH$_2$CO	&	1	&	41	&	1	&	41	&	17	\\
C$^{34}$S	&	1	&	14	&	1	&	14	&	18	\\
DC$_3$N	&	1	&	62	&	-	&	-	&	15	\\
HDO	&	1	&	319	&	1	&	319	&	19	\\
DCN	&	1	&	10	&	1	&	10	&	20	\\
NH$_2$D	&	1	&	183	&	1	&	183	&	21	\\
D$_2$CO	&	1	&	21	&	1	&	21	&	14	\\
\hline
\end{tabular}
\tablecomments{
1:	\citet{Xu2008};
2:	\citet{Ilyushin2009};
3:	\citet{Pearson2012};
4:	\citet{Xu1997};
5:	\citet{Lovas1979};
6:	\citet{Endres2009};
7:	\citet{Cazzoli2006};
8:	\citet{Parise2002};
9:	\citet{Carrol2010};
10:	\citet{Alekseev1996};
11:	\citet{Fukuyama1996};
12:	\citet{Anderson1988};
13:	\citet{Muller2000};
14:	\citet{Dangoisse1978};
15:	\citet{Lafferty1978};
16:	\citet{Johnson1972};
17:	\citet{Fabricant1977};
18:	\citet{Bogey1982};
19:	\citet{Messer1984};
20:	\citet{DeLucia1969};
21:	\citet{Cohen1982};
}
\end{footnotesize}
\label{listmolecules2}
\end{table}

\begin{figure}[h]
\centering
\includegraphics[width=\columnwidth]{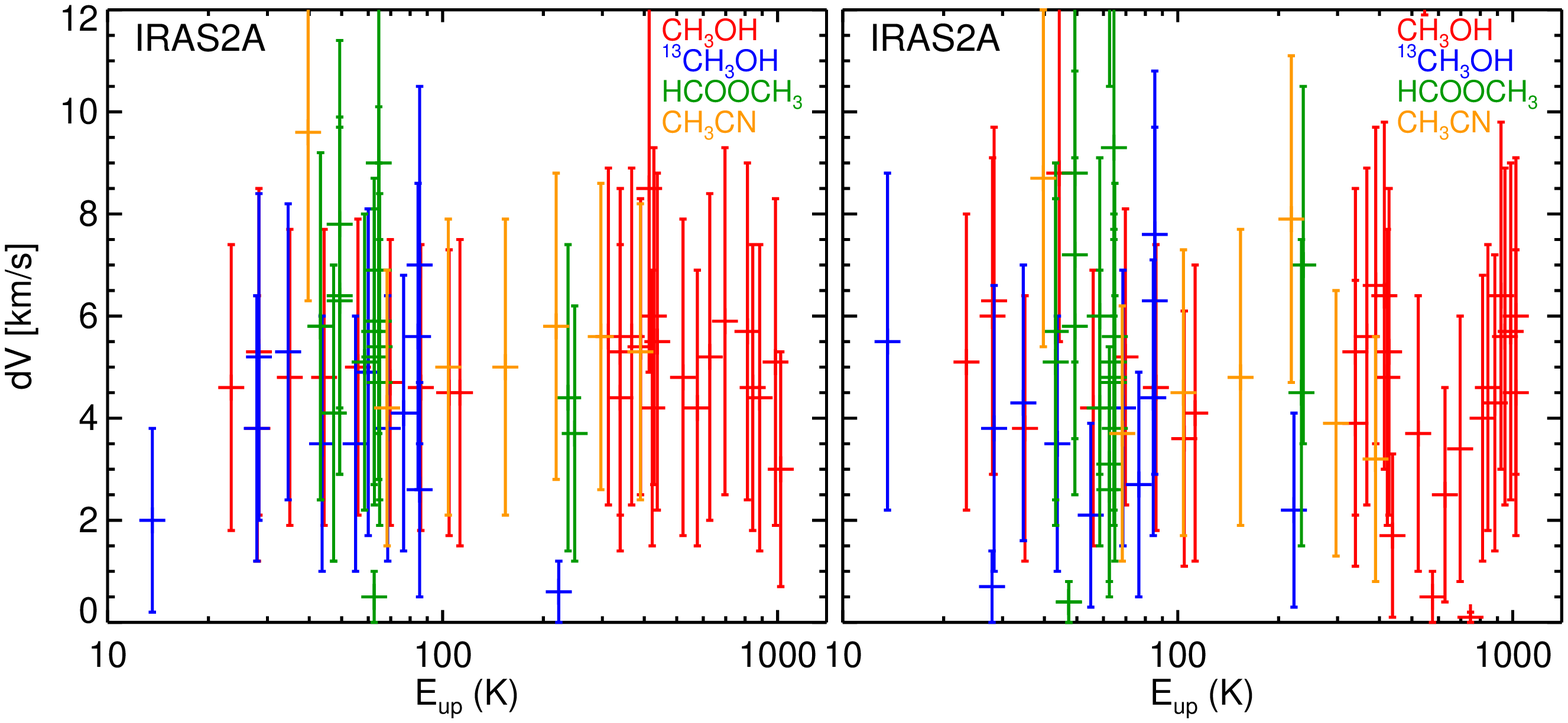}
\includegraphics[width=\columnwidth]{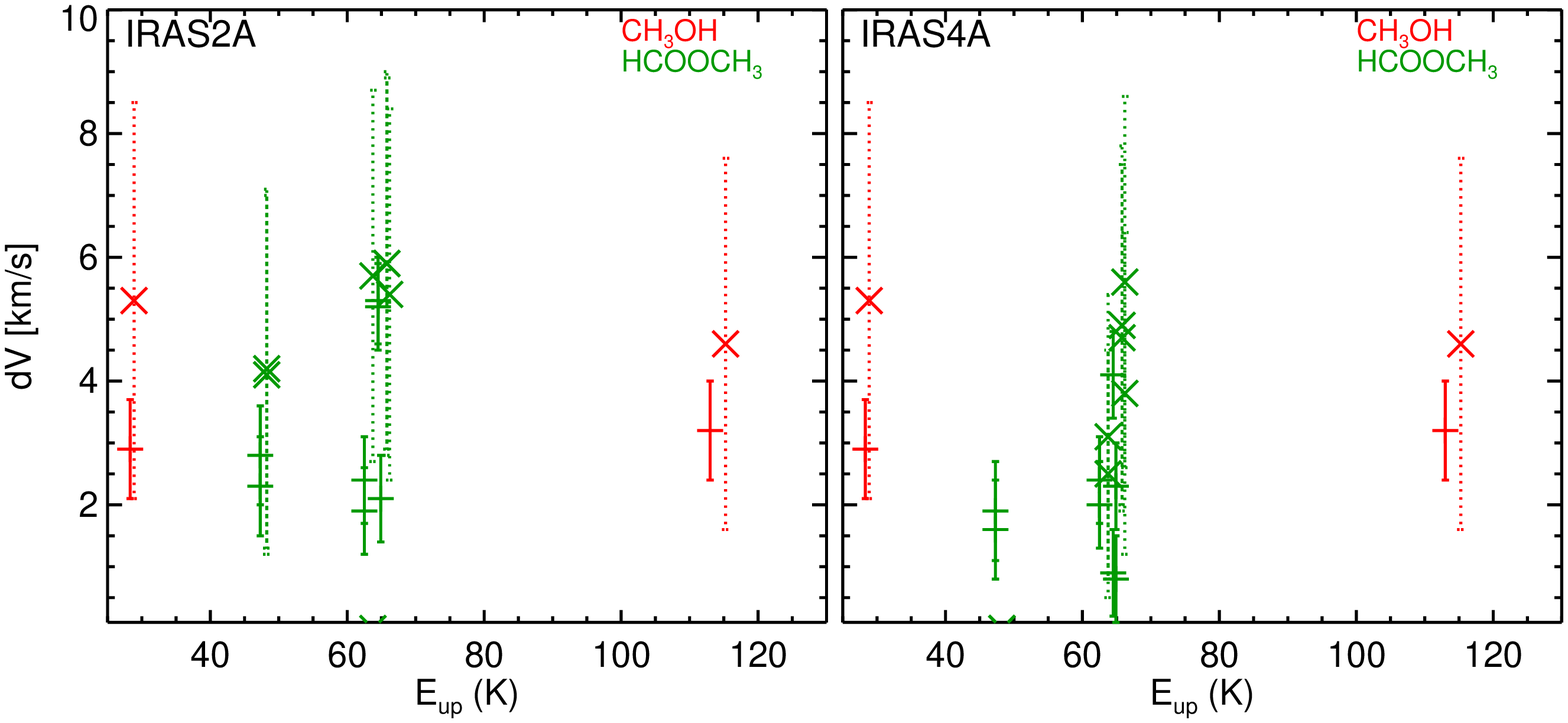}
\caption{
Top) Deconvolved FWHM linewidths of CH$_3$OH, $^{13}$CH$_3$OH,
  HCOOCH$_3$, and CH$_3$CN, obtained from gaussian fits of the WideX
  spectra at the source positions, as a function of $E_{\textrm{up}}$. 
Bottom) Comparison of the FWHM linewidths deduced from the high resolution 
spectra (plus signs with solid error bars) with the deconvolved FWHM
linewidths from the WideX spectra (cross signs with dotted error bars)
for the lines observed with the high spectral resolution correlators. 
} 
\label{linewidth}
\end{figure}

We estimated the Full-Width-at-Half-Maximum (FWHM) of the lines
detected with the narrow band correlators giving a spectral
resolution of 0.4 MHz, and with the WideX correlator providing a spectral
resolution of 1.95 MHz, through a gaussian fit of the spectra obtained
at the coordinates of IRAS2A and IRAS4A-NW. Table \ref{list_dV} lists
the linewidths of the transitions detected with the two correlators. 
The uncertainties in the linewidths are due to the statistical 
errors from the gaussian fit and to the uncertainty from the low spectral 
resolution. In this work, the low spectral resolution dominates the 
uncertainty of the linewidths. 
Tables \ref{lines_ch3oh} to \ref{lines_others} of the Appendix list
the properties of all detected transitions along with the FWHM
linewidths. 
For the two sources, the FWHM of the CH$_3$OH transitions
detected at high spectral resolution are about 3 km/s whilst the widths
of the HCOOCH$_3$ lines vary between 0.9 and 2.8 km/s when the lines
of the -E and the -A states do not overlap. 
The linewidths derived with the PdBi are similar to the linewidths of other
CH$_3$OH and HCOOCH$_3$ transitions derived by \citet{Maret2005} and
\citet{Bottinelli2004a}. 
Due to the low spectral resolution, the linewidths derived from the
WideX correlators result from the convolution of the intrinsic
linewidths with the spectral resolution of 1.95 MHz. 
The top panel of Figure \ref{linewidth} shows the deconvolved FWHM of 
CH$_3$OH, $^{13}$CH$_3$OH, HCOOCH$_3$, and CH$_3$CN, showing a
high number of transitions detected at a high signal-to-noise ratio,
as a function of the energy of the upper level towards the two
sources. 
For this purpose, we excluded several CH$_3$OH and HCOOCH$_3$
transitions that are blended with each other.
No clear trend can be deduced for the two sources. The fluctuation 
of the deconvolved FWHM linewidths between 2 and 8 km/s seems to be 
due to their high uncertainties. 
The bottom panel of Fig. \ref{linewidth} compares the linewidths 
deduced from the narrow band correlators giving a high spectral resolution 
of 0.4 MHz with the deconvolved linewidths from the WideX spectra for the 
lines detected with the narrow band correlators. For all the lines, the 
FWHM linewidth deconvolved from the WideX spectra is higher than the FWHM 
linewidth deduced from the narrow band correlator but the differences 
remain within the uncertainties.


\begin{table}[htp]
\centering
\caption{Linewidths of the transitions detected in the narrow band correlators.}
\begin{tabular}{l c c c c c c}
\hline																					
\hline																					
	&		&		&	\multicolumn{2}{c}{IRAS2A}							&	\multicolumn{2}{c}{IRAS4A}							\\
\cline{4-7}																					
Molecule	&	Frequency	&	$E_{up}$	&	$dV_n$			&	$dV_W$			&	$dV_n$			&	$dV_W$			\\
	&	(GHz)	&	(K)	&	(km/s)			&	(km/s)			&	(km/s)			&	(km/s)			\\
\hline																					
CH$_3$OH	&	143.86580	&	28.3	&	2.9	$\pm$	0.8	&	6.7	$\pm$	4.1	&	3.2	$\pm$	0.8	&	7.5	$\pm$	4.1	\\
CH$_3$OH	&	143.16952	&	113	&	3.2	$\pm$	0.8	&	6.1	$\pm$	4.1	&	2.8	$\pm$	0.8	&	5.8	$\pm$	4.1	\\
HCOOCH$_3$	&	143.23420	&	47.3	&	2.3	$\pm$	0.8	&	5.8	$\pm$	4.1	&	1.6	$\pm$	0.8	&	4.1	$\pm$	4.1	\\
HCOOCH$_3$	&	143.24051	&	47.3	&	2.8	$\pm$	0.8	&	5.8	$\pm$	4.1	&	1.9	$\pm$	0.8	&	4.1	$\pm$	4.1	\\
HCOOCH$_3$	&	163.82968	&	62.5	&	2.4	$\pm$	0.7	&	6.7	$\pm$	3.6	&	2.0	$\pm$	0.7	&	4.7	$\pm$	3.6	\\
HCOOCH$_3$	&	163.83553	&	62.5	&	1.9	$\pm$	0.7	&	3.6	$\pm$	3.6	&	2.4	$\pm$	0.7	&	4.4	$\pm$	3.6	\\
HCOOCH$_3$	&	163.96039	&	64.5	&	5.3	$\pm$	0.7	&	6.9	$\pm$	3.6	&	4.1	$\pm$	0.7	&	5.9	$\pm$	3.6	\\
HCOOCH$_3$	&	163.96188	&	64.5	&				&				&				&				\\
HCOOCH$_3$	&	163.98746	&	64.5	&	5.2	$\pm$	0.7	&	6.9	$\pm$	3.6	&	0.9	$\pm$	0.7	&	6.0	$\pm$	3.6	\\
HCOOCH$_3$	&	163.98891	&	64.5	&			&			&	2.2	$\pm$	0.7	&			\\
HCOOCH$_3$	&	164.20598	&	64.9	&		&	5.9	$\pm$	3.6	&	0.8	$\pm$	0.7	&	6.6	$\pm$	3.6	\\
HCOOCH$_3$	&	164.22382	&	64.9	&	2.1	$\pm$	0.7	&	6.5	$\pm$	3.6	&	2.3	$\pm$	0.7	&	5.2	$\pm$	3.6	\\
\hline																					
																
\end{tabular}
\tablecomments{
$dV_n$ and $dV_W$ are the non-deconvolved linewidths deduced from the spectra obtained with the
narrow band and WideX correlators, respectively.
}
\label{list_dV}
\end{table}

\subsection{Line Maps}

\begin{figure*}[h]
\centering
\includegraphics[width=130mm]{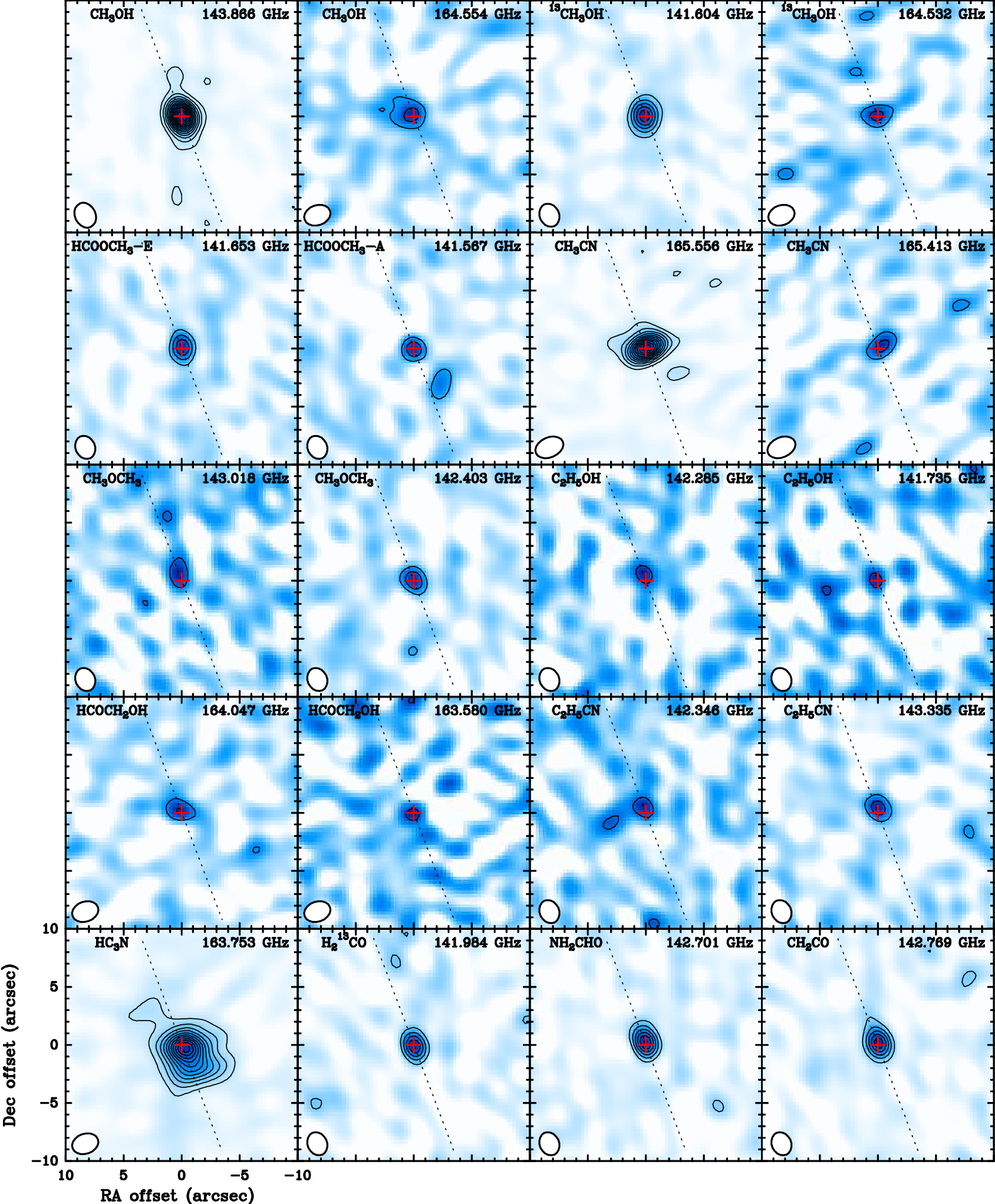}
\caption{Integrated maps of selected lines of complex organic
  molecules towards IRAS2A. For each species
  where several transitions have been detected, a low-excited and a
  high-excited transition are shown. Contour levels are in step of
  3$\sigma$. 1$\sigma$ rms noise levels are the following: 38.5 and 18.2 mJy
  km/s for CH$_3$OH, 13.3 and 28.9 mJy km/s for $^{13}$CH$_3$OH, 13.0
  and 14.6 mJy km/s for for HCOOCH$_3$, 35.3 and 26.5 mJy km/s for
  CH$_3$CN, 15.5 and 9.64 mJy km/s for CH$_3$OCH$_3$, 11.6 and 14.5
  mJy km/s for C$_2$H$_5$OH, 21.7 and 19.5 mJy km/s for HCOCH$_2$OH,
  10.9 and 10.8 mJy km/s for C$_2$H$_5$CN, 24.8 mJy km/s for HC$_3$N,
  13.6 mJy km/s for H$_2^{13}$CO, 17.6 mJy km/s for NH$_2$CHO, 12.0
  mJy km/s for CH$_2$CO. The direction of the outflow is depicted by
  the dotted line.} 
\label{maps_i2a}
\end{figure*}

\begin{figure*}[h]
\centering
\includegraphics[width=130mm]{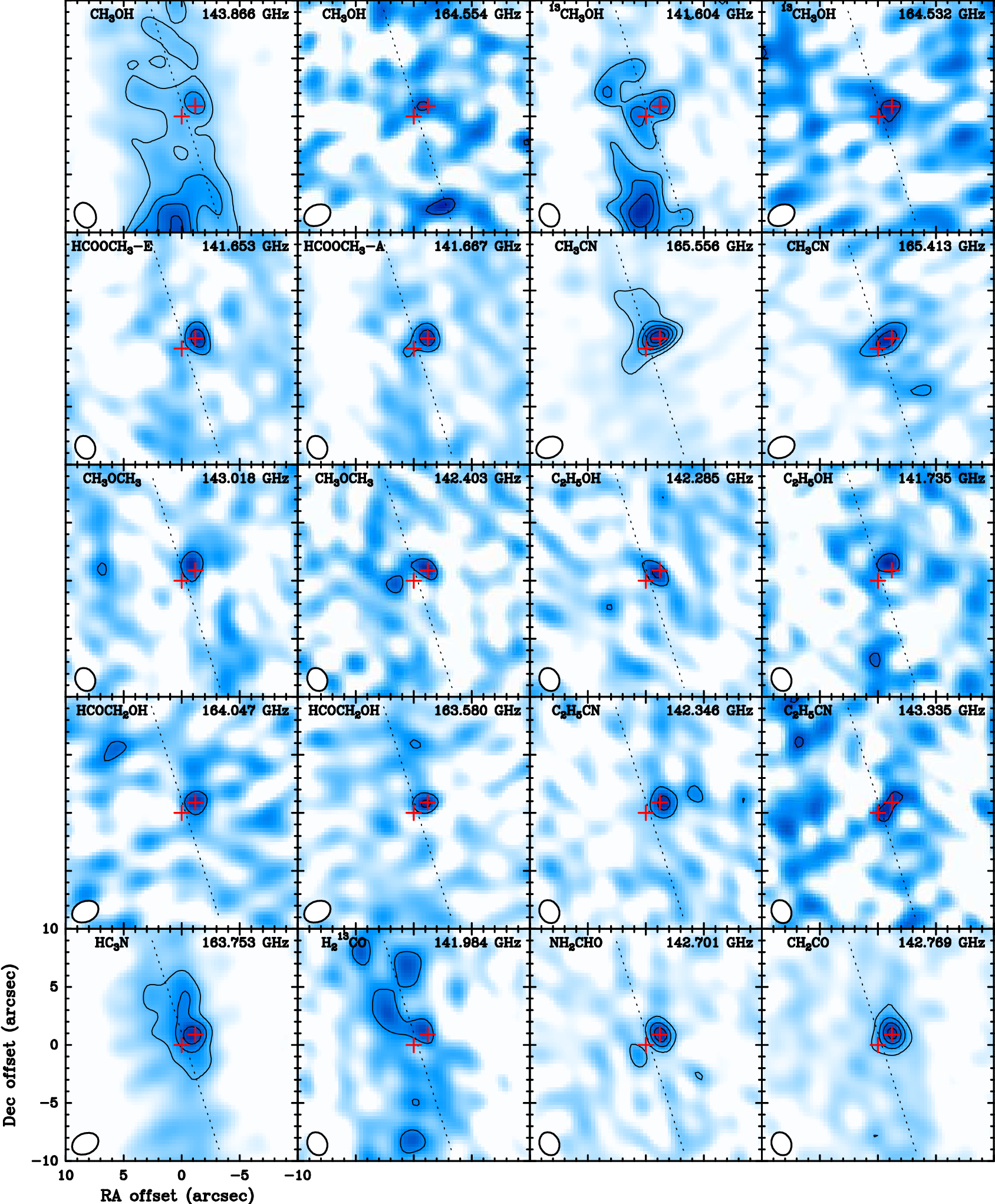}
\caption{Integrated maps of selected lines of complex organic
  molecules towards IRAS4A. For each species
  where several transitions have been detected, a low-excited and a
  high-excited transition are shown. Contour levels are in step of
  3$\sigma$. 1$\sigma$ rms noise levels are the following: 81.9 and 22.8 mJy
  km/s for CH$_3$OH, 27.4 and 32.9 mJy km/s for $^{13}$CH$_3$OH, 25.4
  and 24.5 mJy km/s for for HCOOCH$_3$, 43.6 and 24.0 mJy km/s for
  CH$_3$CN, 16.9 and 11.4 mJy km/s for CH$_3$OCH$_3$, 12.4 and 19.8
  mJy km/s for C$_2$H$_5$OH, 18.8 and 17.2 mJy km/s for HCOCH$_2$OH,
  11.6 and 16.0 mJy km/s for C$_2$H$_5$CN, 109 mJy km/s for HC$_3$N,
  27.8 mJy km/s for H$_2^{13}$CO, 14.3 mJy km/s for NH$_2$CHO, 22.4
  mJy km/s for CH$_2$CO. The direction of the outflow is depicted by
  the dotted line.} 
\label{maps_i4a}
\end{figure*}

For all the transitions, the interferometric maps of the IRAS2A
and IRAS4A protostars have been obtained
by integrating the flux over $V_{\textrm{LSR}} \pm \Delta V$ where
$V_{\textrm{LSR}}$ is the system velocity of the source and $\Delta
V = 3$ km/s following the FWHM linewidths of the CH$_3$OH transitions
listed in Table \ref{list_dV}. In practice, due to the low resolution of the
WideX backends, the line emission is integrated over 3 channels. 
Figures \ref{maps_i2a} and \ref{maps_i4a} show a compilation of the
integrated line maps towards IRAS2A and IRAS4A obtained after natural
weighted cleaning. For species where several transitions were
detected, two maps showing a  low-energy and a high-energy
transition are presented. 
Tables \ref{lines_ch3oh} to \ref{lines_others} of the Appendix list
the properties of all detected transitions along with their FWHM
(full-width-at-half-maximum) sizes of emission and their Position
Angle (PA) derived from the modelling of the visibilities assuming
elliptical gaussians, circular gaussians or point sources if gaussian
fits were not possible.    
   
For the two sources, the emission of most molecules is limited to the 
inner regions near the protostars. 
In IRAS4A, the compact emission of all COMs transitions originates from the 
-NW source although the SE protostar is brighter in the continuum.
In contrast, no molecular lines seem to originate from IRAS4A-SE, as
also observed in previous interferometric observations of H$_2^{18}$O
and other complex organics by \citet{Persson2012}. 
The low-energy transitions of CH$_3$OH (the transition at $\sim
145.094$ GHz including several non-resolved transitions towards IRAS2A 
not shown in this work and all the transitions with $E_{up} \leq
120$ K towards IRAS4A) show an extended emission consistent with the
position of molecular outflows. 
For IRAS2A, the interferometric map of the CH$_3$OH transitions at
$\sim 145.094$ GHz displays a slightly red-shifted emission located 7
$\arcsec$ to the north from IRAS2A, consistent with the direction of an
outflow previously detected at small scales with CO by \citet{Jorgensen2007}.  
For IRAS4A, the emission of the low-energy CH$_3$OH lines extends
in the bipolar outflow along a N-S direction and seems to peak at
$\sim 10$ $\arcsec$ south from the protostars, consistent with the south
lobe detected in SiO by \citet{Choi2005}.  
From Fig. \ref{maps_i4a}, it is clear that the inner hot corino
does not dominate the flux of the weakly excited transitions of
methanol in IRAS4A \citep[see also][]{Maret2005}.  
The emission of the HC$_3$N transition is also spatially resolved and
shows an elongation of its emission towards the SW direction for
IRAS2A and towards the N direction for IRAS4A. 
The spatial distribution and the kinematics of the outflow driven by
IRAS2A and IRAS4A will be analysed in a future work.

The emission originating from all COMs transitions but some
CH$_3$OH and HC$_3$N lines is not spatially resolved by the array
since their FWHM emission size deduced from the fit to the
  visibilities is lower than the synthesized beam of the interferometer.   
Consequently, the source sizes presented in this work can only be used as upper limits.
These observations are qualitatively consistent with previous models and
observations suggesting that methanol and COMs mostly come
from the inner hot corino. In this region, these molecules show a jump
of their abundance when the temperature is higher than the temperature of
ice sublimation ($T \sim 100$ K).   
\citet{Maret2004} estimated a size for the hot corinos of IRAS2A
 and IRAS4A of 0.45 $\arcsec$ by reproducing the formaldehyde
emission observed with single- dish telescopes with an abundance jump
of two orders of magnitude at $r \sim 50$ AU from the central
protostar.
The luminosities assumed for IRAS2A and IRAS4A by \citet{Maret2004}
are lower by a factor of 2.25 and 1.5, respectively, than the
luminosities assumed in this work.  
Assuming that the temperature profile is governed by the Stefan-Boltzmann’s 
law implies a difference in the temperature of a factor of 1.2 and 1.1, 
respectively. The assumption of higher luminosities for the two
sources would therefore increase the size of the corino of a few AU.  
\citet{Maury2014} estimated a FWHM size of 0.4-0.9$\arcsec$ for the hot
corino of IRAS2A through the use of the more extended A configuration of
the PdBi array. 


\section{Chemical Abundances in the Hot Corinos}

\subsection{Rotational Diagrams}

First estimates of the excitation temperatures and the
column densities of observed molecules in the hot corinos of IRAS2A
and IRAS4A have been obtained from the rotational diagram (RD)
analysis by assuming optically thin emission and a LTE population of
the levels.  
Since the emission of most transitions is not spatially resolved with
the PdBi, we measured the flux of all transitions originating from a
circular mask with a diameter equal to the major axis of the synthesized beam
size of the telescope ($\sim$2.1-2.6$\arcsec$).  
Tables \ref{lines_ch3oh} to \ref{lines_others} of the Appendix list
the measured flux for all transitions and in the two sources.   
Astrochemical/dynamical models predict low abundances of COMs in the
the dense regions of protostellar envelopes where the dust temperature
is lower than the temperature of ice sublimation $T_{\textrm{ev}}$
($\sim 100$ K) due to the efficient depletion, while the abundance
profiles show a strong jump once $T=T_{\textrm{ev}}$
\citep[see][]{Aikawa2008, Taquet2014} induced by thermal evaporation
in the hot corino. 
We therefore assumed that all the flux measured in the
$\sim 2 \arcsec$ mask comes from the hot corino region. 
Although the emission sizes of the COMs transitions are not
necessarily similar \citep[in IRAS 16293 for example, methyl formate has a source
size of 4 \arcsec~whereas formamide has a source size of 2
\arcsec][]{Jaber2014} likely due to their different binding energies,
we assumed a hot corino size $\theta_s$ of 0.5 $\arcsec$ for all the COMs
observed towards the two sources \citep{Maret2004}. 
The linewidth at FWHM $\Delta V$ is fixed to 3 km/s, which represents
an average value of Table \ref{list_dV} and previous observations by
\citet{Maret2005} and \citet{Bottinelli2007}.  
%
For molecules with only a few lines detected within a narrow range of
upper energy levels (glycolaldehyde, or ethyl cyanide) or for
molecules showing only one detection, we assumed two values for the
rotational temperature $T_{\textrm{rot}}$: $T_{\textrm{rot}} = T_{\textrm{rot}}$(CH$_3$OH) and
$T_{\textrm{rot}} = T_{\textrm{evap}}$(ice) = 100 K.  
As already shown in several published observational works studying the
emission of methanol towards high-mass and low-mass hot cores
\citep{Parise2006, Bisschop2007, Isokoski2013, Zapata2013}, it is
likely that low-energy transitions of CH$_3$OH are optically thick
towards the center of protostellar envelopes, giving rise to an
underestimation of their population in the rotational
diagrams. Consequently, we excluded the CH$_3$OH transitions with
$E_{\textrm{up}} \leq 200$ K from our RD analysis. 
The rotational temperatures and total column densities of all species
derived towards the two sources are summarized in Table
\ref{recap_RD}. 

For most molecules, observational data can be reasonably well fitted
by a straight line with some scattering, likely due to opacity or non-LTE effects.  
For most of species, the column densities derived towards IRAS2A and
IRAS4A are very similar. Column densities of
CH$_3$OH are about $6 - 12 \times 10^{17}$ cm$^{-2}$ whilst the column
densities of complex organics range between $6 \times 10^{14}$ and $6
\times 10^{16}$ cm$^{-2}$. 
The rotational temperatures derived for most COMs are generally higher than 100
K in IRAS2A and in IRAS4A. 
Although they do not not necessarily reflect the kinetic temperatures, the high
rotational temperatures found in the inner protostellar envelope are
in good agreement with the kinetic temperatures expected in hot
corinos ($T>100$ K).
The CH$_3$CN population distribution can be used to estimate the
kinetic temperature of the warm inner envelope because the CH$_3$CN
population distribution over the different $K$ ladders observed for $J
= 9$ can only be modified by collisions as radiative transitions are
prohibited \citep[see][for a more detailed discussion of the CH$_3$CN
population in hot cores]{Wang2010}.   
If the energy states are thermalized and the transitions are optically
thin, the kinetic temperatures within the hot corinos of IRAS2A and
IRAS4A would be close to 290 and 360 K, respectively. 
As we will see in the next section, the temperatures are probably
overestimated since the CH$_3$CN transitions are likely optically thick.


We derived a $^{12}$C/$^{13}$C abundance ratio of CH$_3$OH 
of 26 and 12 in IRAS2A and IRAS4A, respectively. The two values are
lower than the $^{12}$C/$^{13}$C abundance ratio of $\sim 70$ expected
in the local ISM \citep{Boogert2002, Milam2005, Wirstrom2011} by a
factor 2.7 and 5.8, respectively.  
We verified that using the same rotational temperature for CH$_3$OH and
$^{13}$CH$_3$OH only modifies the $^{12}$C/$^{13}$C abundance ratios
by a few percents at most.
The low $^{12}$C/$^{13}$C abundance ratio might be due to the
different ranges of excitation of the observed transitions used to
derive the column densities. We used excited transitions (with
$E_{\textrm{up}} > 200$ K) to derive $N$($^{12}$CH$_3$OH) whereas
only weakly excited $^{13}$CH$_3$OH transitions (with $E_{\textrm{up}} <
225$ K) have been detected.
The low ratio measured in the two protostars suggests that excited
transitions of CH$_3$OH ($E_{\textrm{up}} > 200$ K) are also
optically thick. An overestimation of the $^{13}$CH$_3$OH column
density by the RD best fit, due to the large uncertainty on the fluxes,
is also possible. An analysis taking the opacities into account is
therefore required to clarify this issue. 

\begin{table*}[htp]
\centering
\caption{Results from the rotational diagram analysis.}
\begin{tiny}
\begin{tabular}{l c c c c c c}
\hline																														
\hline																														
Molecule	&		$N_{hc}$	$^a$						&	$T_{\textrm{rot}}$			&		$X$ $^b$						&	$X_{\textrm{meth}}$ $^c$			&	$N_{hc}$(SD) $^d$	&	$X_{\textrm{meth}}$(SD) $^d$	\\
	&		(cm$^{-2}$)							&	(K)			&								&	(\%)			&	(cm$^{-2}$)	&	(\%)	\\
\hline																														
	&										\multicolumn{4}{c}{IRAS2A}																			\\
\hline																														
CH$_3$OH $^e$	&	$(	1.2	\pm	0.4	)	\times	10^{18}$		&	179	$\pm$	62	&	$(	2.5	\pm	0.9	)	\times	10^{-7}$	&				&	$1.4\times 10^{17}$	&		\\
$^{13}$CH$_3$OH	&	$(	4.8	\pm	1.3	)	\times	10^{16}$		&	164	$\pm$	43	&	$(	9.6	\pm	2.5	)	\times	10^{-9}$	&				&		&		\\
HCOOCH$_3$	&	$(	6.4	\pm	1.9	)	\times	10^{16}$		&	200	$\pm$	61	&	$(	1.3	\pm	0.4	)	\times	10^{-8}$	&	1.9	$\pm$	0.8	&	$< 1.2\times 10^{17}$	&	$< 85$	\\
CH$_3$CN	&	$(	1.0	\pm	0.2	)	\times	10^{16}$		&	289	$\pm$	63	&	$(	2.0	\pm	0.4	)	\times	10^{-9}$	&	0.30	$\pm$	0.10	&	$1.5\times 10^{15}$	&	1	\\
CH$_3$OCH$_3$	&	$(	4.1	\pm	1.6	)	\times	10^{16}$		&	154	$\pm$	62	&	$(	8.2	\pm	3.3	)	\times	10^{-9}$	&	1.2	$\pm$	0.6	&	$< 7.2\times 10^{16}$	&	$<53$	\\
C$_2$H$_5$OH	&	$(	5.1	\pm	2.2	)	\times	10^{16}$		&	325	$\pm$	140	&	$(	1.0	\pm	0.4	)	\times	10^{-8}$	&	1.5	$\pm$	0.8	&	/	&	/	\\
HCOCH$_2$OH	&	$	7.8				\times	10^{15}$		&	179	$^f$		&	$	1.6				\times	10^{-9}$	&	0.23	$\pm$	0.06	&	/	&	/	\\
	&	$	2.5				\times	10^{15}$		&	100	$^f$		&	$	5.0				\times	10^{-10}$	&	0.074	$\pm$	0.019	&		&		\\
C$_2$H$_5$CN	&	$	1.2				\times	10^{15}$		&	179	$^f$		&	$	2.4				\times	10^{-10}$	&	0.036	$\pm$	0.010	&	$< 1.7\times 10^{16}$	&	$<13$	\\
	&	$	6.9				\times	10^{14}$		&	100	$^f$		&	$	1.4				\times	10^{-10}$	&	0.021	$\pm$	0.005	&		&		\\
HC$_3$N	&	$	7.0				\times	10^{14}$		&	179	$^f$		&	$	1.4				\times	10^{-10}$	&	0.021	$\pm$	0.005	&	/	&	/	\\
	&	$	7.1				\times	10^{14}$		&	100	$^f$		&	$	1.4				\times	10^{-10}$	&	0.021	$\pm$	0.006	&		&		\\
H$_2$$^{13}$CO	&	$	6.6				\times	10^{15}$		&	179	$^f$		&	$	1.3				\times	10^{-9}$	&	0.20	$\pm$	0.05	&	/	&	/	\\
	&	$	2.1				\times	10^{15}$		&	100	$^f$		&	$	4.3				\times	10^{-10}$	&	0.063	$\pm$	0.017	&		&		\\
NH$_2$CHO	&	$	1.2				\times	10^{16}$		&	179	$^f$		&	$	2.3				\times	10^{-9}$	&	0.35	$\pm$	0.09	&	/	&	/	\\
	&	$	4.3				\times	10^{15}$		&	100	$^f$		&	$	8.7				\times	10^{-10}$	&	0.13	$\pm$	0.03	&		&		\\
CH$_2$CO	&	$	7.0				\times	10^{15}$		&	179	$^f$		&	$	1.4				\times	10^{-9}$	&	0.21	$\pm$	0.05	&	/	&	/	\\
	&	$	2.6				\times	10^{15}$		&	100	$^f$		&	$	5.2				\times	10^{-10}$	&	0.077	$\pm$	0.020	&		&		\\
\hline																														
	&															\multicolumn{4}{c}{IRAS4A}														\\
\hline																														
CH$_3$OH $^e$	&	$(	6.3	\pm	3.1	)	\times	10^{17}$		&	300	$\pm$	151	&	$(	1.7	\pm	0.9	)	\times	10^{-8}$	&				&	$2.0\times 10^{17}$	&		\\
$^{13}$CH$_3$OH	&	$(	5.1	\pm	1.5	)	\times	10^{16}$		&	197	$\pm$	56	&	$(	1.4	\pm	0.4	)	\times	10^{-9}$	&				&		&		\\
HCOOCH$_3$	&	$(	5.2	\pm	3.3	)	\times	10^{16}$		&	141	$\pm$	90	&	$(	1.4	\pm	0.9	)	\times	10^{-9}$	&	1.5	$\pm$	1.0	&	$1.0\times 10^{17}$	&	52	\\
CH$_3$CN	&	$(	6.5	\pm	2.9	)	\times	10^{15}$		&	360	$\pm$	162	&	$(	1.8	\pm	0.8	)	\times	10^{-10}$	&	0.18	$\pm$	0.10	&	$2.6\times 10^{15}$	&	1	\\
CH$_3$OCH$_3$	&	$(	3.1	\pm	1.0	)	\times	10^{16}$		&	86	$\pm$	27	&	$(	8.5	\pm	2.6	)	\times	10^{-10}$	&	0.87	$\pm$	0.37	&	$< 4.5\times 10^{16}$	&	$< 22$	\\
C$_2$H$_5$OH	&	$(	4.4	\pm	1.4	)	\times	10^{16}$		&	221	$\pm$	69	&	$(	1.2	\pm	3.7	)	\times	10^{-9}$	&	1.2	$\pm$	0.5	&	/	&	/	\\
HCOCH$_2$OH	&	$(	8.9	\pm	3.4	)	\times	10^{15}$		&	124	$\pm$	48	&	$(	2.4	\pm	0.9	)	\times	10^{-10}$	&	0.25	$\pm$	0.12	&	/	&	/	\\
C$_2$H$_5$CN	&	$	2.3				\times	10^{15}$		&	300	$^f$		&	$	6.2				\times	10^{-11}$	&	0.064	$\pm$	0.018	&	$1.9\times 10^{15}$	&	$< 0.92$	\\
	&	$	8.2				\times	10^{14}$		&	100	$^f$		&	$	2.2				\times	10^{-11}$	&	0.023	$\pm$	0.007	&		&		\\
HC$_3$N	&	$	6.7				\times	10^{14}$		&	300	$^f$		&	$	1.8				\times	10^{-11}$	&	0.019	$\pm$	0.005	&	/	&	/	\\
	&	$	6.8				\times	10^{14}$		&	100	$^f$		&	$	1.8				\times	10^{-11}$	&	0.019	$\pm$	0.005	&		&		\\
H$_2$$^{13}$CO	&	$	5.3				\times	10^{15}$		&	300	$^f$		&	$	1.4				\times	10^{-10}$	&	0.15	$\pm$	0.04	&	/	&	/	\\
	&	$	1.1				\times	10^{15}$		&	100	$^f$		&	$	3.1				\times	10^{-11}$	&	0.032	$\pm$	0.009	&		&		\\
NH$_2$CHO	&	$	8.5				\times	10^{15}$		&	300	$^f$		&	$	2.3				\times	10^{-10}$	&	0.24	$\pm$	0.07	&	/	&	/	\\
	&	$	2.1				\times	10^{15}$		&	100	$^f$		&	$	5.7				\times	10^{-11}$	&	0.059	$\pm$	0.017	&		&		\\
CH$_2$CO	&	$	1.6				\times	10^{16}$		&	300	$^f$		&	$	4.2				\times	10^{-10}$	&	0.43	$\pm$	0.12	&	/	&	/	\\
	&	$	4.0				\times	10^{15}$		&	100	$^f$		&	$	1.1				\times	10^{-10}$	&	0.11	$\pm$	0.03	&		&		\\
\hline																														
\end{tabular}
\label{recap_RD}
\tablecomments{ \\
a: Column densities averaged over a source size of 0.5 \arcsec (see text). 
b: The abundances relative to H$_2$ are obtained from $N$(H$_2$)
derived at 145 GHz in Table \ref{prop_cont} assuming an
homogeneous H$_2$ column density within the beam.  
c: The abundance ratios are relative to the $^{13}$CH$_3$OH column
density multiplied by 70 (see text). 
d: Column densities derived from previous single-dish observations and
scaled to a source size of 0.5 \arcsec. CH$_3$OH: \citet{Maret2005},
other COMs in IRAS2A: \citet{Bottinelli2007}, other COMs in IRAS4A:
\citet{Bottinelli2004a}. 
e: $N_{hc}$ and $T_{\textrm{rot}}$ have been derived from the rotation diagram
neglecting the transitions with $E_{up} < 105$ K (see text). 
f: When they could not be derived from the rotation diagram, the
rotational temperatures have been assumed to be equal to the
temperature of evaporation of water ice and the rotational temperature
of CH$_3$OH.  
}
\end{tiny}
\end{table*}

\subsection{Population Diagrams}

We used the so-called Population Diagram (PD) analysis following
the method described by \cite{Goldsmith1999} to investigate the
effect of optical depth on the column densities of each level. We
applied the PD analysis where four or more transitions were detected
for each species. 
Briefly, the PD analysis includes the influence of optical depths 
on the level populations assumed to be at LTE, following the formula
\begin{equation}
\ln(\frac{N_{\textrm{up}}} {g_{\textrm{up}}}) = \ln(\frac{N_{\textrm{tot}}}{Q_{\textrm{rot}}}) - \frac{E_{\textrm{up}}}{kT_{\textrm{rot}}} - \ln(\frac{\Omega_a}{\Omega_s}) - \ln(C_{\tau})
\label{eq_PD}
\end{equation}
where $N_{\textrm{tot}}$ is the total column density of the species 
in question, $N_{\textrm{up}}$ is the observed column density of the upper 
state of the species with an upper energy $E_{\textrm{up}}$ including 
the opacity effect, $Q_{\textrm{rot}}$ is the partition function, $\Omega_a$ 
is the beam solid angle, $\Omega_s$ is the source solid angle. 
$C_{\tau}$ is given by
\begin{equation}
C_{\tau} = \frac{\tau}{1-\exp(-\tau)}
\end{equation} 
with $\tau$ being the optical depth. $\tau$ can be expressed as
\begin{equation}
\tau = \frac{c^3}{8 \pi \nu_0^3} \frac{A_{\textrm{ul}}}{\Delta V} \frac{g_{\textrm{up}} N_{\textrm{tot}}}{Q_{\textrm{rot}}} \exp({\frac{-E_{\textrm{up}}}{k T_{\textrm{rot}}}}) (\exp({\frac{h \nu_0}{k T_{\textrm{rot}}}})-1)
\end{equation} 
where $c$ is the speed of light, $A_{\textrm{ul}}$ is the Einstein-A 
coefficient of spontaneous emission, $\Delta V$ is the FWHM fixed to 3 km/s. 
We performed a reduced $\chi_{\textrm{red}}^2$ minimization by running a grid of 125,000 models 
covering a large parameter space in rotational temperature $T_{\textrm{rot}}$ 
(50 values between 10 and 500 K), total column density in the
source $N_{\textrm{tot}}$ (50 values between $10^{15}$ and $10^{20}$
cm$^{-2}$), and source size $\theta_s$ (50 values between 0.04 and 2 \arcsec). 
At LTE, the column density of every upper state $N_{\textrm{up}}$ can 
be derived for each set of $N_{\textrm{tot}}$, $T_{\textrm{rot}}$, 
and source solid angle $\Omega_s = \theta_s^2$ according to equation 
(\ref{eq_PD}). 
The best-fit model populations are plotted together with the
observed populations of the levels in Figures \ref{PD_meth} and
\ref{PD_coms} and are marked by red cross symbols. Tables \ref{PDmeth}
and \ref{recap_PD} summarise the parameters of the best-fit models.

We started the PD analysis by simultaneously modelling the population distribution of
$^{12}$CH$_3$OH and $^{13}$CH$_3$OH. For this purpose, we assumed a
$^{12}$C/$^{13}$C abundance ratio of 70 following \citet{Boogert2002}
and the same rotational temperature for the two isotopologues. 
The population modelling of high-energy $^{12}$CH$_3$OH ($400 < E_{up} <
1100$ K) and low-energy $^{13}$CH$_3$OH optically thin transitions
allowed us to constrain the rotational temperature through the 
determination of the slope of the level populations, leaving only a 
degeneracy between $N_{\textrm{tot}}$ and $\theta_s$.
Since the optical depth $\tau$ of each level depends on the total 
column density $N_{\textrm{tot}}$, low-energy optically thick 
transitions from $^{12}$CH$_3$OH can be used to constrain 
$N_{\textrm{tot}}$ and $\theta_s$.
 
Table \ref{PDmeth} presents the results of the PD analysis of the
methanol population distribution. 
The methanol emission is relatively well modelled by the PD model
for the two sources. The PD analysis converges towards one single set of
input physical parameters ($N_{\textrm{tot}}$, $\theta_s$, $T_{\textrm{rot}}$) with
$\chi_{\textrm{red}}^2$ of about 1.5-2 and with uncertainties up to 50 \% 
at a one sigma level for the column densities.
The PD model was able to reproduce the population of most transitions 
within their uncertainties except the population of some low-energy
transitions of $^{13}$CH$_3$OH in IRAS2A and of $^{12}$CH$_3$OH in IRAS4A 
which tend to be underestimated.
The rotational temperatures of methanol deduced from the PD analysis
are similar in IRAS2A and IRAS4A ($\sim 140$ K). 
However, the source size derived for IRAS2A is larger than for IRAS4A 
($\sim 0.36 \arcsec$ versus $\sim 0.20 \arcsec$). The source
size deduced for IRAS2A is in good agreement with the size estimated
by \citet{Maret2004} but we found a smaller source size for IRAS4A, by
a factor of 2.5. 
The source size of IRAS2A of 0.36 $\arcsec$, corresponding to a radius
of 42 AU at 235 pc, is also consistent with the FWHM emission size of the
CH$_3$OH  transitions deduced by \citet{Maury2014} from elliptical
gaussian fits that range between 25 AU and 70 AU. 

\begin{table}[htp]
\centering
\caption[Results of the PD analysis for the methanol emission towards
IRAS2A and IRAS4A.]
{Results of the PD analysis of the methanol emission.}
\begin{tabular}{l c c}
\hline					
\hline					
	&	IRAS2A	&	IRAS4A	\\
\hline					
$\chi_{\textrm{red}}^2$	&	1.4	&	2.1	\\
$N$(CH$_3$OH) (cm$^{-2}$)	&	$5.0_{-1.8}^{+2.9}  \times 10^{18}$	&	$1.6_{-0.8}^{+0.6} \times 10^{19}$	\\
$N$($^{13}$CH$_3$OH) (cm$^{-2}$)	&	$7.1_{-2.6}^{+4.2} \times 10^{16}$	&	$2.3_{-1.1}^{+1.3} \times 10^{17}$	\\
$\theta_s$ (\arcsec)	&	$0.36_{-0.04}^{+0.04}$	&	$0.20_{-0.04}^{+0.08}$	\\
$T_{\textrm{rot}}$ (K)	&	$140_{-20}^{+20}$	&	$140_{-30}^{+30}$	\\
\hline					

\end{tabular} \\
\label{PDmeth}
\end{table}

For all other COMs, the low number of observed transitions and the 
relative high uncertainty on the derived column density of each level 
generates a degeneracy between the input parameters, and prevents the 
PD model to converge towards one single set of input parameters:
the observations are overfitted and can be reproduced by a large range
of parameters giving $\chi^2$ lower than 1. Therefore, we decided to
fix the source size for the COMs emission to the size of the methanol
emission, assuming that all COMs will evaporate with methanol in spite
of their slightly different binding energies.
Even by fixing the source size, the analysis of the glycolaldehyde and
ethyl cyanide populations towards the two sources and of the 
ethanol population towards IRAS4A did not allow us to converge
towards one set of $N_{\textrm{tot}}$, and $T_{\textrm{rot}}$.
For other molecules, we were able to deduce a unique column density with
relatively small uncertainties. As seen in Fig. \ref{PD_meth} and
\ref{PD_coms}, the observed population distribution is well reproduced
by the best fit model since most of the column densities of upper energy levels
predicted by the  best fit model lie within the range of uncertainties
of the observed data.  
%
The best fit models of the PDs generally consist in lower rotational
temperatures and higher total column densities than the values derived
with the RDs in order to reproduce the population of the optically
thick transitions. 
For instance, the spread of the population distribution of the
low upper energy levels ($E_{\textrm{up}} \leq 120$ K) of CH$_3$OH, $^{13}$CH$_3$OH,
HCOOCH$_3$, or CH$_3$CN are explained by opacity effects. For these
species, transitions showing a decrease of their population are optically thick
with $\tau$ higher than 1.  

For species where the PD analysis was not able to converge towards 
one set of input parameters (namely glycolaldehyde, ethanol, and ethyl cyanide)
and for molecules showing only one detected transition, we fixed the
source size and the rotational temperature to the values found for
methanol. Results of this analysis are shown in Table \ref{recap_PD}.

Most transitions of COMs whose collision rates have been computed (CH$_3$OH,
CH$_3$CN, and HC$_3$N) have critical densities that range between
$10^5$ and $10^7$ cm$^{-3}$ at 100 K. They are therefore likely lower
than the densities found in the hot corinos of IRAS2A and IRAS4A
  \citep[$n_{\textrm{H2}} > 1.3 \times 10^8$   cm$^{-3}$ following the density
profiles of the two envelopes derived by][]{Maret2004}. Given the
good fit to the observational data with our LTE PD analysis, it is
likely that the observed species are at LTE.  
Most of the scattering of the population distribution can therefore be
attributed to opacity effects only. 

\begin{table*}[htp]
\centering
\caption{Results from the population diagram analysis for the COMs emission.}
\begin{footnotesize}
\begin{tabular}{l c c c c c}
\hline																		
 \hline																		
 Molecule	&		$N_{hc}$			&	$T_{\textrm{rot}}$ &	Source size		&		$X^a$			&	$X_{\textrm{meth}}^b$	\\
	&		(cm$^{-2}$) 			&	(K)	&	(\arcsec)		&					&		\\
\hline																		
	&						\multicolumn{5}{c}{IRAS2A}											\\
\hline																		
CH$_3$OH	&	$	5.0_{-1.8}^{+2.9}	\times	10^{18}$	&	$140_{-20}^{+20}$	&	$0.36_{-0.04}^{+0.04}$		&	$	1.0_{-0.4}^{+0.6}	\times	10^{-6}$	&		\\
$^{13}$CH$_3$OH	&	$	7.1_{-2.6}^{+4.2}	\times	10^{16}$	&	$140_{-20}^{+20}$	&	$0.36_{-0.04}^{+0.04}$		&	$	1.4_{-0.5}^{+0.8}	\times	10^{-8}$	&		\\
HCOOCH$_3$	&	$	7.9_{-1.6}^{+4.6}	\times	10^{16}$	&	$160_{-30}^{+50}$	&	0.36	$^c$	&	$	1.6_{-0.3}^{+0.9}	\times	10^{-8}$	&	$1.6_{-1.0}^{+1.1}$	\\
CH$_3$CN	&	$	2.0_{-0.4}^{+1.2}	\times	10^{16}$	&	$130_{-40}^{+230}$	&	0.36	$^c$	&	$	4.0_{-0.8}^{+2.4}	\times	10^{-9}$	&	$0.40_{-0.25}^{+0.28}$	\\
CH$_3$OCH$_3$	&	$	5.0_{-1.0}^{+2.9}	\times	10^{16}$	&	$110_{-20}^{+60}$	&	0.36	$^c$	&	$	1.0_{-0.2}^{+0.6}	\times	10^{-8}$	&	$1.0_{-0.6}^{+0.7}$	\\
C$_2$H$_5$OH	&	$	7.9_{-4.0}^{+4.6}	\times	10^{16}$	&	$270_{-80}^{+230}$	&	0.36	$^c$	&	$	1.6_{-0.8}^{+0.9}	\times	10^{-8}$	&	$1.6_{-1.2}^{+1.1}$	\\
HCOCH$_2$OH	&	$	6.8	\times	10^{15}$	&	$140 ^d$	&	0.36	$^d$	&	$	1.4	\times	10^{-9}$	&	$0.14_{-0.08}^{+0.05}$	\\
C$_2$H$_5$CN	&	$	1.5	\times	10^{15}$	&	$140 ^d$	&	0.36	$^d$	&	$	3.0	\times	10^{-10}$	&	$0.030_{-0.017}^{+0.011}$	\\
HC$_3$N	&	$	9.3	\times	10^{14}$	&	$140 ^d$	&	0.36	$^d$	&	$	1.9	\times	10^{-10}$	&	$0.019_{-0.011}^{+0.007}$	\\
H$_2$$^{13}$CO	&	$	6.3	\times	10^{15}$	&	$140 ^d$	&	0.36	$^d$	&	$	1.3	\times	10^{-9}$	&	$0.13_{0.07}^{+0.05}$	\\
NH$_2$CHO	&	$	1.2	\times	10^{16}$	&	$140 ^d$	&	0.36	$^d$	&	$	2.4	\times	10^{-9}$	&	$0.24_{-0.14}^{+0.09}$	\\
CH$_2$CO	&	$	6.8	\times	10^{15}$	&	$140 ^d$	&	0.36	$^d$	&	$	1.4	\times	10^{-9}$	&	$0.14_{-0.08}^{+0.05}$	\\
\hline																		
	&						\multicolumn{5}{c}{IRAS4A}											\\
\hline																		
CH$_3$OH	&	$	1.6_{-0.8}^{+0.6}	\times	10^{19}$	&	$140_{-30}^{+30}$	&	$0.20_{-0.04}^{+0.08}$		&	$	4.3_{-2.1}^{+2.5}	\times	10^{-7}$	&		\\
$^{13}$CH$_3$OH	&	$	2.3_{-1.1}^{+1.3}	\times	10^{17}$	&	$140_{-30}^{+30}$	&	$0.20_{-0.04}^{+0.08}$		&	$	6.2_{-3.0}^{+3.5}	\times	10^{-9}$	&		\\
HCOOCH$_3$	&	$	5.0_{-1.8}^{+5.0}	\times	10^{17}$	&	$60_{-10}^{+20}$	&	0.20	$^c$	&	$	1.4_{-0.5}^{+1.4}	\times	10^{-8}$	&	$3.1_{-2.1}^{+3.5}$	\\
CH$_3$CN	&	$	6.3_{-1.3}^{+3.6}	\times	10^{16}$	&	$200_{-40}^{+110}$	&	0.20	$^c$	&	$	1.7_{-0.4}^{+1.0}	\times	10^{-9}$	&	$0.39_{-0.24}^{+0.30}$	\\
CH$_3$OCH$_3$	&	$	1.6_{-0.3}^{+0.9}	\times	10^{17}$	&	$80_{-20}^{+40}$	&	0.20	$^c$	&	$	4.3_{-0.9}^{+2.5}	\times	10^{-9}$	&	$1.0_{-0.6}^{+0.8}$	\\
C$_2$H$_5$OH	&	$	1.6	\times	10^{17}$	&	$140 ^d$	&	0.20	$^d$	&	$	4.3	\times	10^{-9}$	&	$1.0_{-0.6}^{+0.5}$	\\
HCOCH$_2$OH	&	$	4.8	\times	10^{16}$	&	$140 ^d$	&	0.20	$^d$	&	$	1.3	\times	10^{-9}$	&	$0.30_{-0.17}^{+0.15}$	\\
C$_2$H$_5$CN	&	$	6.4	\times	10^{15}$	&	$140 ^d$	&	0.20	$^d$	&	$	1.7	\times	10^{-10}$	&	$0.040_{-0.023}^{+0.020}$	\\
HC$_3$N	&	$	2.9	\times	10^{15}$	&	$140 ^d$	&	0.20	$^d$	&	$	7.8	\times	10^{-11}$	&	$0.018_{-0.010}^{+0.009}$	\\
H$_2$$^{13}$CO	&	$	1.1	\times	10^{16}$	&	$140 ^d$	&	0.20	$^d$	&	$	3.0	\times	10^{-10}$	&	$0.069_{0.040}^{+0.034}$	\\
NH$_2$CHO	&	$	1.9	\times	10^{16}$	&	$140 ^d$	&	0.20	$^d$	&	$	5.1	\times	10^{-10}$	&	$0.12_{-0.07}^{+0.06}$	\\
CH$_2$CO	&	$	3.4	\times	10^{16}$	&	$140 ^d$	&	0.20	$^d$	&	$	9.2	\times	10^{-10}$	&	$0.21_{-0.12}^{+0.10}$	\\
\hline																	
\end{tabular}
\label{recap_PD}
\tablecomments{
a: The abundances relative to H$_2$ are obtained from $N$(H$_2$)
derived at 145 GHz in Table \ref{prop_cont} assuming an
homogeneous H$_2$ column density within the beam.  
b: The abundances relative to CH$_3$OH were computed from
$N$(CH$_3$OH) derived from the PD analysis of the CH$_3$OH and
$^{13}$CH$_3$OH emissions and adapted for the same source size. 
c: The source size was assumed to be equal to that of methanol when
the size could not be constrained.
d: The source size and rotational temperatures were assumed to be
equal to those of methanol.
}
\end{footnotesize}
\end{table*}

\section{Discussion}

\subsection{Abundances in IRAS2A and IRAS4A}

Methanol is likely the most abundant complex organics, and is believed
to be the precursor molecule of several COMs.
It is therefore worth comparing the abundance of COMs with respect to
methanol to quantify the efficiency of their formation. 
Moreover, column densities of COMs and methanol have been derived with
similar methods and from the same observational data, the estimates of
the abundance ratios are therefore more accurate than the absolute abundances
derived with respect to H$_2$. 
Tables \ref{recap_RD} and \ref{recap_PD} list the abundance ratios of
the COMs with respect to methanol for the RD and PD analyses. 
%
 %
The two targeted sources seem to have a similar chemical composition since
the COMs abundance ratios differ by only a factor of two at
maximum. Methyl formate is the most abundant COM of our sample, with
an abundance of 1.5-3 \% followed by ethanol (1-1.5 \%) and di-methyl
ether (1 \%). Other COMs are detected with abundances lower than 1
\%: glycolaldehyde and methyl cyanide show abundances of 0.15 and
0.40 \% respectively whilst ethyl cyanide is detected with an abundance
of 0.03-0.04 \%. 

Table \ref{recap_RD} compares the column densities and abundance 
ratios deduced from the RD analysis with previous single-dish studies by 
\citet{Bottinelli2004a, Maret2005, Bottinelli2007} carried out towards 
IRAS2A and IRAS4A. 
The column densities obtained from these previous observations suffer from
several limitations: most detected transitions have low upper
energy levels and the large beam of single-dish telescopes encompasses the
cold envelope where weakly excited lines may have contaminated the hot
corino emission (see the interferometric maps in Figures
\ref{maps_i2a} and \ref{maps_i4a}).
Consequently, the rotational temperatures and column densities derived with 
the PdBi are higher ($T_{\textrm{rot}}$ = 80-290 K and 300-360 K in this work towards 
IRAS2A and IRAS4A versus 100 and 25 K respectively and higher column densities
up to one order of magnitude), since the interferometric observations
probe material closer to the central protostars.
The abundance ratios deduced from previous single-dish studies have also 
higher relative uncertainties than in this work due to the different 
telescope calibrations since the observations of methanol and COMs have been
carried out separately.
The abundances relative to methanol derived from our interferometric observations therefore
differ from the abundances obtained with the single-dish observations,
the latter being usually overestimated. 
The methyl formate abundance derived in IRAS2A  of 2 \% is consistent
with the upper limit of 85 \% by \citet{Bottinelli2007}. However, 
 the abundance derived in IRAS4A of $\sim$ 3 \% is 18 times
lower than the value derived by \citet{Bottinelli2004a} and using the
column density of CH$_3$OH derived by \citet{Maret2005}.
The higher abundance of methyl formate with respect to methanol in IRAS4A derived from
single-dish observations is explained by the lower column density of
methanol derived in \citet{Maret2005} assuming optically thin emission.
For the same reasons, the abundances of methyl cyanide derived in the
two sources by our PD analysis are also lower, by a factor of 3 to 6,
than the abundances obtained by \citet{Bottinelli2007}. 
Di-methyl ether, ethanol and ethyl cyanide have not been detected
with single-dish telescopes towards IRAS2A and IRAS4A but their upper
limits agree well with our observations. 

We report here the first detection of glycolaldehyde in low-mass
protostars other than IRAS 16293. 
Glycolaldehyde co-exists with its isomer methyl formate with a
[HCOOCH$_3$]/[HCOCH$_2$OH] abundance ratio of $12_{-2}^{+7}$ towards
IRAS2A and of $10_{-4}^{+10}$ towards IRAS4A.
These abundance ratios are similar to the ratios of $\sim 13$
found in the sources A and B of the IRAS 16293 protostellar binary
system by \citet{Jorgensen2012} from high angular resolution ALMA
observations. They are also consistent with the ratios derived towards
SgrB2(N) ranging from 52 in the hot core \citep{Hollis2001} to 5 found
on more extended scales \citep{Hollis2000}.   

\begin{figure*}[htp]
\centering
\includegraphics[width=\columnwidth]{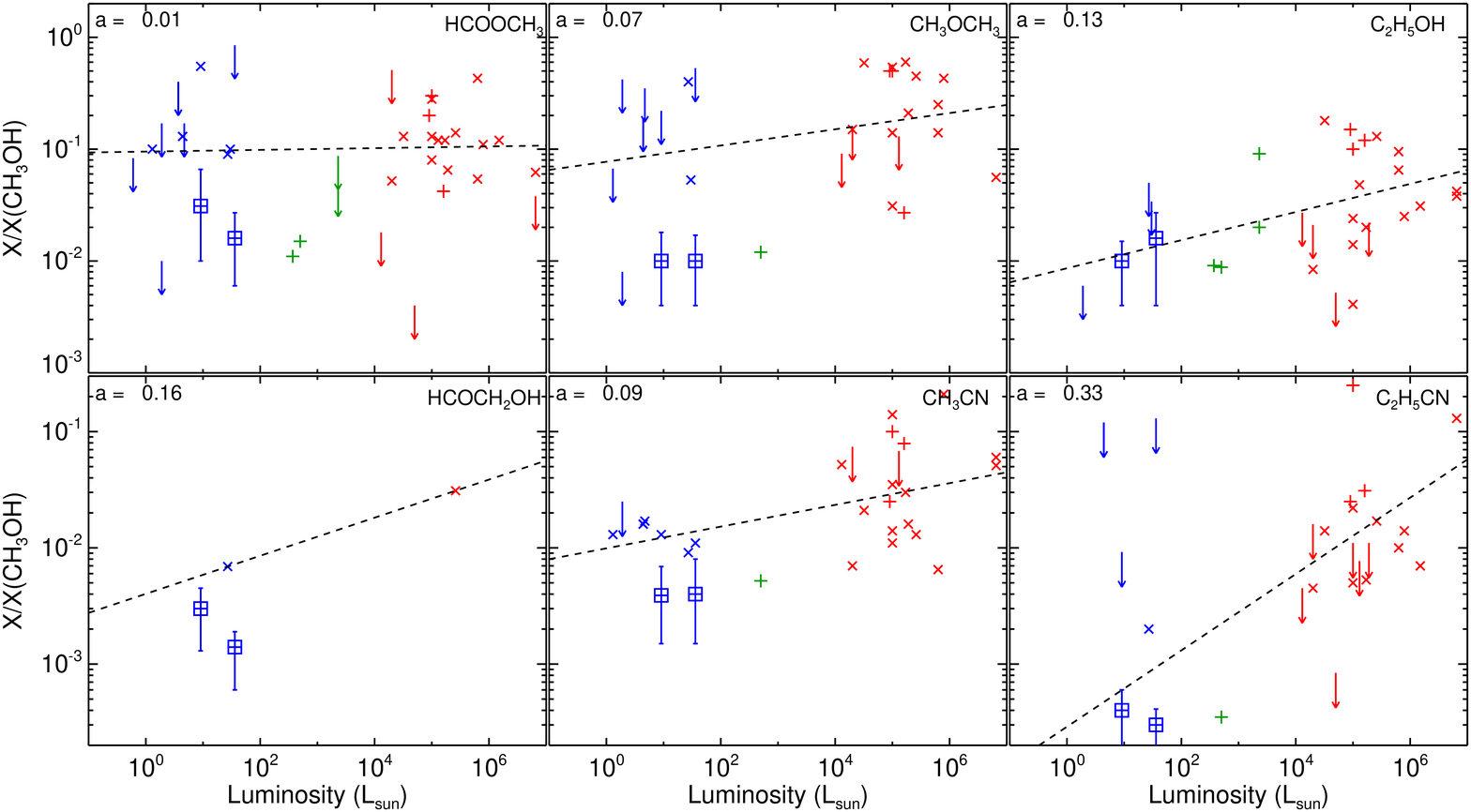}
\caption{Abundances of COMs detected in several transitions with
  respect to CH$_3$OH towards a sample of low-mass ($L < 100 L_{\odot}$; blue),
  intermediate-mass ($100 L_{\odot} < L < 10^4 L_{\odot}$; green), and
  high-mass ($L > 10^4 L_{\odot}$; red) hot cores as
  function of the protostar luminosity. Cross and plus symbols represent
  abundance ratios derived with single-dish telescopes and
  interferometers, respectively. The abundance ratios derived in this
  work with the Population Diagram analyses  are denoted by squares
  with error bars. The sample of hot  cores is listed in Table
  \ref{list_hc} along with their physical and  chemical properties.}  
\label{sum_Xmeth}
\end{figure*}

\subsection{Comparison with Other Sources}

The COMs abundances derived in this work are compared with other
published data of low-mass, intermediate-mass, and
high-mass hot cores obtained with single dish and interferometric sub-mm
telescopes in Figure \ref{sum_Xmeth} showing the COMs abundances as a
function of the protostar luminosity. Table \ref{list_hc} lists the
abundances of selected COMs towards the hot cores shown in
Fig. \ref{sum_Xmeth} along with the references. For this purpose, we
only selected observational studies where several transitions of
methanol were detected. 
For most of the works, COMs and methanol were detected simultaneously
and the abundance ratios were derived from the 
main isotopologue CH$_3$OH, either by assuming optically thin emission
and LTE population \citep{Macdonald1996, Ikeda2001, Beuther2007,
  Beuther2009, Palau2011, Oberg2011, Oberg2014}, by 
neglecting optically thick lines in the RD analysis
\citep{Bisschop2007, Isokoski2013} or by taking the opacity of the
lines into account in their model \citep{Nummelin2000, Qin2010,
  Crockett2014, Neill2014}.   
\citet{Fuente2014} derived the methanol abundances from the
$^{13}$CH$_3$OH isotopologue. 
For IRAS 16293, we combined the CH$_3$OH absolute abundance derived by
\citet{Schoier2002} with the COMs abundances obtained by \citet{Jaber2014},
both from a radiative transfer modelling, to obtain the abundance
ratios. We also derived the glycolaldehyde abundance from
  \citet{Jorgensen2012}.

Overall, the abundance ratios of COMs estimated towards IRAS2A and
IRAS4A in this work tend to be lower than the abundances derived in
other low-mass protostars from single-dish observations. 
For example, methyl formate and methyl cyanide have been detected
towards 5 other low-mass sources and show abundances of 5-50\% and
0.8-1.7\% respectively, representing a factor of 1.5-15 and 1.5-3
higher than in our work. 
Since these observations suffer from the same limitations than the single-dish
observations by \citet{Bottinelli2004a, Bottinelli2007} presented in the previous
section (low number of detected transitions, low upper energy levels
of detected transitions, large beam encompassing the external
envelope), the discrepancy likely comes from the differences in the
observational methods and does not necessarily reflect differences in
the chemistry between the sources.
 

Table \ref{mean_abu} summarises the mean abundance ratios of COMs in
low-mass, intermediate-mass and high-mass protostars. Along with
Fig. \ref{sum_Xmeth}, they allow us to investigate any possible
correlation of the COMs abundances with the protostar luminosity.  
For each molecule, the data has been fitted by a linear curve depicted
by the dashed curve whose slope $a$ is shown at the top left of each
panel of Fig. \ref{sum_Xmeth}, by considering detected ratios only.
It can be noticed that the abundances of the six COMs tend to slightly
increase with the protostar luminosity. 
However, for all species but C$_2$H$_5$CN, the increase remains
negligible compared to the dispersion of the abundance ratio values. 
We can conclude that the abundance ratio of these COMs stays
relatively constant with the protostar luminosity within six orders of 
magnitude. 
In spite of their lower luminosities, inducing lower temperatures, and
smaller sizes, low-mass protostars seem to be as chemically complex as
high-mass protostars.

\begin{table}[htp]
\centering
\caption{Averaged COMs abundances (in \%) with respect to methanol.}
\begin{tabular}{l c c c c c}
\hline
\hline
Molecule	&	LMP	&	IMP	&	HMP	&	G13	&	R01	\\
\hline											
HCOOCH$_3$	&	$14_{-7}^{+41}$	&	$1.3 \pm 0.2$	&	$14_{-6}^{+18}$	&	0.082 - 0.84	&	0.013 - 0.67	\\
CH$_3$OCH$_3$	&	$12_{-10}^{+28}$	&	$1.2$	&	$30_{-18}^{+23}$	&	0.44 - 0.74	&	0.62 - 18	\\
HCOCH$_2$OH	&	$0.38_{-0.18}^{+0.31}$	&	/	&	$3.1$	&	0.54 - 1.3	&	-	\\
C$_2$H$_5$OH	&	$1.3_{-0.3}^{+0.3}$	&	$3.2_{-2.0}^{+5.9}$	&	$6.4_{-4.1}^{+6.6}$	&	0.50 - 2.6	&	15 - 23	\\
CH$_3$CN	&	$1.1_{-0.5}^{+0.4}$	&	$0.52$	&	$5.1_{-3.3}^{+7.8}$	&	0.034 - 0.45	&	3.0 - 3.7	\\
C$_2$H$_5$CN	&	$0.090_{-0.055}^{+0.11}$	&	/	&	$4.1_{-2.8}^{+16}$	&	0.052 - 0.79	&	-	\\
\hline																	
\end{tabular}
\label{mean_abu}
\tablecomments{~LMP, IMP, and HMP stand for Low-Mass ($L < 100 L_{\odot}$),
  Intermediate-Mass ($100 L_{\odot} < L < 10^4 L_{\odot}$), and
  High-Mass ($10^4 L_{\odot} < L$) Protostars. The two abundances in
  the column ``G13'' are the peak abundances of the ``Fast'' and ``Slow''
  models of \citet{Garrod2013}. The two abundances in the column ``R01''
  are the peak abundances of the models including ammonia at $T=100$ and
  $T=300$ K of \citet{Rodgers2001}. 
}
\end{table}

\subsection{Comparison with Chemical Model Predictions}

We compare the observed abundance ratios in low-mass,
intermediate-mass, and high-mass protostars with the results of two
astrochemical models in Table \ref{mean_abu}. 
The model of \citet{Garrod2013} is a multilayer gas-grain
astrochemical model in which COMs are assumed to be mostly formed at the surface of
interstellar ices during the warm-up protostellar phase ($30$ K $< T < 100$ K)
through radical recombination induced by the UV photodissociation of the
main ice components.
In the model of \citet{Rodgers2001}, COMs are only formed by warm gas phase
chemistry for a set of constant physical parameters representative of hot
cores ($n_{\textrm{H}} = 10^7$ cm$^{-3}$; $T = 100$ and 300 K) after the
sublimation of interstellar ices with typical ice composition, already
  containing C$_2$H$_5$OH with an abundance of 20 \% with 
respect to CH$_3$OH. 
Although the rates of some key reactions for the formation of COMs
have been lowered meanwhile, such as the methyl cation transfer
reaction between H$_2$CO and CH$_3$OH$_2^+$ or the electronic
recombination of protonated COMs ions, this model still provides a good
basis to estimate the formation efficiency of methyl formate, di-methyl
ether and methyl cyanide in the gas phase. 

Both models tend to underpredict the abundance of methyl formate
relative to methanol observed in low-mass and high-mass protostars by
at least 1-2 orders of magnitude. Moreover, the model of \citet{Garrod2013}
also underpredicts the [HCOOCH$_3$]/[HCOCH$_2$OH] abundance ratio by two
orders of magnitude since it seems to reproduce well the observed
abundance of glycolaldehyde. This comparison suggests that the
chemical network forming methyl formate either in the gas phase or on
ices is still incomplete. Possible alternative branching ratios for the
photodissociation of CH$_3$OH on ices or other gas phase reactions
involving HCOOH could enhance its formation.
The observed abundances of di-methyl ether and methyl cyanide with
respect to methanol can be reproduced by the gas phase model of
\citet{Rodgers2001} only. The 
absolute abundances of these two molecules are similar in the two
models ($10^{-8}-10^{-7}$ for di-methyl ether and $10^{-9}-10^{-8}$ for
methyl cyanide), showing that warm gas phase chemistry tends to be as efficient
as surface chemistry to produce these COMs. However, the difference in
the abundances comes from the efficient destruction of methanol
in the warm gas in the model of \citet{Rodgers2001} increasing the
abundance ratio of COMs.
No efficient formation routes in the gas phase have been proposed for
glycolaldehyde and ethanol but their formation at the surface of
interstellar ices seem to be efficient enough to reproduce the
observations towards low-mass and high-mass protostars.
The abundance of ethyl cyanide shows an increase of almost two
  orders of magnitude between low-mass and high-mass protostars. Grain
  surface chemistry is able to reproduce the observations towards
  low-mass protostars but not towards high-mass hot cores.  
It is also possible that models also missed gas phase
reactions, as it was the case for methyl formate, where a new gas
phase reaction has been recently recognised by \citet{Balucani2015}.

\section{Conclusions}

In this work, we have presented interferometric multi-line observations of
methanol (CH$_3$OH, $^{13}$CH$_3$OH) and various COMs towards the two
bright low-mass protostars NGC1333-IRAS2A and -IRAS4A, carried out 
with the Plateau de Bure interferometer at an angular resolution 
of $\sim 2 \arcsec$. 
We summarize here the main conclusions of this work: \\
- Except for methanol, none of the transitions from complex organics
are spatially resolved with the interferometer, confirming that most
of the emission comes from the inner arcsec from the central protostars. \\ 
- Low-energy transitions ($E_{\textrm{up}} \leq 120$ K) of methanol
show extended emission around IRAS4A and trace the outflows driven by
the central protostar. \\ 
- We detected a high number of transitions (up to 45 for methanol)
from COMs with a wide range of upper energy levels (up to 1000 K for methanol)
allowing us to accurately derive their column densities with the use
of LTE methods. \\ 
- The abundances derived in this work, of a few percents for methyl
formate and di-methyl ether and of $\sim 0.5 \%$ for methyl cyanide for
instance, seem to be slightly lower than the abundances found towards
other low-mass protostars. However, the difference likely comes from
differences in the observational methods rather than different
chemistries taking place in these protostars. \\ 
- We  report the first detection of glycolaldehyde in low-mass
protostars other than IRAS 16293, with a methyl formate-to-glycol
aldehyde abundance ratio of 12 and 10 in IRAS2A and IRAS4A,
respectively. \\
- The analysis of the variation of the COMs abundance ratios with the
protostellar luminosity shows that low-mass hot corinos seem to be as
chemically rich as high-mass hot cores, in spite of their lower
temperatures and their smaller sizes. \\ 
- Comparison with theoretical models shows that the two theories of
COMs formation, either in the gas phase or at the surface of
interstellar ices, tend to underpredict the formation of methyl
formate by one to orders of magnitude. 
The comparison of the abundance ratios of other molecules favours a
gas phase formation for di-methyl ether and methyl cyanide and a
formation on ices for ethanol, ethyl cyanide, and glycolaldehyde. 

\begin{acknowledgements}

The authors are grateful to the anonymous referee whose comments
contributed to improve the quality of the present paper. 
This work was supported by NASA's Origins of Solar Systems and
Exobiology Programs.
V. T. acknowledges support from the NASA postdoctoral program. 
A. L-S and C. C acknowledge financing from the french space agency
CNES. 

\end{acknowledgements}


\newpage
\appendix



\begin{deluxetable}{l c c c c c c c c c c c c}
\centering
\rotate
\tabletypesize
\tiny
\tablecaption{Line parameters of CH$_3$OH lines observed towards IRAS2A and
  IRAS4A-NW.}
\startdata
\hline
\hline
	&		&		&		&		&	\multicolumn{4}{c}{IRAS2A}											&	\multicolumn{4}{c}{IRAS4A}											\\
\cline{6-13}																																	
$N$	&	Frequency	&	Transition	&	E$_{up}$	&	A$_{ul}$	&	Beam size		&	Source size $^a$	&		Flux $^b$			&	$dV_W$ $^c$	&	Beam size		&	Source size (PA) $^a$	&		Flux $^b$			&	$dV_W$  $^c$	\\
	&	(GHz)	&		&	(K)	&	(s$^{-1}$)	&	(\arcsec$\times$\arcsec, $^o$)		&	(\arcsec$\times$\arcsec, $^o$)	&		(Jy km/s)			&	(km/s)	&	(\arcsec$\times$\arcsec, $^o$)		&	(\arcsec$\times$\arcsec, $^o$)	&		(Jy km/s)			&	(km/s)	\\
\hline																																	
1	&	145.093754	&	$3_0$-$2_0$ E1, $v_t$=0	&	27.1	&	1.23(-5)	&	2.08$\times$1.65 (30)	&	1.65$\times$1.04	(19)	&		2.62$\pm$0.53	&	-	&	2.16$\times$1.73 (25)	&	outflow	&		2.54$\pm$0.51	&	-	\\
	&	145.097435	&	$3_0$-$2_0$ E2, $v_t$=0	&	19.5	&	1.10(-5)	&			&		&					&		&			&		&					&		\\
	&	145.103185	&	$3_0$-$2_0$ A$^+$, $v_t$=0	&	13.9	&	1.23(-5)	&			&		&					&		&			&		&					&		\\
2	&	165.050175	&	$1_1$-$1_0$ E1, $v_t$=0	&	23.4	&	2.35(-5)	&	2.31$\times$1.72 (110)	&	1.01$\times$0.87	(11)	&		1.28$\pm$0.26	&	5.8	&	2.39$\times$1.77 (114)	&	outflow	&		0.38$\pm$0.08	&	6.2	\\
3	&	165.061130	&	$2_1$-$2_0$ E1, $v_t$=0	&	28.0	&	2.34(-5)	&	2.31$\times$1.72 (110)	&	1.18$\times$0.83	(50)	&		1.49$\pm$0.30	&	5.2	&	2.39$\times$1.77 (114)	&	outflow	&		0.56$\pm$0.12	&	7.0	\\
4	&	143.865795	&	$3_1$-$2_1$ A$^+$, $v_t$=0	&	28.3	&	1.07(-5)	&	2.24$\times$1.77	(25)	&	0.83$\times$0.63	(10)	&		1.02$\pm$0.21	&	6.7	&	2.24$\times$1.77	(25)	&	outflow	&		0.44$\pm$0.13	&	7.5	\\
5	&	165.099240	&	$3_1$-$3_0$ E1, $v_t$=0	&	35.0	&	2.33(-5)	&	2.31$\times$1.72 (110)	&	1.19$\times$1.07	(30)	&		1.44$\pm$0.29	&	6.0	&	2.39$\times$1.77 (114)	&	outflow	&		0.39$\pm$0.08	&	5.2	\\
6	&	145.124332	&	$3_0$-$2_0$ A$^-$, $v_t$=0	&	51.6	&	6.89(-6)	&	2.08$\times$1.65 (30)	&	1.05$\times$0.71	(32)	&		3.18$\pm$0.64	&	-	&	2.16$\times$1.73 (25)	&	outflow	&		1.65$\pm$0.33	&	-	\\
	&	145.126191	&	$3_2$-$2_2$ E1, $v_t$=0	&	36.2	&	6.77(-6)	&			&		&					&		&			&		&					&		\\
	&	145.126386	&	$3_2$-$2_2$ E2, $v_t$=0	&	39.8	&	6.86(-6)	&			&		&					&		&			&		&					&		\\
	&	145.131864	&	$3_1$-$2_1$ E1, $v_t$=0	&	35.0	&	1.12(-5)	&			&		&					&		&			&		&					&		\\
	&	145.133415	&	$3_2$-$2_2$ A$^+$, $v_t$=0	&	51.6	&	6.89(-6)	&			&		&					&		&			&		&					&		\\
7	&	165.190475	&	$4_1$-$4_0$ E1, $v_t$=0	&	44.3	&	2.32(-5)	&	2.31$\times$1.72 (110)	&	1.26$\times$1.02	(76)	&		1.96$\pm$0.40	&	6.0	&	2.39$\times$1.77 (114)	&	outflow	&		0.64$\pm$0.13	&	9.5	\\
8	&	165.369341	&	$5_1$-$5_0$ E1, $v_t$=0	&	55.9	&	2.31(-5)	&	2.54$\times$1.71 (113)	&	1.10$\times$0.92	(42)	&		1.59$\pm$0.32	&	6.1	&	2.38$\times$1.76 (114)	&	outflow	&		0.68$\pm$0.14	&	5.5	\\
9	&	165.678649	&	$6_1$-$6_0$ E1, $v_t$=0	&	69.8	&	2.30(-5)	&	2.54$\times$1.71 (113)	&	1.08$\times$0.99	(54)	&		1.52$\pm$0.31	&	5.9	&	2.38$\times$1.76 (114)	&	outflow	&		0.55$\pm$0.11	&	6.3	\\
10	&	166.169098	&	$7_1$-$7_0$ E1, $v_t$=0	&	86.1	&	2.28(-5)	&	2.54$\times$1.71 (113)	&	1.06$\times$0.99	(30)	&		1.46$\pm$0.29	&	5.8	&	2.38$\times$1.76 (114)	&	outflow	&		0.54$\pm$0.11	&	5.8	\\
11	&	166.898566	&	$8_1$-$8_0$ E1, $v_t$=0	&	104.6	&	2.28(-5)	&	2.54$\times$1.71 (113)	&	1.02$\times$0.56	(44)	&		1.45$\pm$0.29	&	5.7	&	2.38$\times$1.76 (114)	&	outflow	&		0.51$\pm$0.11	&	5.0	\\
12	&	143.169517	&	$7_3$-$8_2$ E1, $v_t$=0 	&	112.7	&	4.13(-6)	&	2.31$\times$1.82 (26)	&	point	&		0.63$\pm$0.22	&	6.1	&	2.31$\times$1.82 (26)	&	outflow	&		0.24$\pm$0.08	&	5.8	\\
13	&	144.728359	&	$3_2$-$2_2$ A$^+$, $v_t = 1$	&	312.6	&	6.78(-6)	&	2.08$\times$1.65 (30)	&	0.68$\times$0.64	(-73)	&		0.80$\pm$0.16	&	6.9	&	2.16$\times$1.73 (25)	&	point	&		0.23$\pm$0.05	&	-	\\
	&	144.728782	&	$3_2$-$2_2$ E2, $v_t$=1	&	378.5	&	6.83(-6)	&			&		&					&		&			&		&					&		\\
	&	144.729074	&	$3_2$-$2_2$ A$^-$, $v_t$=1	&	312.6	&	6.78(-6)	&			&		&					&		&			&		&					&		\\
14	&	144.733262	&	$3_2$-$2_2$ E1, $v_t$=1	&	413.8	&	6.80(-6)	&	2.08$\times$1.65 (30)	&	0.38	&		0.96$\pm$0.19	&	9.4	&	2.16$\times$1.73 (25)	&	0.48	&		0.42$\pm$0.09	&	7.6	\\
	&	144.734429	&	$3_1$-$2_1$ E1, $v_t$=1 	&	305.4	&	1.09(-5)	&			&		&					&		&			&		&					&		\\
	&	144.736349	&	$3_0$-$2_0$ E1, $v_t$=1 	&	314.5	&	1.22(-5)	&			&		&					&		&			&		&					&		\\
15	&	144.589854	&	$3_1$-$2_1$ A$^+$, $v_t$=1 	&	339.1	&	1.09(-5)	&	2.08$\times$1.65 (30)	&	0.59	&		0.37$\pm$0.08	&	6.0	&	2.16$\times$1.73 (25)	&	0.3	&		0.21$\pm$0.06	&	6.7	\\
16	&	144.878576	&	$3_1$-$2_1$ A$^-$, $v_t$=1	&	339.2	&	1.09(-5)	&	2.08$\times$1.65 (30)	&	0.35	&		0.43$\pm$0.09	&	6.7	&	2.16$\times$1.73 (25)	&	0.91$\times$0.75	(-5)	&		0.23$\pm$0.06	&	5.6	\\
17	&	143.108385	&	$17_0$-$17_{-1}$ E, $v_t$=0	&	366.8	&	6.44(-6)	&	2.06$\times$1.67 (26)	&	0.61x0.50	(0)	&		0.52$\pm$0.11	&	6.9	&	2.06$\times$1.67 (26)	&	0.50	&		0.18$\pm$0.04	&	6.9	\\
18	&	166.773281	&	$5_1$-$5_0$ A$^+$, $v_t$=1	&	390.0	&	1.85(-5)	&	2.54$\times$1.71 (113)	&	0.91$\times$0.46	(67)	&		0.90$\pm$0.19	&	6.4	&	2.38$\times$1.76 (114)	&	0.58	&		0.45$\pm$0.10	&	7.5	\\
19	&	165.074355	&	$14_6$-$15_5$ E1, $v_t$=0	&	422.4	&	5.36(-6)	&	2.31$\times$1.72 (110)	&	0.90$\times$0.62	(37)	&		0.47$\pm$0.10	&	5.5	&	2.39$\times$1.77 (114)	&	2.02$\times$1.01	(-53)	&		0.30$\pm$0.06	&	6.0	\\
20	&	144.750264	&	$3_1$-$2_1$ E2, $v_t$=1  	&	427.3	&	1.08(-5)	&	2.08$\times$1.65 (30)	&	point	&		0.28$\pm$0.06	&	7.2	&	2.16$\times$1.73 (25)	&	0.15	&		0.12$\pm$0.04	&	6.7	\\
21	&	144.768193	&	$3_0$-$2_0$ A$^+$, $v_t$=1	&	437.5	&	1.22(-5)	&	2.08$\times$1.65 (30)	&	0.46$\times$0.40	(49)	&		0.28$\pm$0.06	&	6.8	&	2.16$\times$1.73 (25)	&	1.23	&		0.09$\pm$0.03	&	4.4	\\
22	&	144.572025	&	$3_0$-$2_0$ A$^+$, $v_t$=2	&	522.1	&	1.42(-5)	&	2.08$\times$1.65 (30)	&	0.52	&		0.21$\pm$0.04	&	6.3	&	2.16$\times$1.73 (25)	&	1.61$\times$1.03	(-60)	&		0.15$\pm$0.04	&	5.5	\\
	&	144.571262	&	$3_2$-$2_2$ E2, $v_t$=2	&	658.8	&	6.74(-6)	&			&		&					&		&			&		&					&		\\
23	&	144.583961	&	$3_1$-$2_1$ E2. $v_t$=2 	&	545.9	&	1.09(-5)	&	2.08$\times$1.65 (30)	&		&		0.20$\pm$0.04	&	4.0	&	2.16$\times$1.73 (25)	&		&		0.15$\pm$0.05	&	16.3	\\
24	&	166.843662	&	$11_2$-$10_3$ E1, $v_t$=1	&	553.0	&	1.19(-6)	&	2.54$\times$1.71 (113)	&	0.48	&		0.07$\pm$0.02	&	3.5	&	2.38$\times$1.76 (114)	&	-	&	$<$	0.04			&	-	\\
25	&	164.299104	&	$15_2$-$14_1$ A$^-$, $v_t$=1	&	576.0	&	1.44(-5)	&	2.31$\times$1.72 (110)	&	0.94$\times$0.64	(-86)	&		0.43$\pm$0.10	&	5.5	&	2.39$\times$1.77 (114)	&	1.95$\times$0.94	(-83)	&		0.21$\pm$0.05	&	3.6	\\
26	&	142.276432	&	$9_7$-$10_6$ E2, $v_t$=1	&	627.5	&	1.11(-6)	&	2.06$\times$1.67 (26)	&	1.66$\times$0.50	(-42)	&		0.03$\pm$0.01	&	6.6	&	2.06$\times$1.67 (26)	&	0.41	&		0.02$\pm$0.01	&	4.8	\\
27	&	144.281736	&	$3_1$-$2_1$ A$^+$, $v_t$=2	&	696.6	&	1.07(-5)	&	2.08$\times$1.65 (30)	&	point	&		0.08$\pm$0.02	&	7.2	&	2.16$\times$1.73 (25)	&	0.78	&		0.06$\pm$0.03	&	5.3	\\
28	&	144.530553	&	$3_0$-$2_0$ E1, $v_t$=2	&	748.1	&	1.22(-5)	&	2.08$\times$1.65 (30)	&	0.34	&		0.04$\pm$0.01	&	4.0	&	2.16$\times$1.73 (25)	&	1.15	&		0.05$\pm$0.01	&	4.1	\\
29	&	144.499723	&	$3_1$-$2_1$ E1, $v_t$=2	&	812.8	&	1.05(-5)	&	2.08$\times$1.65 (30)	&	0.96	&		0.03$\pm$0.01	&	7.0	&	2.16$\times$1.73 (25)	&	point	&		0.03$\pm$0.02	&	5.7	\\
30	&	164.486238	&	$26_2$-$26_1$ E2, $v_t$=0	&	843.0	&	2.69(-5)	&	2.31$\times$1.72 (110)	&	0.85$\times$0.43	(38)	&		0.30$\pm$0.07	&	5.8	&	2.39$\times$1.77 (114)	&	1.25	&		0.26$\pm$0.05	&	5.8	\\
31	&	144.195014	&	$24_6$-$25_5$ E2, $v_t$=0	&	884.4	&	5.16(-6)	&	2.08$\times$1.65 (30)	&	1.76$\times$0.69	(-59)	&		0.05$\pm$0.01	&	6.0	&	2.16$\times$1.73 (25)	&	point	&		0.07$\pm$0.03	&	5.9	\\
32	&	144.437702	&	$10_9$-$10_{a0}$ E2, $v_t$=1	&	922.9	&	2.82(-6)	&	2.08$\times$1.65 (30)	&		&	$<$	0.05			&	-	&	2.16$\times$1.73 (25)	&	1.39	&		0.04$\pm$0.01	&	7.6	\\
33	&	144.807264	&	$11_9$-$11_{a0}$ E1, $v_t$=1	&	948.3	&	4.94(-6)	&	2.08$\times$1.65 (30)	&		&	$<$	0.05			&	-	&	2.16$\times$1.73 (25)	&	1.39$\times$0.67	(-2)	&		0.07$\pm$0.02	&	6.9	\\
34	&	145.068727	&	$19_5$-$20_6$ E2, $v_t$=1	&	985.9	&	8.88(-6)	&	2.08$\times$1.65 (30)	&	1.33	&		0.07$\pm$0.02	&	6.5	&	2.16$\times$1.73 (25)	&	0.6	&		0.10$\pm$0.04	&	7.0	\\
35	&	163.526070	&	$28_4$-$27_5$ A$^-$, $v_t$=0	&	1021.9	&	8.13(-6)	&	2.31$\times$1.72 (110)	&	1	&		0.04$\pm$0.02	&	4.7	&	2.39$\times$1.77 (114)	&	point	&		0.10$\pm$0.04	&	7.0	\\
36	&	164.554640	&	$28_4$-$27_5$ A$^+$, $v_t$=0	&	1021.9	&	8.28(-6)	&	2.31$\times$1.72 (110)	&	2.57$\times$1.42	(28)	&		0.08$\pm$0.03	&	-	&	2.39$\times$1.77 (114)	&	point	&		0.04$\pm$0.02	&	5.7	\\
\hline
\enddata
\tablecomments{
a: Size of the best gaussian fit to the visibilities. Gaussian fits
were performed on the channel showing the peak emission. 
When an elliptical gaussian fit was not successful, a circular
gaussian was attempted. Gaussian fits resulting in a size of 0 \arcsec
are marked by ``point''. Lines dominated by a molecular outflow are
marked by ``outflow''. See text for more details. \\
b: Flux derived from a circular mask with a diameter equal to the
major axis of the beam of the telescope given for each transition in
the Table.
The errors on the fluxes were computed as the quadratic sum of the
statistical error and the calibration uncertainty (about $\sim 20$
\%). \\
c: Non-deconvolved FWHM linewidths derived from a gaussian fit of the
WideX spectra towards the source positions. }
\label{lines_ch3oh}
\end{deluxetable}

\begin{deluxetable}{l c c c c c c c c c c c c}
\centering
\rotate
\tabletypesize
\tiny
\tablecaption{Same as Table \ref{lines_ch3oh} but for $^{13}$CH$_3$OH.}
\startdata 
\hline
\hline
	&		&		&		&		&	\multicolumn{4}{c}{IRAS2A}										&	\multicolumn{4}{c}{IRAS4A}										\\
\cline{6-13}																															
$N$	&	Frequency	&	Transition	&	E$_{up}$	&	A$_{ul}$	&	Beam size		&	Source size	&	Flux			&	$dV_W$	&	Beam size		&	Source size	&	Flux			&	$dV_W$	\\
	&	(GHz)	&		&	(K)	&	(s$^{-1}$)	&	(\arcsec$\times$\arcsec, $^o$)		&	(\arcsec$\times$\arcsec, $^o$)	&	(Jy km/s)			&	(km/s)	&	(\arcsec$\times$\arcsec, $^o$)		&	(\arcsec$\times$\arcsec, $^o$)	&	(Jy km/s)			&	(km/s)	\\
\hline																															
1	&	141.603710	&	$3_0$-$2_0$, A$^+$, $v_t$=0	&	13.6	&	1.15(-5)	&	2.06$\times$1.67 (26)	&	1.26$\times$0.84	(-40)	&	0.142$\pm$0.032	&	4.6	&	2.06$\times$1.67 (26)	&	2.15$\times$1.13	(-33)	&	0.099$\pm$0.026	&	6.9	\\
2	&	165.575639	&	$2_1$-$2_0$, E1, $v_t$=0	&	27.9	&	2.37(-5)	&	2.54$\times$1.71 (113)	&	0.48	&	0.167$\pm$0.044	&	5.2	&	2.38$\times$1.76 (114)	&	2.27$\times$1.39	(-62)	&	0.167$\pm$0.037	&	3.6	\\
3	&	142.807657	&	$3_1$-$2_1$ A$^-$, $v_t$=0	&	28.3	&	1.05(-5)	&	2.06$\times$1.67 (26)	&	1.20$\times$0.78	(18)	&	0.112$\pm$0.026	&	6.6	&	2.06$\times$1.67 (26)	&	point	&	0.069$\pm$0.020	&	5.6	\\
4	&	165.609427	&	$3_1$-$3_0$ E1, $v_t$=0	&	34.6	&	2.36(-5)	&	2.54$\times$1.71 (113)	&	0.56	&	0.293$\pm$0.073	&	6.4	&	2.38$\times$1.76 (114)	&	0.56	&	0.191$\pm$0.042	&	5.6	\\
5	&	165.690996	&	$4_1$-$4_0$ E1, $v_t$=0	&	43.7	&	2.34(-5)	&	2.54$\times$1.71 (113)	&	1.04$\times$0.71	(83)	&	0.270$\pm$0.068	&	5.0	&	2.38$\times$1.76 (114)	&	0.91	&	0.172$\pm$0.040	&	5.0	\\
6	&	165.851224	&	$5_1$-$5_0$ E1, $v_t$=0	&	55.0	&	2.33(-5)	&	2.54$\times$1.71 (113)	&	0.5	&	0.294$\pm$0.067	&	5.0	&	2.38$\times$1.76 (114)	&	2.43$\times$0.73	(-55)	&	0.165$\pm$0.036	&	4.1	\\
7	&	142.173740	&	$5_2$-$6_1$ E2, $v_t$=0	&	60	&	4.85(-6)	&	2.06$\times$1.67 (26)	&	1.25	&	0.033$\pm$0.010	&	6.4	&	-		&	-	&	-			&	-	\\
8	&	166.128782	&	$6_1$-$6_0$ E1, $v_t$=0	&	68.6	&	2.32(-5)	&	2.54$\times$1.71 (113)	&	0.99$\times$0.42	(25)	&	0.327$\pm$0.078	&	5.2	&	2.38$\times$1.76 (114)	&	1.74$\times$0.75	(-88)	&	0.166$\pm$0.041	&	5.5	\\
9	&	163.872900	&	$7_0$-$6_1$ E1, $v_t$=0	&	76.5	&	9.91(-6)	&	2.31$\times$1.72 (110)	&	point	&	0.156$\pm$0.051	&	5.4	&	2.39$\times$1.77 (114)	&	1.80$\times$0.74	(-45)	&	0.151$\pm$0.037	&	4.5	\\
10	&	166.569486	&	$7_1$-$7_0$ E1, $v_t$=0	&	84.5	&	2.31(-5)	&	2.54$\times$1.71 (113)	&	0.99$\times$0.95	(-68)	&	0.419$\pm$0.093	&	6.6	&	2.38$\times$1.76 (114)	&	1.60$\times$1.11	(-60)	&	0.233$\pm$0.050	&	5.6	\\
11	&	142.896760	&	$6_2$-$7_1$ A$^-$, $v_t$=0	&	85.4	&	5.31(-6)	&	2.06$\times$1.67 (26)	&	2.24	&	0.085$\pm$0.021	&	8.1	&	2.06$\times$1.67 (26)	&	1.45$\times$0.77	(0)	&	0.090$\pm$0.023	&	7.5	\\
12	&	165.280537	&	$6_2$-$7_1$ A$^+$, $v_t$=0	&	85.4	&	7.77(-6)	&	2.54$\times$1.71 (113)	&	point	&	0.067$\pm$0.020	&	4.4	&	2.38$\times$1.76 (114)	&	0.76	&	0.079$\pm$0.018	&	8.4	\\
13	&	164.531587	&	$13_1$-$12_2$ A$^-$, $v_t$=0	&	222.3	&	9.96(-6)	&	2.31$\times$1.72 (110)	&	1.01	&	0.121$\pm$0.047	&	3.6	&	2.39$\times$1.77 (114)	&	2.15	&	0.105$\pm$0.028	&	4.2	\\
\hline
\enddata
\label{lines_13ch3oh}
\end{deluxetable}

\begin{deluxetable}{l c c c c c c c c c c c c}
\centering
\rotate
\tabletypesize
\tiny
\tablecaption{Same as Table \ref{lines_ch3oh} but for HCOOCH$_3$.}
\startdata 
\hline
\hline
	&		&		&		&		&	\multicolumn{4}{c}{IRAS2A}											&	\multicolumn{4}{c}{IRAS4A}											\\
\cline{6-13}																																	
$N$	&	Frequency	&	Transition	&	E$_{up}$	&	A$_{ul}$	&	Beam size		&	Source size	&		Flux			&	$dV_W$	&	Beam size		&	Source size	&		Flux			&	$dV_W$	\\
	&	(GHz)	&		&	(K)	&	(s$^{-1}$)	&	(\arcsec$\times$\arcsec, $^o$)		&	(\arcsec$\times$\arcsec, $^o$)	&		(Jy km/s)			&	(km/s)	&	(\arcsec$\times$\arcsec, $^o$)		&	(\arcsec$\times$\arcsec, $^o$)	&		(Jy km/s)			&	(km/s)	\\
\hline																																	
1	&	141.652995 	&	E, $11_{2,9}$-$10_{2,8}$	&	43.2	&	4.06(-5)	&	2.06$\times$1.67 (26)	&	point	&		0.090$\pm$0.024	&	7.2	&	2.06$\times$1.67 (26)	&	1.65$\times$0.50	(-14)	&		0.112$\pm$0.029	&	7.0	\\
2	&	141.667012	&	A, $11_{2,9}$-$10_{2,8}$	&	43.2	&	4.06(-5)	&	2.06$\times$1.67 (26)	&	0.2	&		0.069$\pm$0.018	&	7.2	&	2.06$\times$1.67 (26)	&	0.93	&		0.119$\pm$0.027	&	6.6	\\
3	&	143.234201	&	E, $12_{1,11}$-$11_{1,10}$	&	47.3	&	4.22(-5)	&	2.31$\times$1.82 (26)	&	point	&		0.073$\pm$0.028	&	5.8	&	2.31$\times$1.82 (26)	&	0.44	&		0.091$\pm$0.035	&	4.1	\\
4	&	143.240505	&	A, $12_{1,11}$-$11_{1,10}$	&	47.3	&	4.23(-5)	&	2.31$\times$1.82 (26)	&	point	&		0.072$\pm$0.023	&	5.8	&	2.31$\times$1.82 (26)	&	0.52	&		0.111$\pm$0.036	&	4.1	\\
5	&	142.733524	&	E, $13_{1,13}$-$12_{1,12}$	&	49.3	&	4.28(-5)	&	2.06$\times$1.67 (26)	&	0.43	&		0.190$\pm$0.042	&	7.5	&	2.06$\times$1.67 (26)	&	1.13	&		0.195$\pm$0.044	&	7.1	\\
	&	142.735139	&	A, $13_{1,13}$-$12_{1,12}$	&	49.3	&	4.28(-5)	&			&		&					&		&			&		&					&		\\
6	&	142.815476	&	E, $13_{0,13}$-$12_{0,12}$	&	49.3	&	4.28(-5)	&	2.06$\times$1.67 (26)	&	0.30	&		0.211$\pm$0.047	&	7.6	&	2.06$\times$1.67 (26)	&	1.20$\times$0.70	(32)	&		0.213$\pm$0.050	&	8.3	\\
	&	142.817021	&	A, $13_{0,13}$-$12_{0,12}$	&	49.3	&	4.28(-5)	&			&		&					&		&			&		&					&		\\
7	&	142.924506	&	E, $13_{1,13}$-$12_{0,12}$	&	49.3	&	6.59(-6)	&	2.06$\times$1.67 (26)	&	point	&		0.045$\pm$0.012	&	8.8	&	2.06$\times$1.67 (26)	&	point	&		0.112$\pm$0.027	&	9.7	\\
	&	142.925911	&	A, $13_{1,13}$-$12_{0,12}$	&	49.3	&	6.59(-6)	&			&		&					&		&			&		&					&		\\
8	&	164.955703	&	E, $13_{2,11}$-$12_{2,10}$	&	58.5	&	6.46(-5)	&	2.31$\times$1.72 (110)	&	0.85	&		0.159$\pm$0.051	&	6.2	&	2.39$\times$1.77 (114)	&	1.88	&		0.236$\pm$0.058	&	5.5	\\
9	&	164.968638	&	A, $13_{2,11}$-$12_{2,10}$	&	58.5	&	6.46(-5)	&	2.31$\times$1.72 (110)	&	0.4	&		0.124$\pm$0.039	&	3.5	&	2.39$\times$1.77 (114)	&	0.38	&		0.209$\pm$0.047	&	7.0	\\
10	&	163.829677	&	E, $14_{1,13}$-$13_{1,12}$	&	62.5	&	6.37(-5)	&	2.31$\times$1.72 (110)	&	1.13$\times$0.46	(66)	&		0.131$\pm$0.056	&	6.7	&	2.28$\times$1.76	&	0.88	&		0.152$\pm$0.048	&	4.7	\\
11	&	163.835525	&	A, $14_{1,13}$-$13_{1,12}$	&	62.5	&	6.37(-5)	&	2.31$\times$1.72 (110)	&	1.82$\times$1.36	(14)	&		0.129$\pm$0.039	&	3.6	&	2.28$\times$1.76	(-70)	&	1.64$\times$0.73	&		0.149$\pm$0.043	&	4.4	\\
12	&	165.653657	&	E, $14_{2,13}$-$13_{1,12}$	&	62.6	&	7.63(-6)	&	2.54$\times$1.71 (113)	&	1.6	&		0.057$\pm$0.017	&	6.3	&	2.38$\times$1.76 (114)	&	1.65	&		0.196$\pm$0.026	&	14.3	\\
	&	165.657529	&	A, $14_{2,13}$-$13_{1,12}$	&	62.6	&	7.63(-6)	&			&		&					&		&			&		&					&		\\
13	&	163.925845	&	E, $15_{0,15}$-$14_{1,14}$	&	64.5	&	1.02(-5)	&	2.31$\times$1.73	&	1.89$\times$1.30	(28)	&		0.070$\pm$0.033	&	7.8	&	2.39$\times$1.77 (114)	(-61)	&	2.40$\times$0.86	&		0.141$\pm$0.035	&	6.2	\\
	&	163.927369	&	A, $15_{0,15}$-$14_{1,14}$	&	64.5	&	1.02(-5)	&			&		&					&		&			&		&					&		\\
14	&	163.960387	&	A, $15_{1,15}$-$14_{1,14}$	&	64.5	&	6.53(-5)	&	2.31$\times$1.72 (110)	&	1.48$\times$0.56	(79)	&		0.301$\pm$0.072	&	6.9	&	2.28$\times$1.76	&	0.96	&		0.329$\pm$0.080	&	5.9	\\
	&	163.961884	&	E, $15_{1,15}$-$14_{1,14}$	&	64.5	&	6.53(-5)	&			&		&					&		&			&		&					&		\\
15	&	163.987455	&	E, $15_{0,15}$-$14_{0,14}$	&	64.5	&	6.54(-5)	&	2.31$\times$1.72 (110)	&	0.78$\times$0.33	(19)	&		0.299$\pm$0.072	&	6.9	&	2.22$\times$1.94	&	0.93	&		0.086$\pm$0.027	&	6.0	\\
15	&	163.988912	&	A, $15_{0,15}$-$14_{0,14}$	&	64.5	&	6.54(-5)	&			&		&					&		&	2.22$\times$1.94	&	0.55	&		0.136$\pm$0.034	&		\\
16	&	164.022026	&	E, $15_{1,15}$-$14_{0,14}$	&	64.5	&	1.03(-5)	&	2.31$\times$1.72 (110)	&	1.81$\times$0.76	(90)	&		0.118$\pm$0.046	&	9.7	&	2.39$\times$1.77 (114)	&	point	&		0.202$\pm$0.065	&	10.0	\\
	&	164.023416	&	A, $15_{1,15}$-$14_{0,14}$	&	64.5	&	1.03(-5)	&			&		&					&		&			&		&					&		\\
17	&	164.205978	&	E, $13_{4,9}$-$12_{4,8}$	&	64.9	&	5.98(-5)	&	2.31$\times$1.72 (110)	&	1.17	&		0.126$\pm$0.039	&	5.9	&	2.28$\times$1.76	&	0.65	&		0.193$\pm$0.088	&	6.6	\\
18	&	164.223815	&	A, $13_{4,9}$-$12_{4,8}$	&	64.9	&	5.98(-5)	&	2.31$\times$1.72 (110)	&	1.29$\times$0.90	(73)	&		0.138$\pm$0.041	&	6.5	&	2.28$\times$1.76	(0)	&	3.02$\times$1.74	&		0.202$\pm$0.074	&	5.2	\\
19	&	142.664676	&	E, $12_{1,11}$-$11_{1,10}, v_t$=1	&	234.0	&	4.19(-5)	&	2.06$\times$1.67 (26)	&	-	&	$<$	0.040			&	-	&	2.06$\times$1.67 (26)	&	2.65$\times$2.24	(0)	&		0.037$\pm$0.009	&	6.1	\\
20	&	142.125411	&	E, $13_{0,13}$-$12_{0,12}, v_t$=1	&	236.9	&	4.24(-5)	&	2.06$\times$1.67 (26)	&	1.19	&		0.033$\pm$0.011	&	6.0	&	2.06$\times$1.67 (26)	&	-	&	$<$	0.048			&	-	\\
21	&	142.052774	&	A, $13_{0,13}$-$12_{0,12}, v_t$=1	&	236.9	&	4.21(-5)	&	2.06$\times$1.67 (26)	&	-	&	$<$	0.040			&	-	&	2.06$\times$1.67 (26)	&	2.24$\times$0.56	(-29)	&		0.038$\pm$0.018	&	8.1	\\
22	&	166.388878	&	$13_{3,10}$-$12_{3,9}, v_t$=1	&	248.0	&	6.51(-5)	&	2.54$\times$1.71 (113)	&	point	&		0.072$\pm$0.032	&	5.1	&	2.63$\times$1.73	&	-	&	$<$	0.043			&	-	\\
\hline
\enddata
\label{lines_hcooch3}
\end{deluxetable}

\begin{deluxetable}{l c c c c c c c c c c c c}
\centering
\rotate
\tabletypesize
\tiny
\tablecaption{Same as Table \ref{lines_ch3oh} but for CH$_3$CN.}
\startdata 
\hline
\hline
	&		&		&		&		&	\multicolumn{4}{c}{IRAS2A}										&	\multicolumn{4}{c}{IRAS4A}										 \\
\cline{6-13}																															
$N$	&	Frequency	&	Transition	&	E$_{up}$	&	A$_{ul}$	&	Beam size		&	Source size	&	Flux			&	$dV_W$	&	Beam size		&	Source size	&	Flux			&	$dV_W$	\\
	&	(GHz)	&		&	(K)	&	(s$^{-1}$)	&	(\arcsec$\times$\arcsec, $^o$)		&	(\arcsec$\times$\arcsec, $^o$)	&	(Jy km/s)			&	(km/s)	&	(\arcsec$\times$\arcsec, $^o$)		&	(\arcsec$\times$\arcsec, $^o$)	&	(Jy km/s)			&	(km/s)	\\
\hline																															
1	&	165.569082	&	$9_0$-$8_0$	&	39.7	&	2.65(-4)	&	2.54$\times$1.71 (113)	&	0.99$\times$0.78	(45)	&	1.845$\pm$0.373	&	10.2	&	2.38$\times$1.76 (114)	&	1.68$\times$1.27	(-67)	&	1.058$\pm$0.217	&	9.4	\\
	&	165.565891	&	$9_1$-$8_1$	&	46.9	&	2.62(-4)	&			&		&				&		&			&		&				&		\\
2	&	165.556322	&	$9_2$-$8_2$	&	68.3	&	2.52(-4)	&	2.54$\times$1.71 (113)	&	1.07$\times$0.70	(-57)	&	0.818$\pm$0.170	&	5.5	&	2.38$\times$1.76 (114)	&	0.55	&	0.533$\pm$0.116	&	5.1	\\
3	&	165.540377	&	$9_3$-$8_3$	&	104.0	&	2.36(-4)	&	2.54$\times$1.71 (113)	&	0.98$\times$0.77	(65)	&	0.986$\pm$0.205	&	6.1	&	2.38$\times$1.76 (114)	&	0.51	&	0.493$\pm$0.104	&	5.7	\\
4	&	165.518064	&	$9_4$-$8_4$	&	154.0	&	2.13(-4)	&	2.54$\times$1.71 (113)	&	0.87$\times$0.79	(-26)	&	0.541$\pm$0.116	&	6.1	&	2.38$\times$1.76 (114)	&	1.22$\times$0.5	(-56)	&	0.371$\pm$0.081	&	6.0	\\
5	&	165.489391	&	$9_5$-$8_5$	&	218.3	&	1.83(-4)	&	2.54$\times$1.71 (113)	&	0.44	&	0.426$\pm$0.097	&	6.8	&	2.38$\times$1.76 (114)	&	0.52	&	0.482$\pm$0.109	&	8.7	\\
6	&	165.454370	&	$9_6$-$8_6$	&	296.8	&	1.47(-4)	&	2.54$\times$1.71 (113)	&	0.74$\times$0.45	(-80)	&	0.331$\pm$0.078	&	6.6	&	2.38$\times$1.76 (114)	&	1.29$\times$0.65	(-74)	&	0.310$\pm$0.070	&	5.3	\\
7	&	165.413015	&	$9_7$-$8_7$	&	389.5	&	1.04(-4)	&	2.54$\times$1.70	&	0.44	&	0.111$\pm$0.049	&	6.4	&	2.38$\times$1.76 (114)	&	1.88$\times$0.75	(-20)	&	0.075$\pm$0.032	&	4.8	\\
\hline
\enddata
\label{lines_ch3cn}
\end{deluxetable}

\begin{deluxetable}{l c c c c c c c c c c c c}
\centering
\rotate
\tabletypesize
\tiny
\tablecaption{Same as Table \ref{lines_ch3oh} but for CH$_3$OCH$_3$.}
\startdata 
\hline
\hline
	&		&		&		&		&	\multicolumn{4}{c}{IRAS2A}										&	\multicolumn{4}{c}{IRAS4A}											\\
\cline{6-13}																																
$N$	&	Frequency	&	Transition	&	E$_{up}$	&	A$_{ul}$	&	Beam size		&	Source size	&	Flux			&	$dV_W$	&	Beam size		&	Source size	&		Flux			&	$dV_W$	\\
	&	(GHz)	&		&	(K)	&	(s$^{-1}$)	&	(\arcsec$\times$\arcsec, $^o$)		&	(\arcsec$\times$\arcsec, $^o$)	&	(Jy km/s)			&	(km/s)	&	(\arcsec$\times$\arcsec, $^o$)		&	(\arcsec$\times$\arcsec, $^o$)	&		(Jy km/s)			&	(km/s)	\\
\hline																																
1	&	143.017994	&	$3_{2,2}$-$2_{1,1}$, EA	&	11.1	&	1.09e-5	&	2.06$\times$1.67 (26)	&	0.81	&	0.044$\pm$0.023	&	10.1	&	2.06$\times$1.67 (26)	&	3.81$\times$1.66	(89)	&		0.060$\pm$0.022	&	8.5	\\
	&	143.018373	&	$3_{2,2}$-$2_{1,1}$, AE	&	11.1	&	1.09e-5	&			&		&				&		&			&		&					&		\\
	&	143.020764	&	$3_{2,2}$-$2_{1,1}$, EE	&	11.1	&	1.09e-5	&			&		&				&		&			&		&					&		\\
	&	143.023345	&	$3_{2,2}$-$2_{1,1}$, AA	&	11.1	&	1.09e-5	&			&		&				&		&			&		&					&		\\
2	&	144.856766	&	$6_{3,3}$-$6_{2,4}$ EA	&	31.8	&	1.04(-5)	&	2.08$\times$1.65 (30)	&	0.38	&	0.070$\pm$0.016	&	5.1	&	2.16$\times$1.73 (25)	&	1.29$\times$1.28	(0)	&		0.161$\pm$0.045	&	19.6	\\
	&	144.858984	&	$6_{3,3}$-$6_{2,4}$ EE	&	31.8	&	1.06(-5)	&			&		&				&		&			&		&					&		\\
	&	144.855091	&	$6_{3,3}$-$6_{2,4}$ AE	&	31.8	&	1.06(-5)	&			&		&				&		&			&		&					&		\\
3	&	143.599420	&	$7_{3,4}$-$7_{2,5}$ AE	&	38.2	&	1.10(-5)	&	2.08$\times$1.65 (30)	&	1.39$\times$0.94	(-11)	&	0.083$\pm$0.030	&	16.1	&	2.16$\times$1.73 (25)	&	point	&		0.136$\pm$0.044	&	14.0	\\
	&	143.600084	&	$6_{3,3}$-$6_{2,4}$ EA	&	38.2	&	1.10(-5)	&			&		&				&		&			&		&					&		\\
	&	143.602993	&	$6_{3,3}$-$6_{2,4}$ EE	&	38.2	&	1.10(-5)	&			&		&				&		&			&		&					&		\\
	&	143.606232	&	$6_{3,3}$-$6_{2,4}$ AA	&	38.2	&	1.10(-5)	&			&		&				&		&			&		&					&		\\
4	&	141.828855	&	$8_{3,5}$-$8_{2,6}$, AE 	&	45.5	&	1.11e-5	&	2.06$\times$1.67 (26)	&	0.65	&	0.149$\pm$0.045	&	14.3	&	2.06$\times$1.67 (26)	&	0.46	&		0.140$\pm$0.049	&	16.7	\\
	&	141.829146	&	$8_{3,5}$-$8_{2,6}$, EA 	&	45.5	&	1.11e-5	&			&		&				&		&			&		&					&		\\
	&	141.832261	&	$8_{3,5}$-$8_{2,6}$, EE	&	45.5	&	1.11e-5	&			&		&				&		&			&		&					&		\\
	&	141.835521	&	$8_{3,5}$-$8_{2,6}$, AA	&	45.5	&	1.11e-5	&			&		&				&		&			&		&					&		\\
5	&	143.159951	&	$13_{2,12}$-$13_{1,13}$, EA 	&	88	&	7.75e-6	&	2.06$\times$1.67 (26)	&	0.31	&	0.050$\pm$0.012	&	7.1	&	2.06$\times$1.67 (26)	&	point	&		0.079$\pm$0.022	&	13.0	\\
	&	143.159952	&	$13_{2,12}$-$13_{1,13}$, AE	&	88	&	7.75e-6	&			&		&				&		&			&		&					&		\\
	&	143.162986	&	$13_{2,12}$-$13_{1,13}$, EE	&	88	&	7.75e-6	&			&		&				&		&			&		&					&		\\
	&	143.166020	&	$13_{2,12}$-$13_{1,13}$, AA	&	88	&	7.75e-6	&			&		&				&		&			&		&					&		\\
6	&	165.208844	&	$15_{3,13}$-$15_{2,14}$ EA	&	122	&	1.75(-5)	&	2.31$\times$1.72 (110)	&	point	&	0.121$\pm$0.046	&	13.8	&	2.39$\times$1.77 (114)	&	2.47$\times$1.20	(-26)	&		0.115$\pm$0.025	&	9.5	\\
	&	165.208848	&	$15_{3,13}$-$15_{2,14}$ AE	&	122	&	1.75(-5)	&			&		&				&		&			&		&					&		\\
	&	165.211731	&	$15_{3,13}$-$15_{2,14}$ EE	&	122	&	1.75(-5)	&			&		&				&		&			&		&					&		\\
	&	165.214617	&	$15_{3,13}$-$15_{2,14}$ AA	&	122	&	1.75(-5)	&			&		&				&		&			&		&					&		\\
7	&	164.988708	&	$20_{3,18}$-$19_{4,15}$ AA	&	204	&	4.60(-6)	&	2.31$\times$1.72 (110)	&	point	&	0.074$\pm$0.033	&	5.5	&	2.39$\times$1.77 (114)	&	-	&	$<$	0.072			&	-	\\
	&	164.990831	&	$20_{3,18}$-$19_{4,15}$ EE	&	204	&	4.60(-6)	&			&		&				&		&			&		&					&		\\
	&	164.992951	&	$20_{3,18}$-$19_{4,15}$ EA	&	204	&	4.60(-6)	&			&		&				&		&			&		&					&		\\
	&	164.992958	&	$20_{3,18}$-$19_{4,15}$ AE	&	204	&	4.60(-6)	&			&		&				&		&			&		&					&		\\
8	&	142.403201	&	$25_{1,24}$-$25_{2,23}$, EA 	&	313.5	&	1.57e-5	&	2.06$\times$1.67 (26)	&	point	&	0.051$\pm$0.013	&	7.2	&	2.06$\times$1.67 (26)	&	0.41	&		0.035$\pm$0.011	&	6.0	\\
	&	142.403201	&	$25_{1,24}$-$25_{2,23}$, AE 	&	313.5	&	1.57e-5	&			&		&				&		&			&		&					&		\\
	&	142.404442	&	$25_{1,24}$-$25_{2,23}$, EE	&	313.5	&	1.57e-5	&			&		&				&		&			&		&					&		\\
	&	142.405682	&	$25_{1,24}$-$25_{2,23}$, AA	&	313.5	&	1.57e-5	&			&		&				&		&			&		&					&		\\
\hline
\enddata
\label{lines_ch3och3}
\end{deluxetable}

\begin{deluxetable}{l c c c c c c c c c c c c}
\centering
\rotate
\tabletypesize
\tiny
\tablecaption{Same as Table \ref{lines_ch3oh} but for C$_2$H$_5$OH.}
\startdata 
\hline
\hline
	&		&		&		&		&	\multicolumn{4}{c}{IRAS2A}											&		\multicolumn{4}{c}{IRAS4A}								\\		
\cline{6-13}																																	
$N$	&	Frequency	&	Transition	&	E$_{up}$	&	A$_{ul}$	&	Beam size		&	Source size	&		Flux			&	$dV_W$	&	Beam size		&	Source size	&		Flux			&	$dV_W$	\\
	&	(GHz)	&		&	(K)	&	(s$^{-1}$)	&	(\arcsec$\times$\arcsec, $^o$)		&	(\arcsec$\times$\arcsec, $^o$)	&		(Jy km/s)			&	(km/s)	&	(\arcsec$\times$\arcsec, $^o$)		&	(\arcsec$\times$\arcsec, $^o$)	&		(Jy km/s)			&	(km/s)	\\
\hline																																	
1	&	142.285054	&	$9_{0,9,2}$-$8_{1,8,2}$ 	&	37.2	&	1.51(-5)	&	2.06$\times$1.67 (26)	&	2.74$\times$0.47	(1)	&		0.027$\pm$0.011	&	4.1	&	2.06$\times$1.67 (26)	&	0.52	&		0.045$\pm$0.015	&	8.2	\\
2	&	164.900973	&	$6_{0,6,1}$-$5_{1,4,0}$   	&	78.8	&	1.22(-5)	&	2.31$\times$1.72 (110)	&	-	&	$<$	0.066			&	-	&	2.39$\times$1.77 (114)	&	-	&	$<$	0.072			&	-	\\
3	&	144.057496	&	$13_{3,11,2}$-$13_{2,12,2}$	&	87.9	&	1.80(-5)	&	2.08$\times$1.65 (30)	&	point	&		0.031$\pm$0.010	&	-	&	2.16$\times$1.73 (25)	&	point	&		0.067$\pm$0.018	&	4.1	\\
4	&	141.820317 	&	$8_{1,7,0}$-$7_{1,6,0}$ 	&	88.8	&	2.46e-5	&	2.06$\times$1.67 (26)	&	1.16	&		0.034$\pm$0.013	&	4.1	&	2.06$\times$1.67 (26)	&	point	&		0.039$\pm$0.015	&	6.4	\\
5	&	164.626167	&	$5_{4,1,0}$-$4_{3,1,1}$	&	88.8	&	2.06(-5)	&	2.31$\times$1.72 (110)	&	-	&	$<$	0.066			&	-	&	2.39$\times$1.77 (114)	&	-	&	$<$	0.072			&	-	\\
6	&	164.630894	&	$5_{4,2,0}$-$4_{3,2,1}$	&	88.8	&	2.06(-5)	&	2.31$\times$1.72 (110)	&	-	&	$<$	0.066			&	-	&	2.39$\times$1.77 (114)	&	-	&	$<$	0.081			&	-	\\
7	&	144.493107	&	$14_{2,13,2}$-$14_{11,4,2}$	&	92.6	&	1.86(-5)	&	2.08$\times$1.65 (30)	&	-	&	$<$	0.046			&	-	&	2.16$\times$1.73 (25)	&	1.19	&		0.045$\pm$0.021	&	4.1	\\
8	&	166.259891	&	$10_{1,10,0}$-$9_{1,9,0}$	&	102.1	&	4.03(-5)	&	2.54$\times$1.71 (113)	&	point	&		0.114$\pm$0.049	&	7.3	&	2.38$\times$1.76 (114)	&	1.08	&		0.118$\pm$0.053	&	4.2	\\
9	&	166.758214	&	$10_{1,10,1}$-$9_{1,9,1}$	&	106.8	&	4.07(-5)	&	2.54$\times$1.71 (113)	&	0.95	&		0.043$\pm$0.030	&	-	&	2.38$\times$1.76 (114)	&	point	&		0.053$\pm$0.020	&	-	\\
10	&	142.083012	&	$11_{2,10,1}$-$11_{1,10,0}$  	&	121	&	9.96(-6)	&	2.06$\times$1.67 (26)	&	point	&		0.028$\pm$0.009	&	16.3	&	2.06$\times$1.67 (26)	&	-	&	$<$	0.040			&	-	\\
11	&	164.511879	&	$18_{3,15,2}$-$17_{4,14,2}$	&	156.8	&	8.73(-6)	&	2.31$\times$1.72 (110)	&	point	&		0.036$\pm$0.021	&	3.6	&	2.39$\times$1.77 (114)	&	-	&	$<$	0.072			&	-	\\
12	&	164.429108	&	$14_{3,12,1}$-$13_{4,10,0}$	&	160.0	&	4.25(-6)	&	2.31$\times$1.72 (110)	&	-	&	$<$	0.066			&	-	&	2.39$\times$1.77 (114)	&	-	&	$<$	0.072			&	-	\\
13	&	142.046310	&	$20_{2,18,2}$-$20_{1,19,2}$	&	185.5	&	1.77(-5)	&	2.06$\times$1.67 (26)	&	-	&	$<$	0.040			&	-	&	2.06$\times$1.67 (26)	&	-	&	$<$	0.040			&	-	\\
13	&	163.515766	&	$21_{3,19,2}$-$20_{4,16,2}$	&	205.4	&	9.22(-6)	&	2.31$\times$1.72 (110)	&	-	&	$<$	0.066			&	-	&	2.39$\times$1.77 (114)	&	-	&	$<$	0.072			&	-	\\
14	&	141.735490	&	$21_{4,17,2}$-$21_{3,18,2}$	&	215.5	&	1.72e-5	&	2.06$\times$1.67 (26)	&	point	&		0.030$\pm$0.015	&	8.0	&	2.06$\times$1.67 (26)	&	point	&		0.048$\pm$0.018	&	7.6	\\
\hline
\enddata
\label{lines_c2h5oh}
\end{deluxetable}

\begin{deluxetable}{l c c c c c c c c c c c c}
\centering
\rotate
\tabletypesize
\tiny
\tablecaption{Same as Table \ref{lines_ch3oh} but for HCOCH$_2$OH.}
\startdata 
\hline
\hline
	&		&		&		&		&	\multicolumn{4}{c}{IRAS2A}											&	\multicolumn{4}{c}{IRAS4A}											\\
\cline{6-13}																																	
$N$	&	Frequency	&	Transition	&	E$_{up}$	&	A$_{ul}$	&	Beam size		&	Source size	&		Flux			&	$dV_W$	&	Beam size		&	Source size	&		Flux			&	$dV_W$	\\
	&	(GHz)	&		&	(K)	&	(s$^{-1}$)	&	(\arcsec$\times$\arcsec, $^o$)		&	(\arcsec$\times$\arcsec, $^o$)	&		(Jy km-s)			&	(km/s)	&	(\arcsec$\times$\arcsec, $^o$)		&	(\arcsec$\times$\arcsec, $^o$)	&		(Jy km-s)			&	(km/s)	\\
\hline																																	
1	&	163.951686	&	$8_{3,5}$-$7_{2,6}, v=0$	&	25.6	&	3.68(-5)	&	2.31$\times$1.72 (110)	&	-	&	$<$	0.066			&	-	&	2.39$\times$1.77 (114)	&		&		-			&	-	\\
2	&	143.640947	&	$14_{0,14}$-$13_{1,13}, v=0$	&	53.1	&	8.04(-5)	&	2.08$\times$1.65 (30)	&	point	&		0.037$\pm$0.020	&	-	&	2.06$\times$1.67 (26)	&	-	&	$<$	0.129			&	-	\\
3	&	143.765755	&	$14_{1,14}$-$13_{0,13}, v=0$	&	53.1	&	8.06(-5)	&	2.08$\times$1.65 (30)	&	-	&	$<$	0.046			&	-	&	2.06$\times$1.67 (26)	&	point	&		0.069$\pm$0.035	&	5.6	\\
4	&	164.047038	&	$15_{2,14}$-$14_{1,13}, v=0$	&	66.5	&	9.25(-5)	&	2.31$\times$1.72 (110)	&	point	&		0.076$\pm$0.034	&	7.4	&	2.39$\times$1.77 (114)	&	1.89	&		0.132$\pm$0.051	&	4.5	\\
5	&	163.542260	&	$16_{0,16}$-$15_{1,15}, v=0$	&	68.3	&	1.21(-4)	&	2.31$\times$1.72 (110)	&	point	&		0.037$\pm$0.023	&	7.8	&	2.39$\times$1.77 (114)	&	-	&		0.172$\pm$0.050	&	8.0	\\
6	&	163.580057	&	$16_{1,16}$-$15_{0,15}, v=0$	&	68.3	&	1.21(-4)	&	2.31$\times$1.72 (110)	&	point	&		0.038$\pm$0.021	&	8.5	&	2.39$\times$1.77 (114)	&	point	&		0.103$\pm$0.038	&	5.0	\\
7	&	163.697251	&	$12_{7,5}$-$12_{6,6}, v=0$	&	73.1	&	5.60(-5)	&	2.31$\times$1.72 (110)	&	-	&	$<$	0.066			&	-	&	2.39$\times$1.77 (114)	&	point	&		0.071$\pm$0.028	&	3.6	\\
8	&	163.709163	&	$12_{7,6}$-$12_{6,7}, v=0$	&	73.1	&	5.60(-5)	&	2.31$\times$1.72 (110)	&	-	&	$<$	0.066			&	-	&	2.39$\times$1.77 (114)	&	1.55	&		0.076$\pm$0.045	&	3.8	\\
9	&	142.784665	&	$23_{6,18}$-$23 _{5,19}$	&	177	&	5.19(-5)	&	2.08$\times$1.65 (30)	&	-	&	$<$	0.040			&	-	&	2.06$\times$1.67 (26)	&	point	&		0.039$\pm$0.014	&	5.2	\\
\hline
\enddata
\label{lines_hcoch2oh}
\end{deluxetable}

\begin{deluxetable}{l c c c c c c c c c c c c}
\centering
\rotate
\tabletypesize
\tiny
\tablecaption{Same as Table \ref{lines_ch3oh} but for C$_2$H$_5$CN.}
\startdata 
\hline
\hline
	&		&		&		&		&	\multicolumn{4}{c}{IRAS2A}											&	\multicolumn{4}{c}{IRAS4A}											\\
\cline{6-13}																																	
$N$	&	Frequency	&	Transition	&	E$_{up}$	&	A$_{ul}$	&	Beam size		&	Source size	&		Flux			&	$dV_W$	&	Beam size		&	Source size	&		Flux			&	$dV_W$	\\
	&	(GHz)	&		&	(K)	&	(s$^{-1}$)	&	(\arcsec$\times$\arcsec, $^o$)		&	(\arcsec$\times$\arcsec, $^o$)	&		(Jy km/s)			&	(km/s)	&	(\arcsec$\times$\arcsec, $^o$)		&	(\arcsec$\times$\arcsec, $^o$)	&		(Jy km/s)			&	(km/s)	\\
\hline																																	
1	&	142.346330	&	$16_{2,15}$-$15_{2,14}$	&	62.7	&	2.37(-4)	&	2.06$\times$1.67 (26)	&	2.33	&		0.029$\pm$0.010	&	12.0	&	2.06$\times$1.67 (26)	&	2.51$\times$1.17 (-40)	&		0.045$\pm$0.013	&	6.8	\\
2	&	143.529200	&	$16_{3,14}$-$15_{3,13}$	&	68.5	&	2.39(-4)	&	2.08$\times$1.65 (30)	&	point	&		0.025$\pm$0.009	&	4.1	&	2.16$\times$1.73 (25)	&	-	&	$<$	0.081			&	-	\\
3	&	144.104740	&	$16_{3,13}$-$15_{3,12}$	&	68.6	&	2.42(-4)	&	2.08$\times$1.65 (30)	&	-	&	$<$	0.046			&	-	&	2.16$\times$1.73 (25)	&	-	&	$<$	0.081			&	-	\\
4	&	143.506970	&	$16_{4,13}$-$15_{4,12}$	&	76.3	&	2.32(-4)	&	2.08$\times$1.65 (30)	&	-	&	$<$	0.046			&	-	&	2.16$\times$1.73 (25)	&	-	&	$<$	0.081			&	-	\\
5	&	143.535290	&	$16_{4,12}$-$15_{4,11}$	&	76.3	&	2.32(-4)	&	2.08$\times$1.65 (30)	&	-	&	$<$	0.046			&	-	&	2.16$\times$1.73 (25)	&	-	&	$<$	0.081			&	-	\\
6	&	164.584755	&	$19_{0,19}$-$18_{0,18}$	&	80	&	3.74(-4)	&	2.31$\times$1.72 (110)	&	-	&	$<$	0.066			&	-	&	2.39$\times$1.77 (114)	&	-	&	$<$	0.072			&	-	\\
8	&	163.948705	&	$19_{1,19}$-$18_{1,18}$	&	80.1	&	3.69(-4)	&	2.31$\times$1.72 (110)	&	-	&	$<$	0.066			&	-	&	2.39$\times$1.77 (114)	&	point	&		0.110$\pm$0.044	&	8.5	\\
7	&	143.406554	&	$16_{5,12}$-$15_{5,11}$	&	86.3	&	2.23(-4)	&	2.08$\times$1.65 (30)	&	point	&		0.024$\pm$0.016	&	5.8	&	2.16$\times$1.73 (25)	&	-	&	$<$	0.081			&	-	\\
	&	143.407188	&	$16_{5,11}$-$15_{5,10}$	&	86.3	&	2.23(-4)	&	2.08$\times$1.65 (30)	&		&					&		&			&		&					&	-	\\
9	&	143.335284	&	$16_{8,8}$-$15_{8,7}$	&	129.6	&	1.85(-4)	&	2.06$\times$1.67 (26)	&	point	&		0.045$\pm$0.011	&	7.4	&	2.06$\times$1.67 (26)	&	0.17	&		0.037$\pm$0.012	&	11.2	\\
	&	143.335284	&	$16_{8,9}$-$15_{8,8}$	&	129.6	&	1.85(-4)	&			&		&					&		&			&		&					&		\\
	&	143.337710	&	$16_{7,10}$-$15_{7,9}$	&	112.9	&	1.99(-4)	&			&		&					&		&			&		&					&		\\
	&	143.337710	&	$16_{7,9}$-$15_{7,8}$	&	112.9	&	1.99(-4)	&			&		&					&		&			&		&					&		\\
10	&	143.343925	&	$16_{9,7}$-$15_{9,6}$	&	148.4	&	1.68(-4)	&	2.06$\times$1.67 (26)	&	point	&		0.034$\pm$0.010	&	7.1	&	2.06$\times$1.67 (26)	&	1.43	&		0.045$\pm$0.012	&	4.1	\\
	&	143.343925	&	$16_{9,8}$-$15_{9,7}$	&	148.4	&	1.68(-4)	&			&		&					&		&			&		&					&		\\
\hline
\enddata
\label{lines_c2h5cn}
\end{deluxetable}

\begin{deluxetable}{l c c c c c c c c c c c c}
\centering
\rotate
\tabletypesize
\tiny
\tablecaption{Same as Table \ref{lines_ch3oh} but for other molecules.}
\startdata 
\hline
\hline
	&		&		&		&		&	\multicolumn{4}{c}{IRAS2A}										&	\multicolumn{4}{c}{IRAS4A}											\\
\cline{6-13}																																
Molecule	&	Frequency	&	Transition	&	E$_{up}$	&	A$_{ul}$	&	Beam size		&	Source size	&	Flux			&	$dV_W$	&	Beam size		&	Source size	&		Flux			&	$dV_W$	\\
	&	(GHz)	&		&	(K)	&	(s$^{-1}$)	&	(\arcsec$\times$\arcsec, $^o$)		&	(\arcsec$\times$\arcsec, $^o$)	&	(Jy km/s)			&	(km/s)	&	(\arcsec$\times$\arcsec, $^o$)		&	(\arcsec$\times$\arcsec, $^o$)	&		(Jy km/s)			&	(km/s)	\\
\hline																																
HC$_3$N	&	163.753389	&	$18$-$17$	&	74.7	&	3.45(-4)	&	2.31$\times$1.72 (110)	&	2.66$\times$1.80 (34)	&	0.572$\pm$0.116	&	6.2	&	2.39$\times$1.77 (114)	&	6.03$\times$1.96 (17)	&		0.553$\pm$0.121	&	7.4	\\
H$_2$$^{13}$CO	&	141.98374	&	$2_{0,2}$-$1_{0,1}$	&	10.2	&	7.25(-5)	&	2.06$\times$1.67 (26)	&	0.60	&	0.115$\pm$0.028	&	6.3	&	2.06$\times$1.67 (26)	&	2.45$\times$1.25 (0)	&		0.062$\pm$0.022	&	7.3	\\
H$_2$C$^{18}$O	&	143.213062	&	$2_{1,1}$-$1_{1,0}$	&	22.2	&	5.57(-5)	&	2.06$\times$1.67 (26)	&	1.94$\times$1.05 (1)	&	0.028$\pm$0.007	&	-	&	-		&	-	&		-			&	-	\\
NH$_2$CHO	&	142.701479	&	$7_{7,1,7,8}$-$6_{1,6,7}$	&	30.4	&	2.02(-4)	&	2.06$\times$1.67 (26)	&	point	&	0.185$\pm$0.043	&	8.2	&	2.06$\times$1.67 (26)	&	point	&		0.090$\pm$0.025	&	7.4	\\
CH$_2$CO	&	142.76892	&	$7_{1,6}$-$6_{1,5}$	&	40.5	&	3.10(-5)	&	2.06$\times$1.67 (26)	&	1.03$\times$0.61 (30)	&	0.119$\pm$0.028	&	5.8	&	2.06$\times$1.67 (26)	&	1.80$\times$1.34 (34)	&		0.184$\pm$0.040	&	5.1	\\
\hline
\enddata
\label{lines_others}
\end{deluxetable}



\begin{figure}[htp]
\centering
\includegraphics[width=60mm]{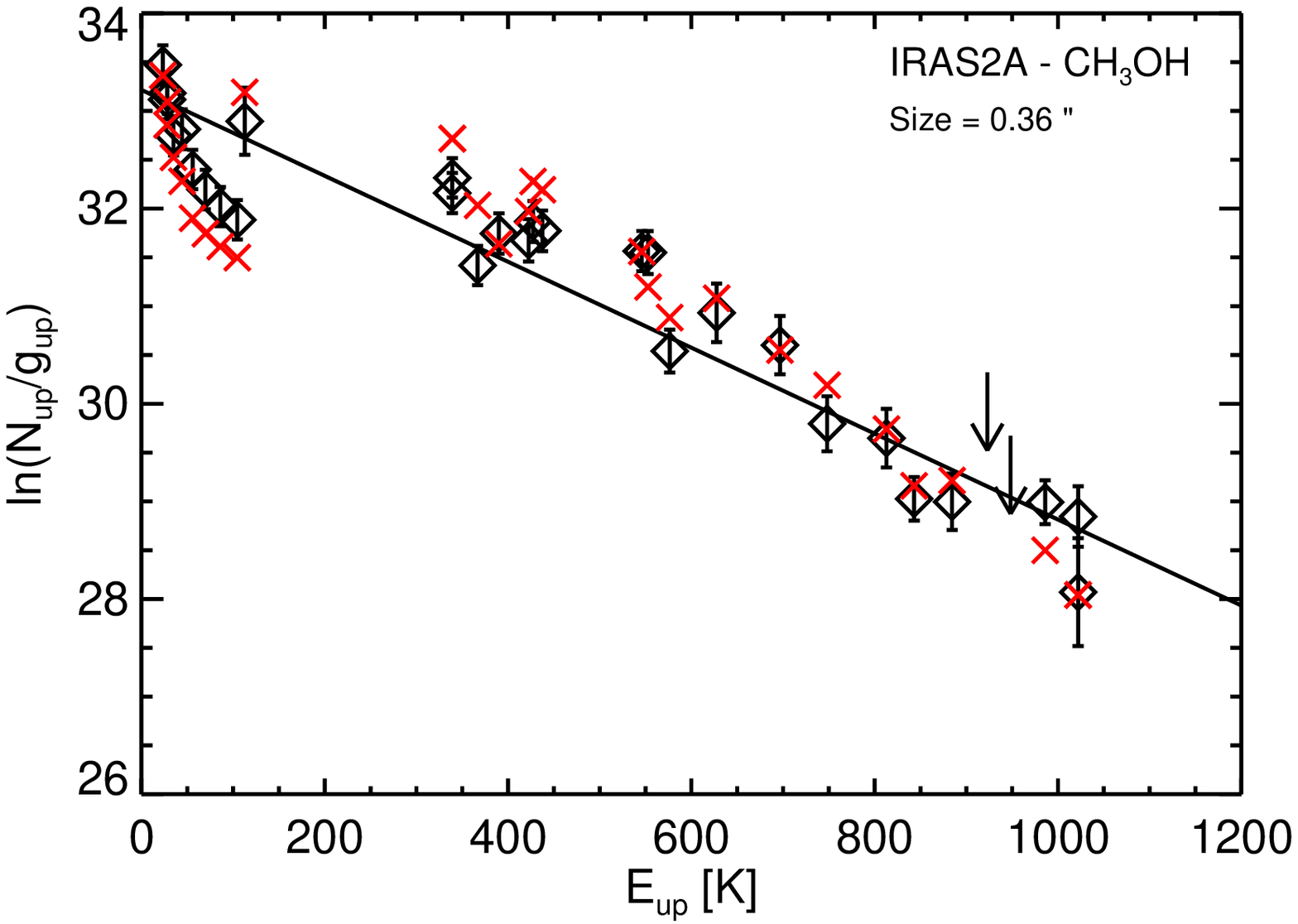} 
\includegraphics[width=60mm]{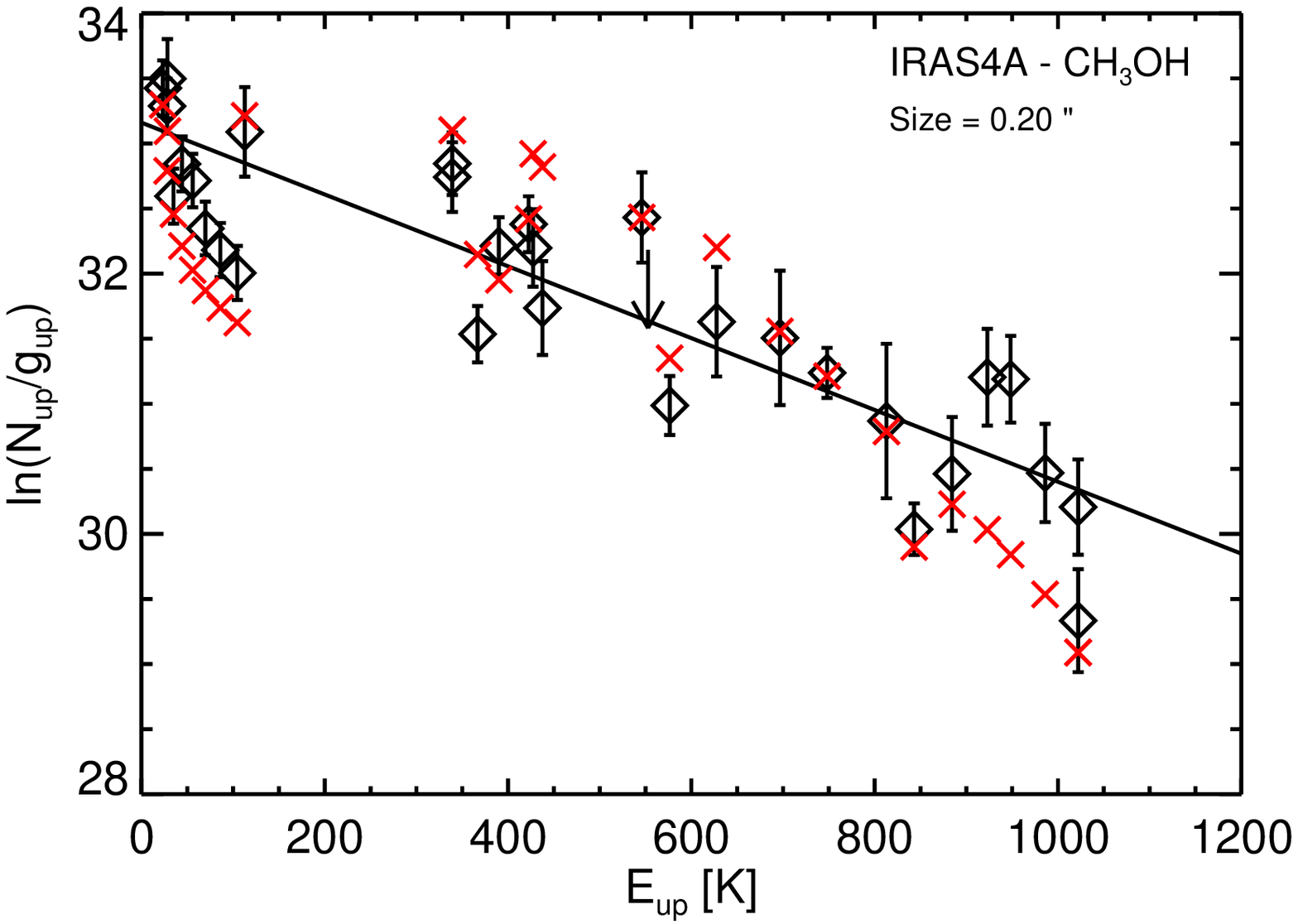} \\
\includegraphics[width=60mm]{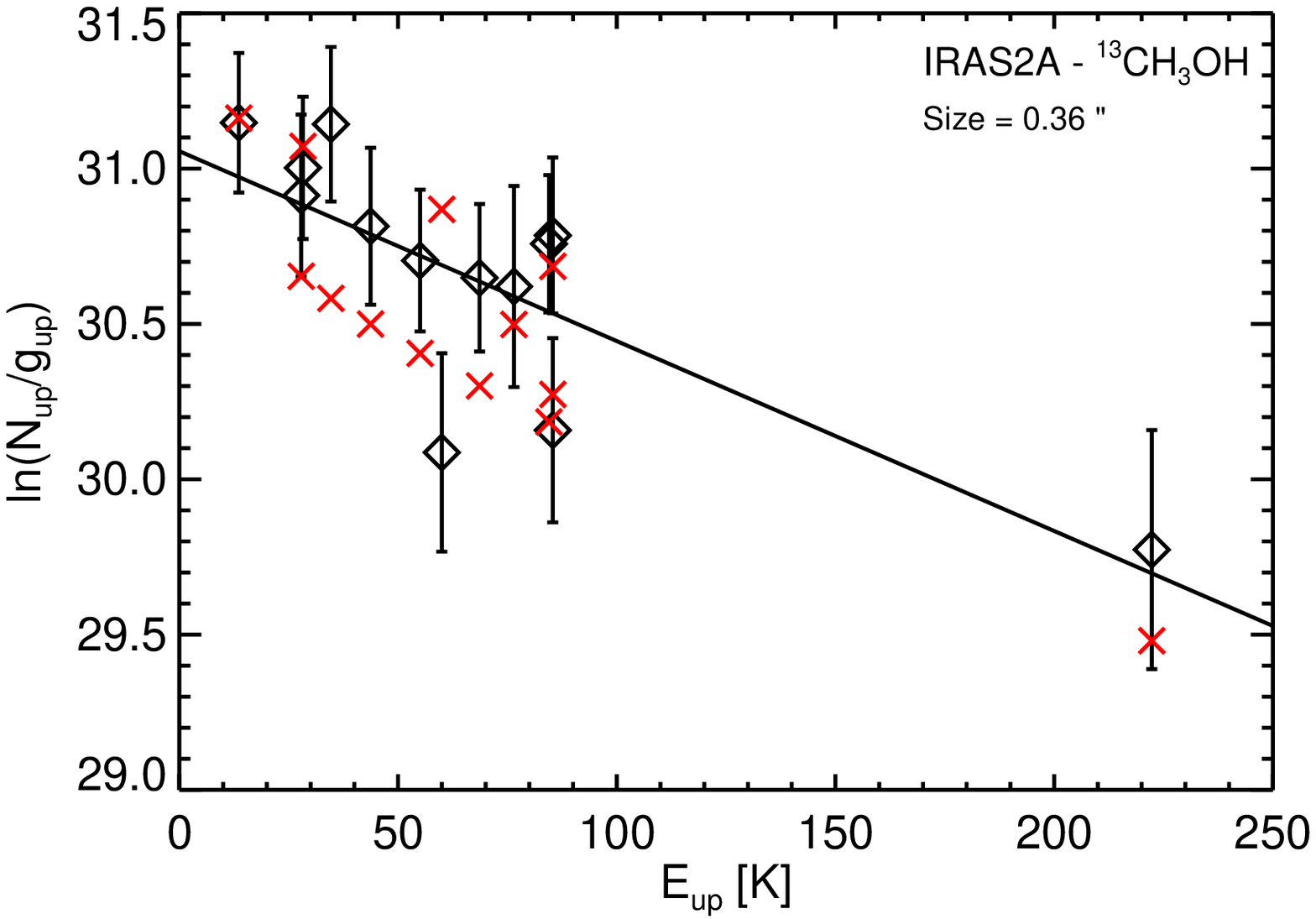} 
\includegraphics[width=60mm]{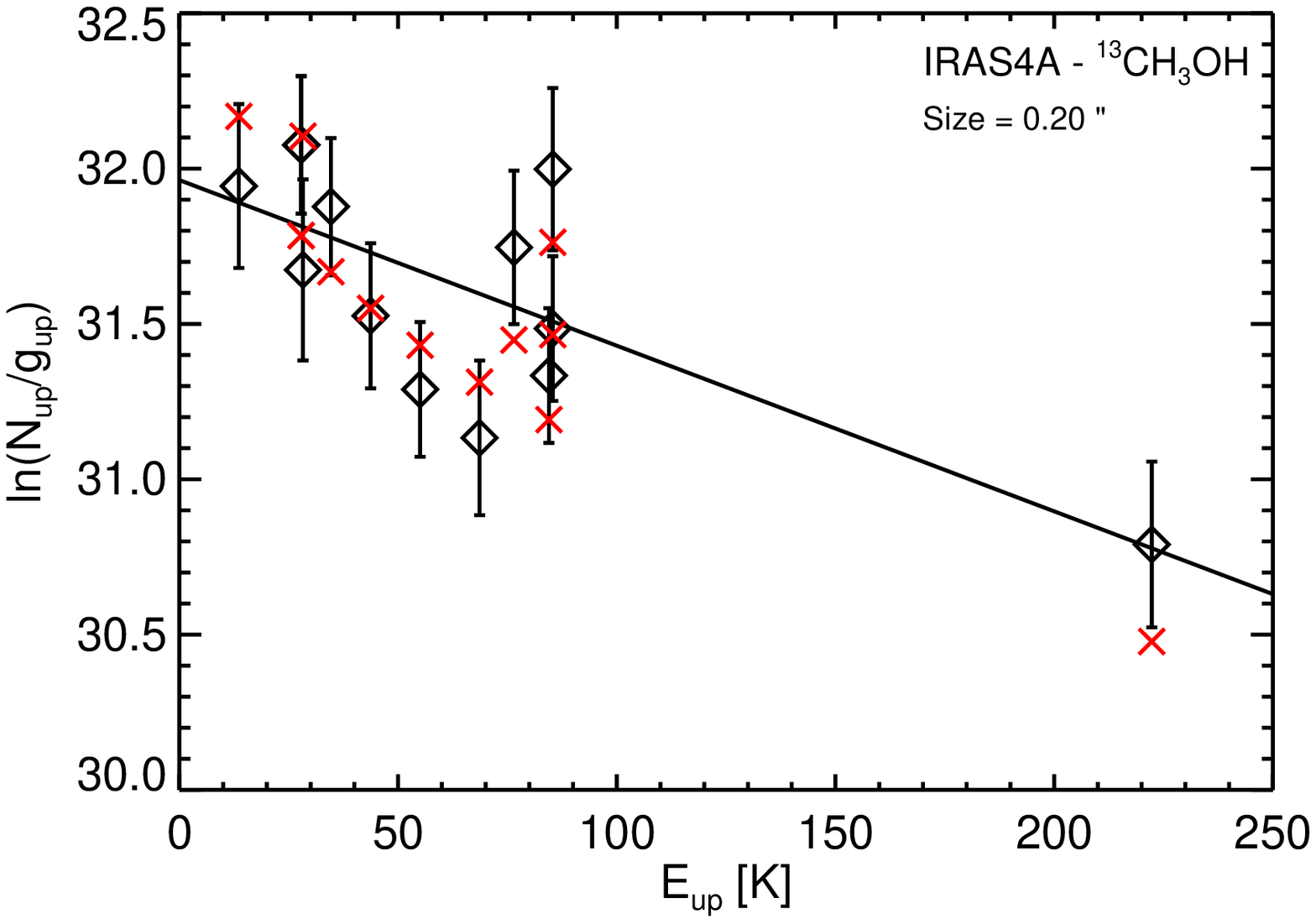} \\
\includegraphics[width=60mm]{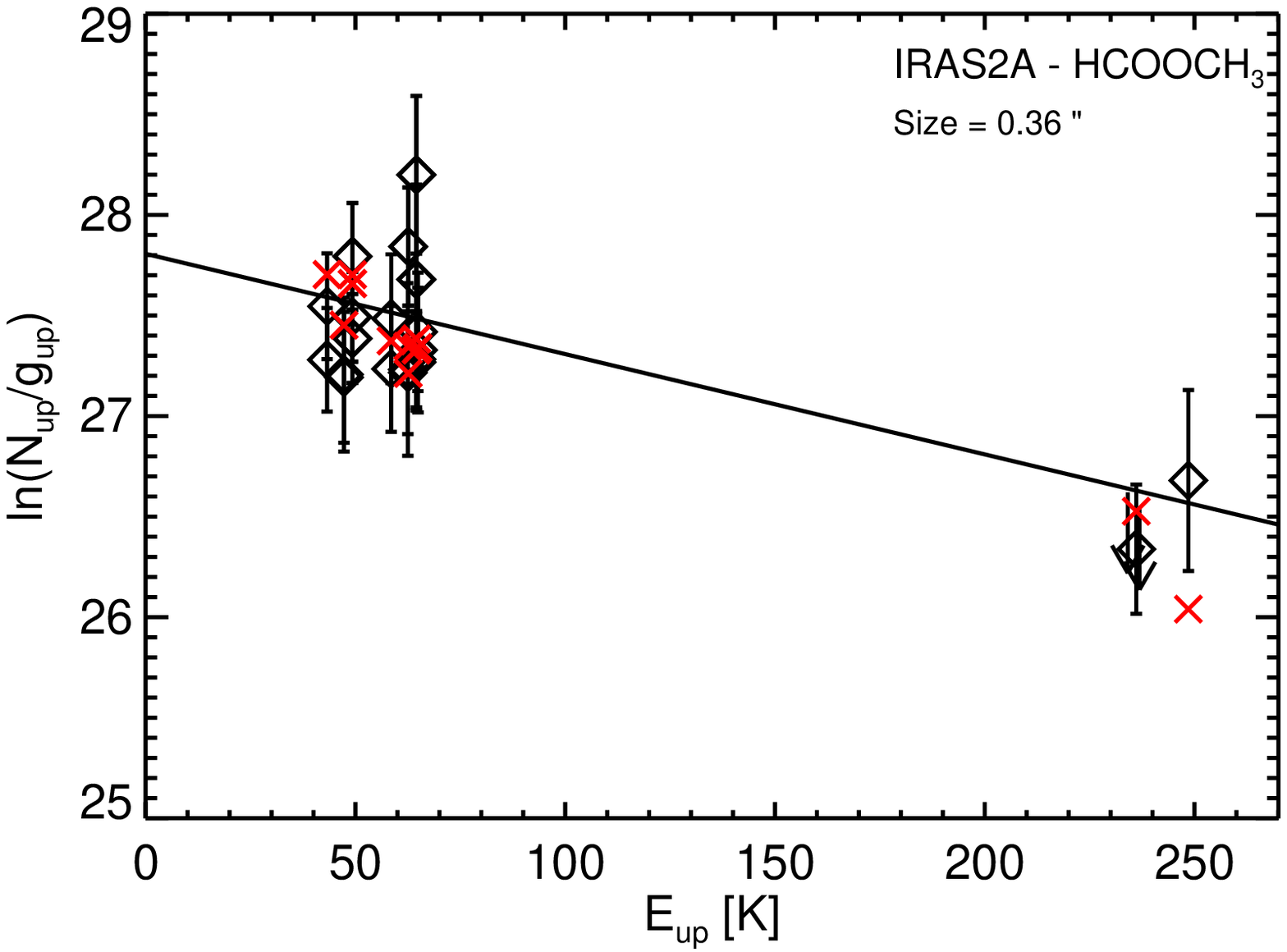} 
\includegraphics[width=60mm]{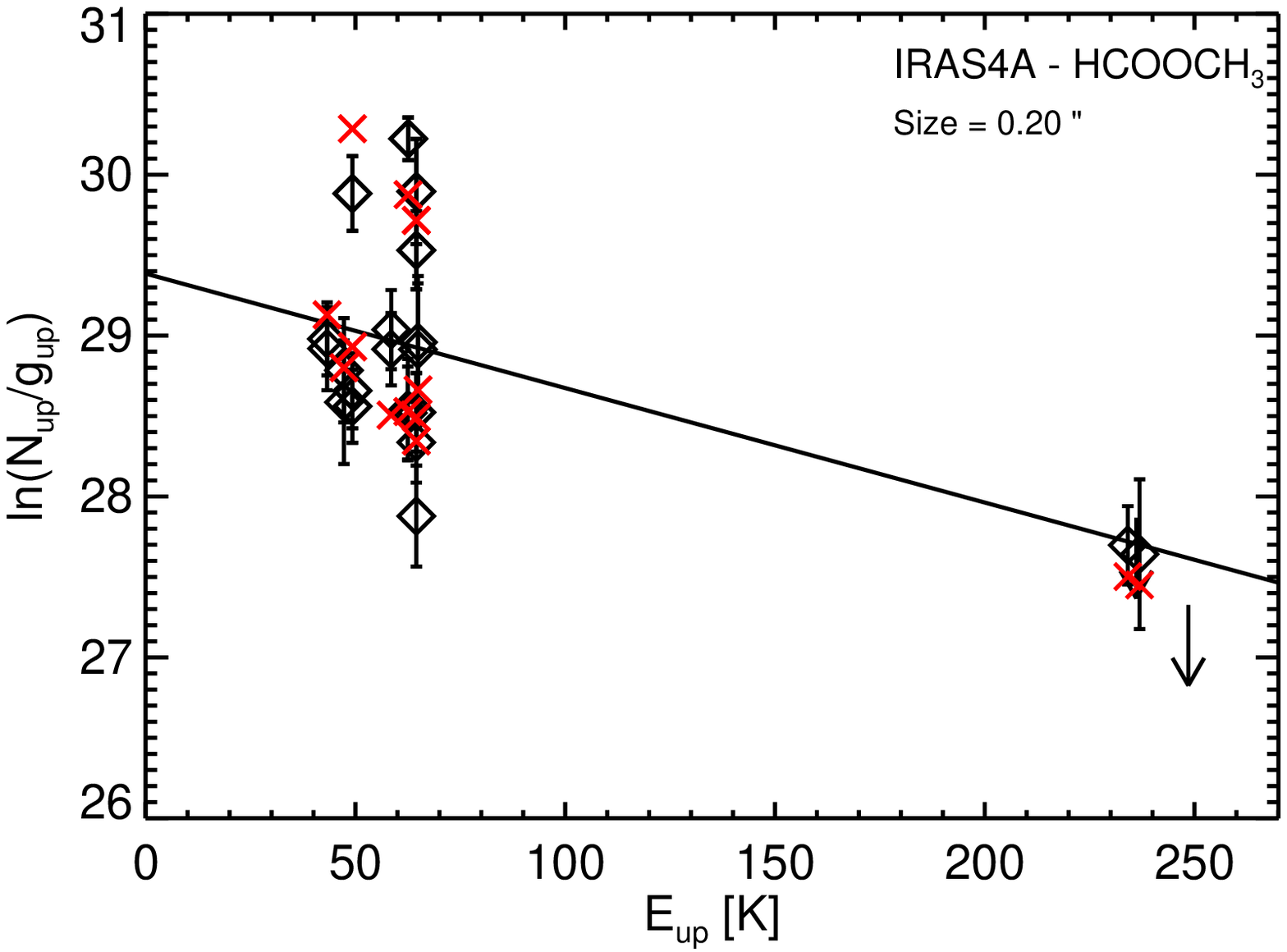} \\
\includegraphics[width=60mm]{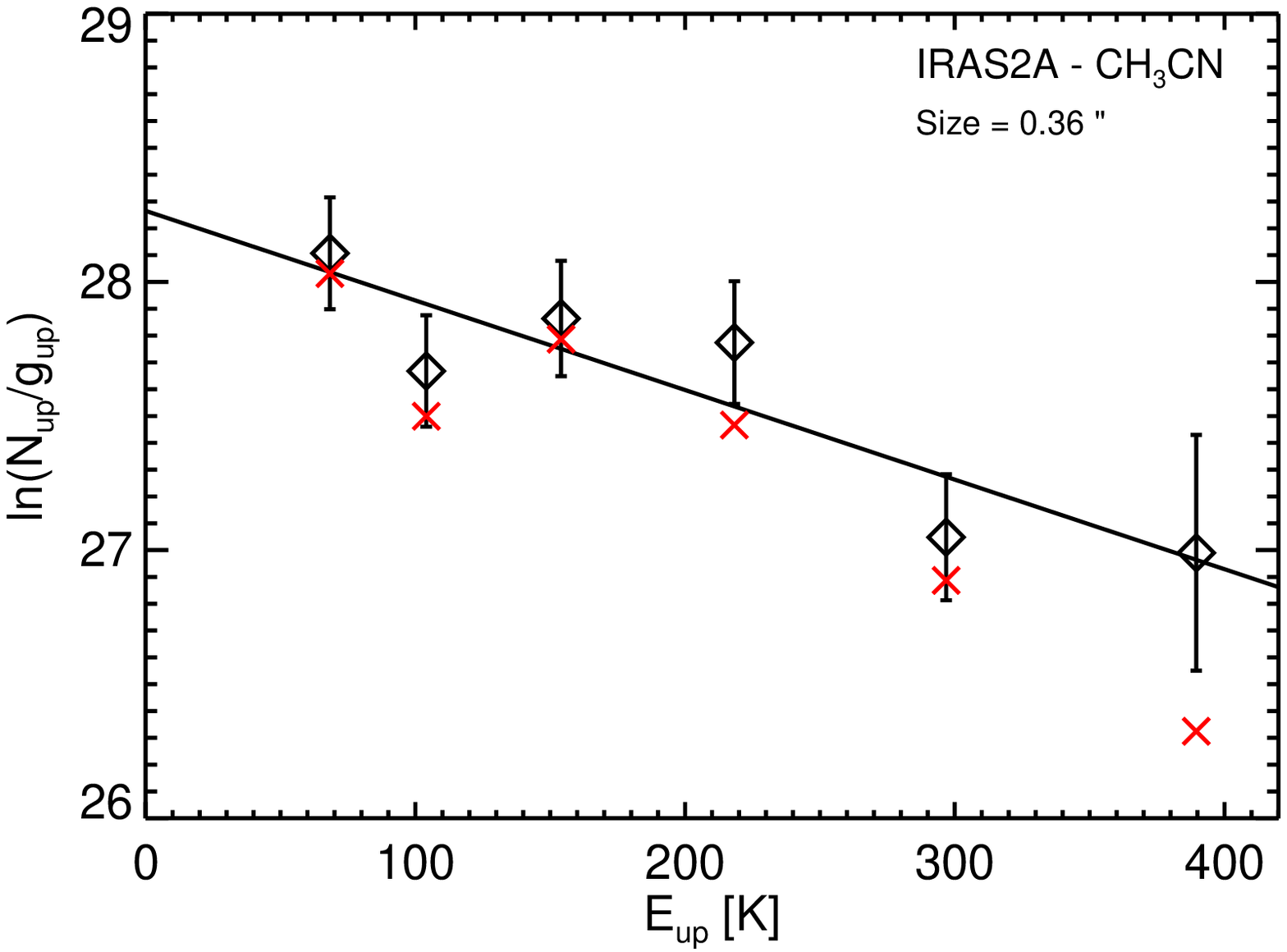} 
\includegraphics[width=60mm]{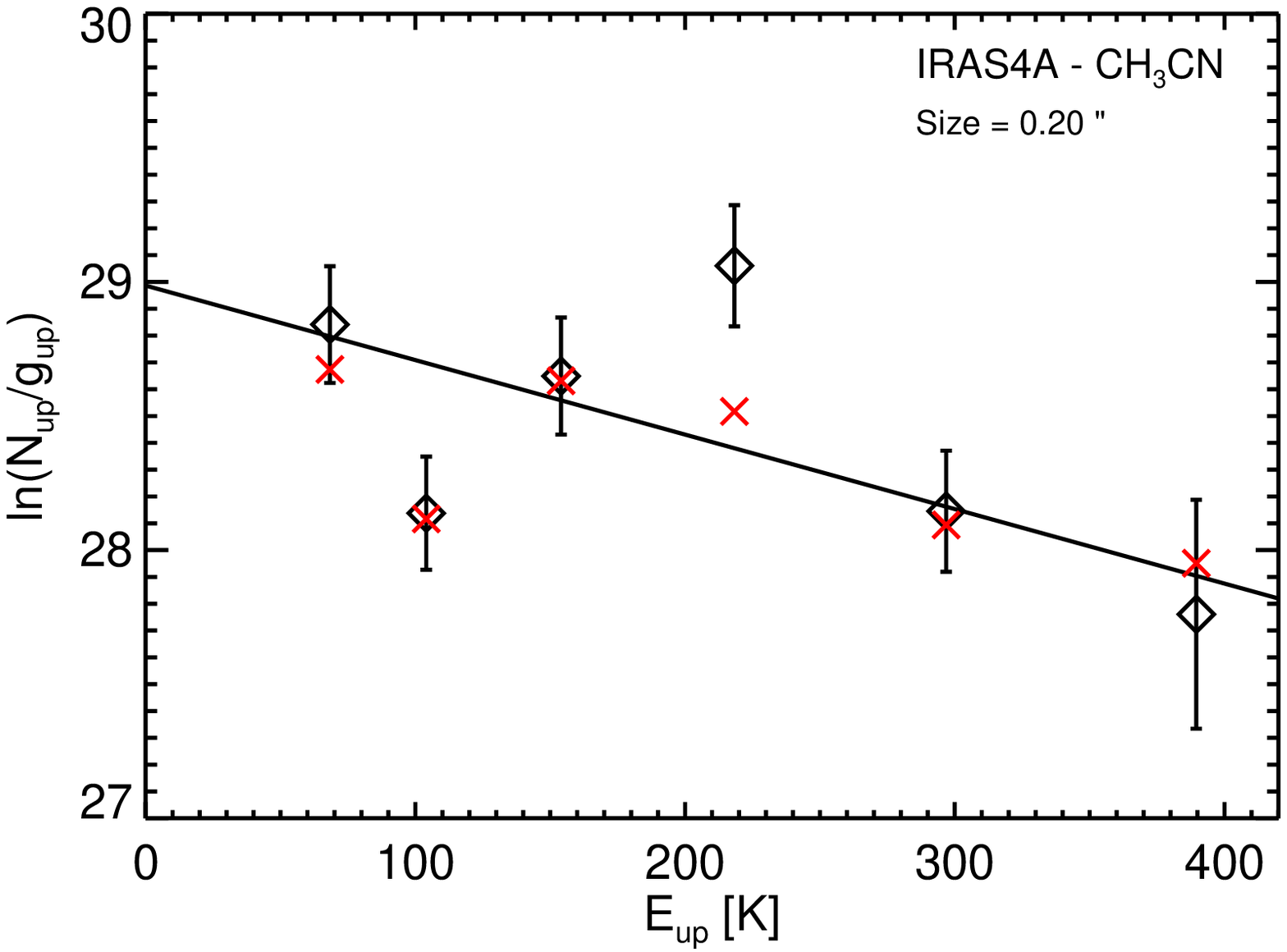} \\
\caption{Rotational and population diagrams of methanol isotopologues
  (CH$_3$OH, $^{13}$CH$_3$OH), HCOOCH$_3$, and CH$_3$CN for 
  source sizes derived from the PD analysis of the methanol population
  distribution (0.36 $\arcsec$ for IRAS2A   and 0.20 $\arcsec$ for
  IRAS4A). 
  Observational data is depicted by the black
  diamonds. Error bars are derived assuming a calibration
  uncertainty of 20 \% on top of the statistical error.
  Straight lines represent the best fit of the RD analysis to the data. Red
  crosses show the best fit of the PD to the data. }
\label{PD_meth}
\end{figure}

\begin{figure}[htp]
\centering
\includegraphics[width=60mm]{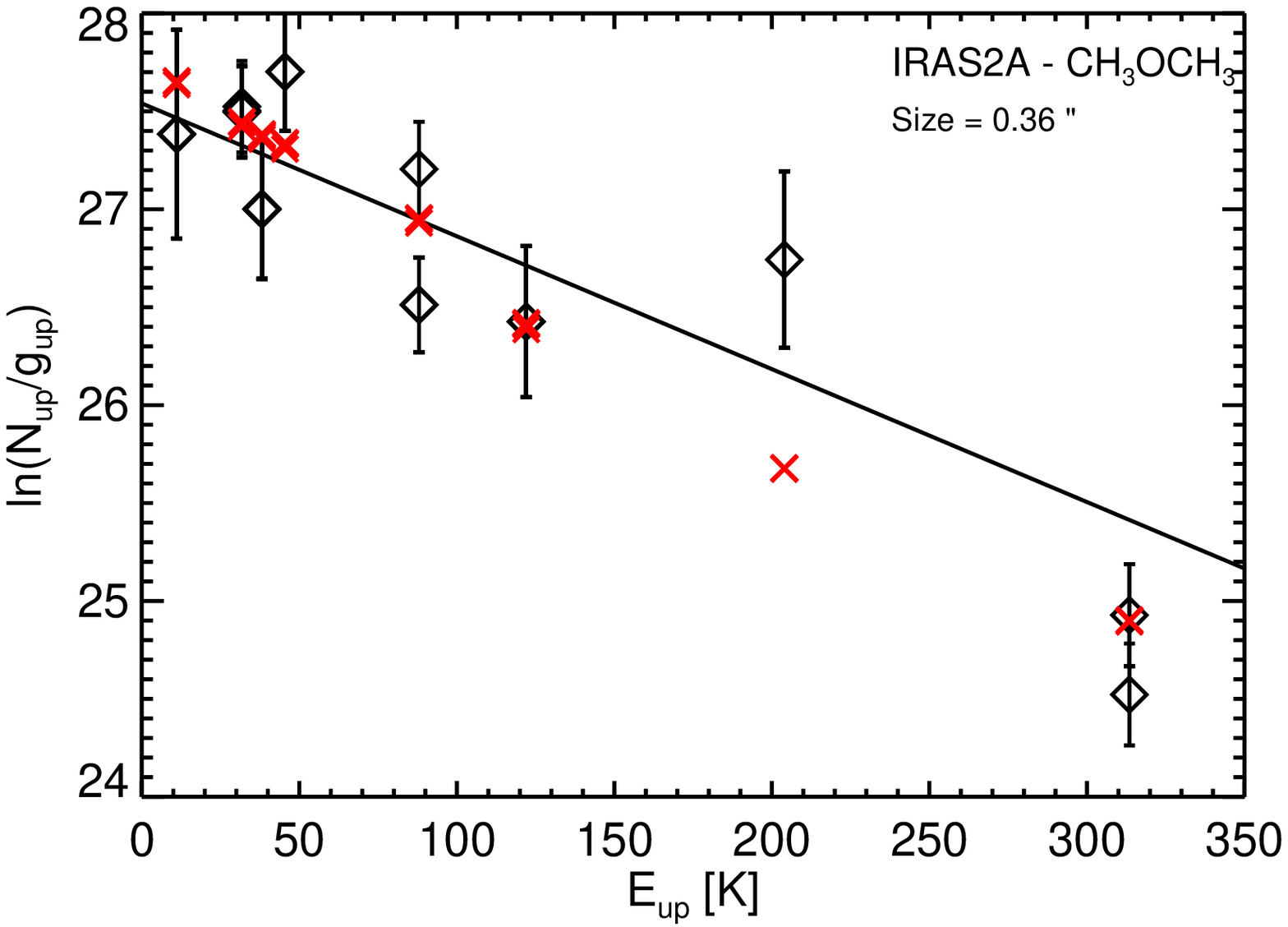} 
\includegraphics[width=60mm]{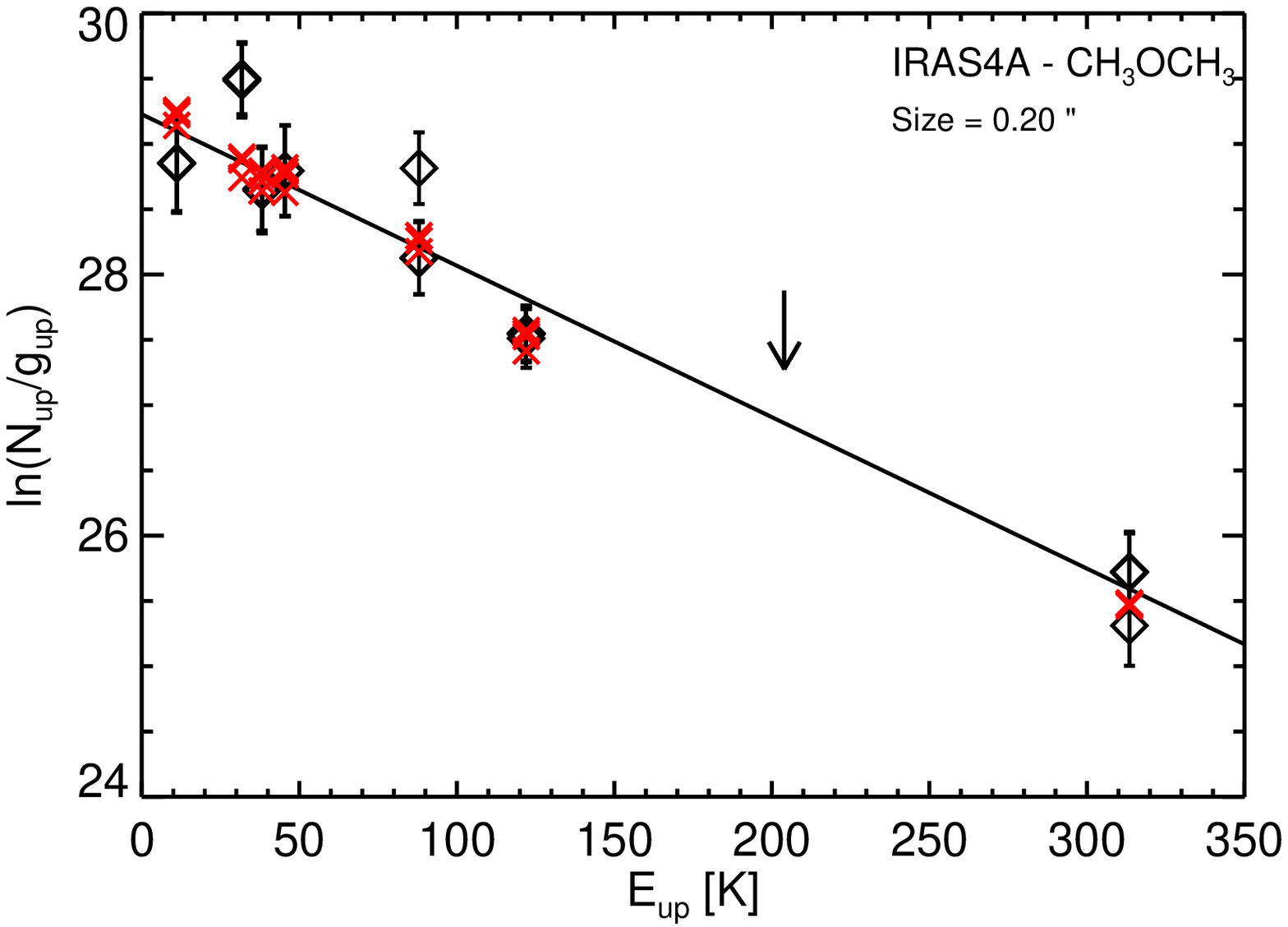} \\
\includegraphics[width=60mm]{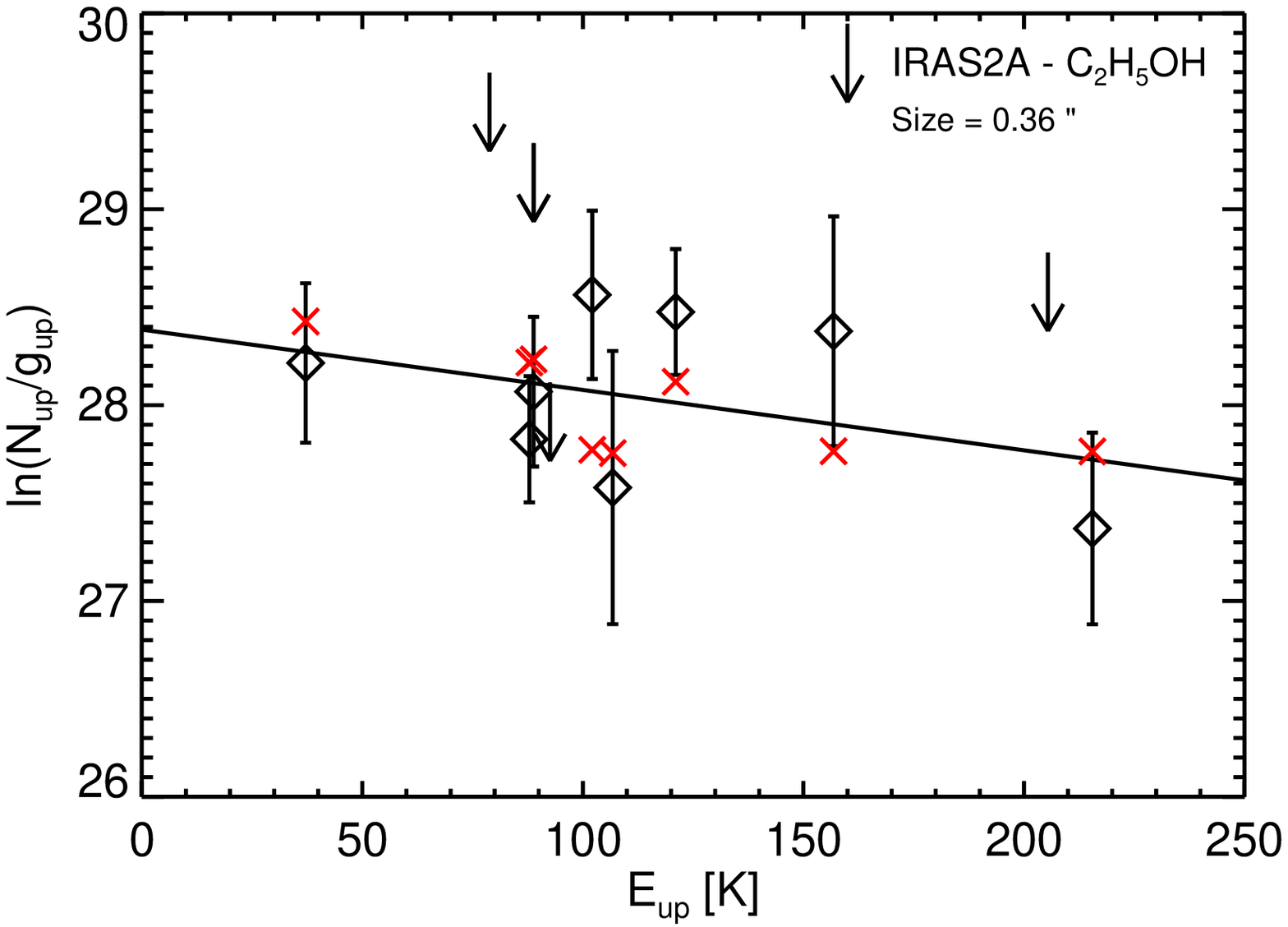} 
\includegraphics[width=60mm]{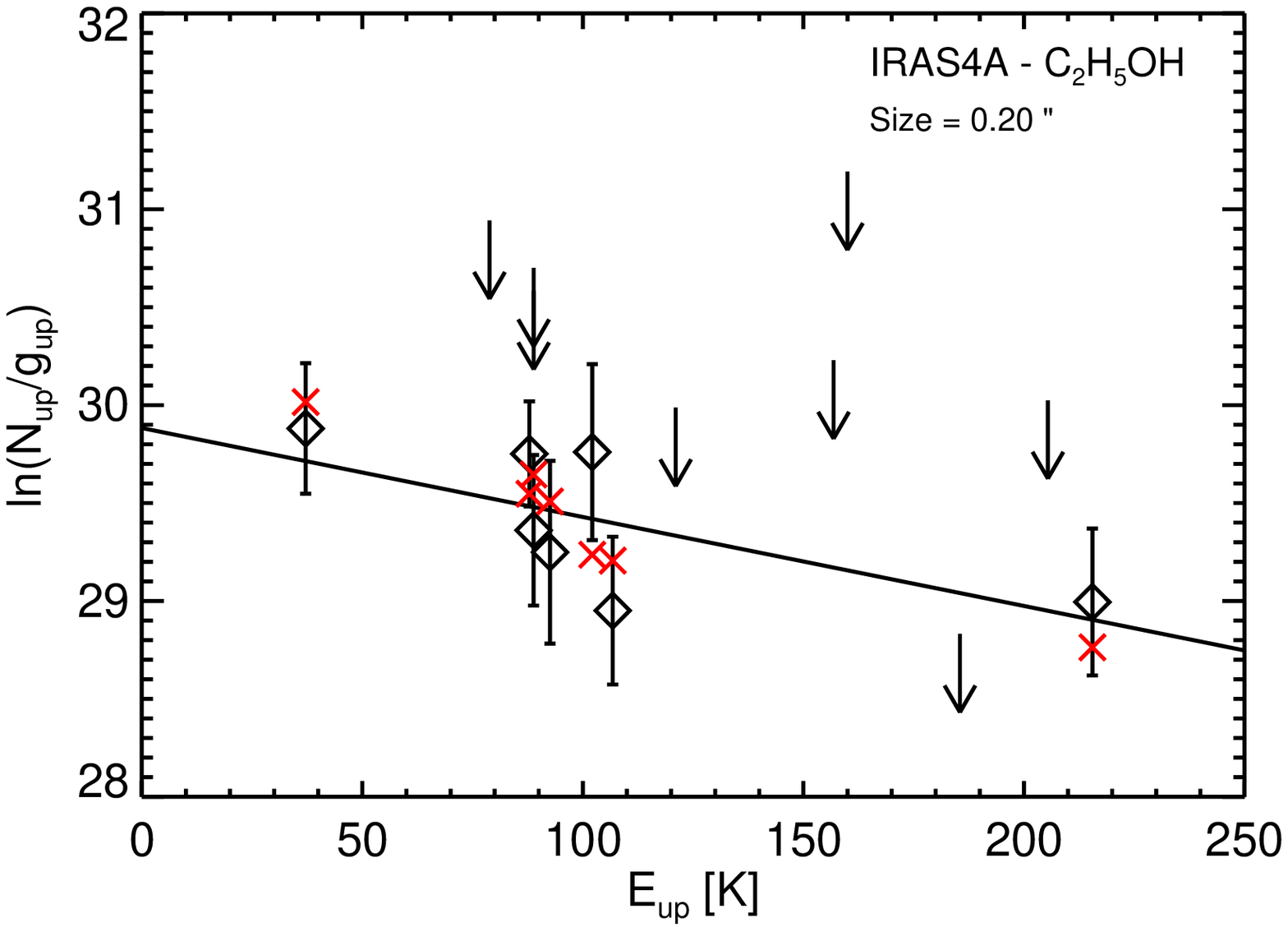} \\
\includegraphics[width=60mm]{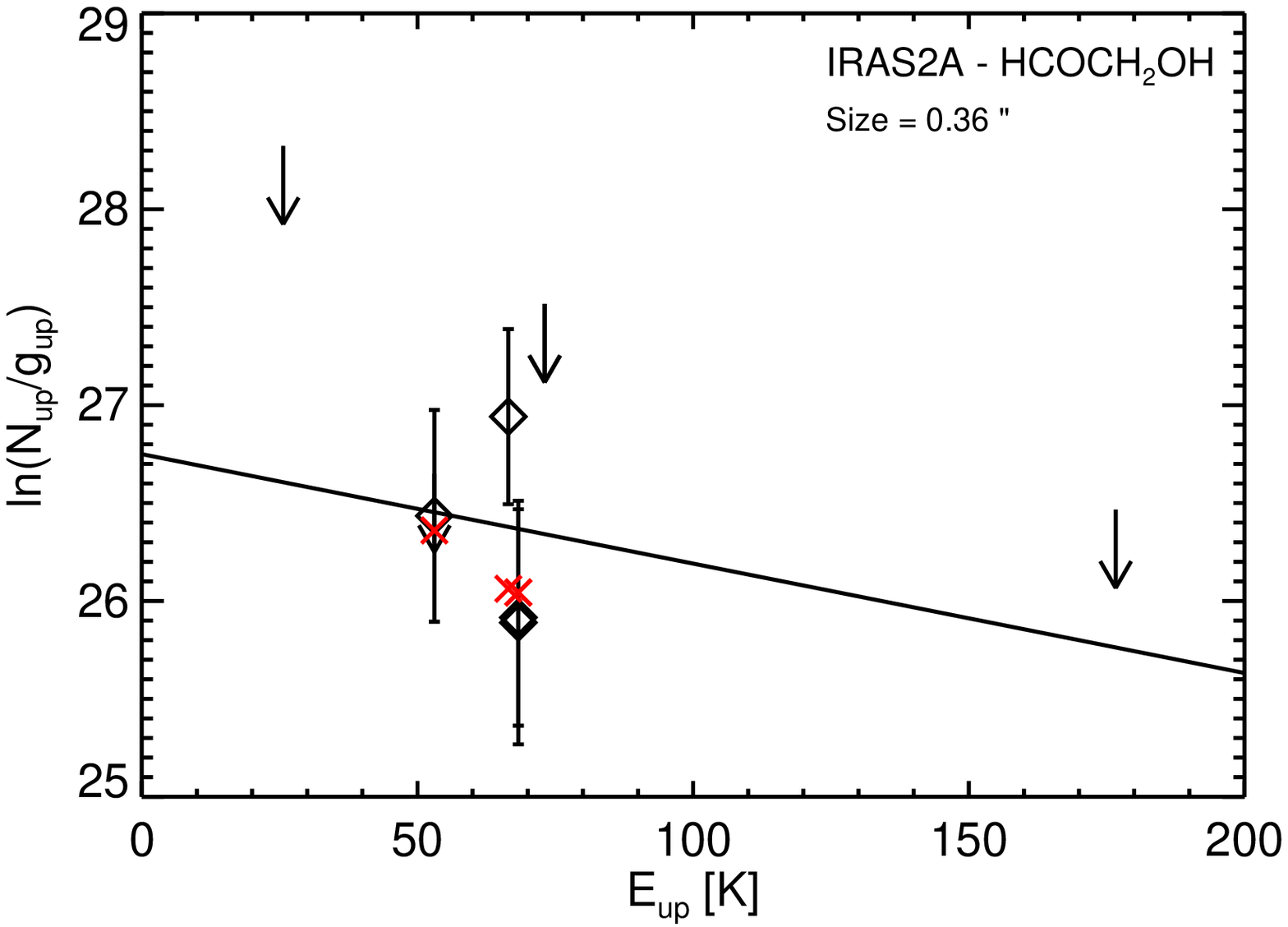} 
\includegraphics[width=60mm]{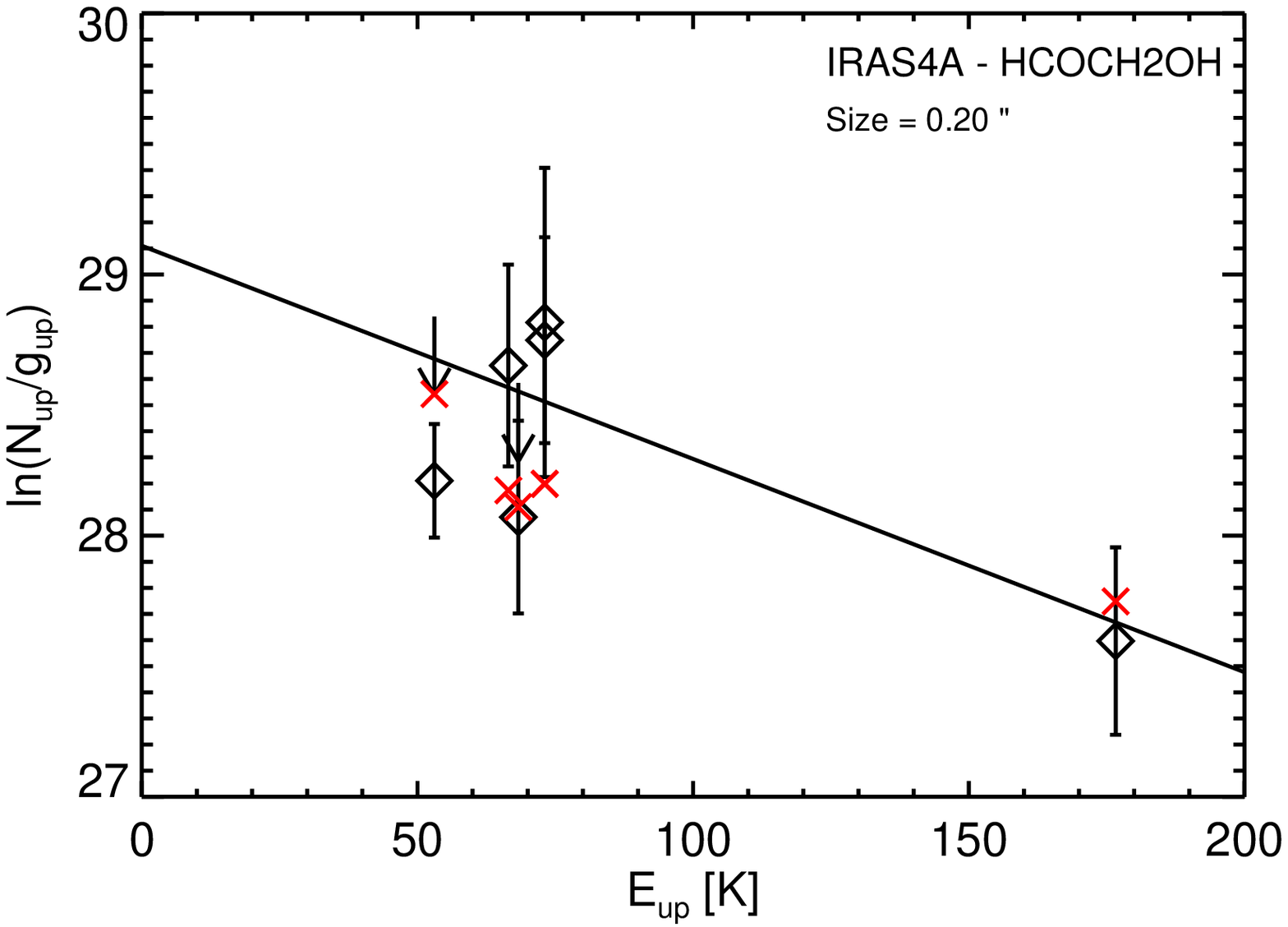} \\
\includegraphics[width=60mm]{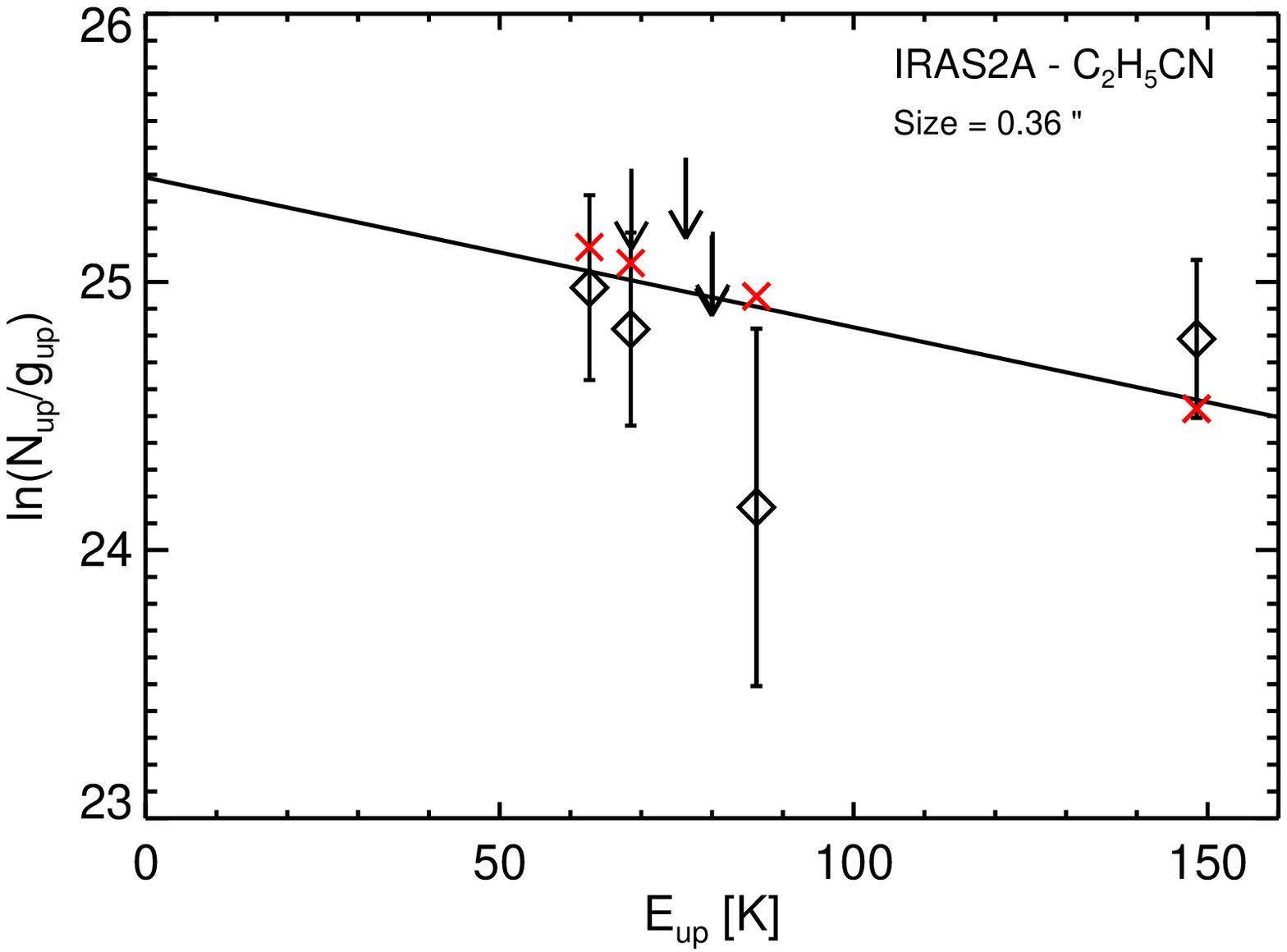} 
\includegraphics[width=60mm]{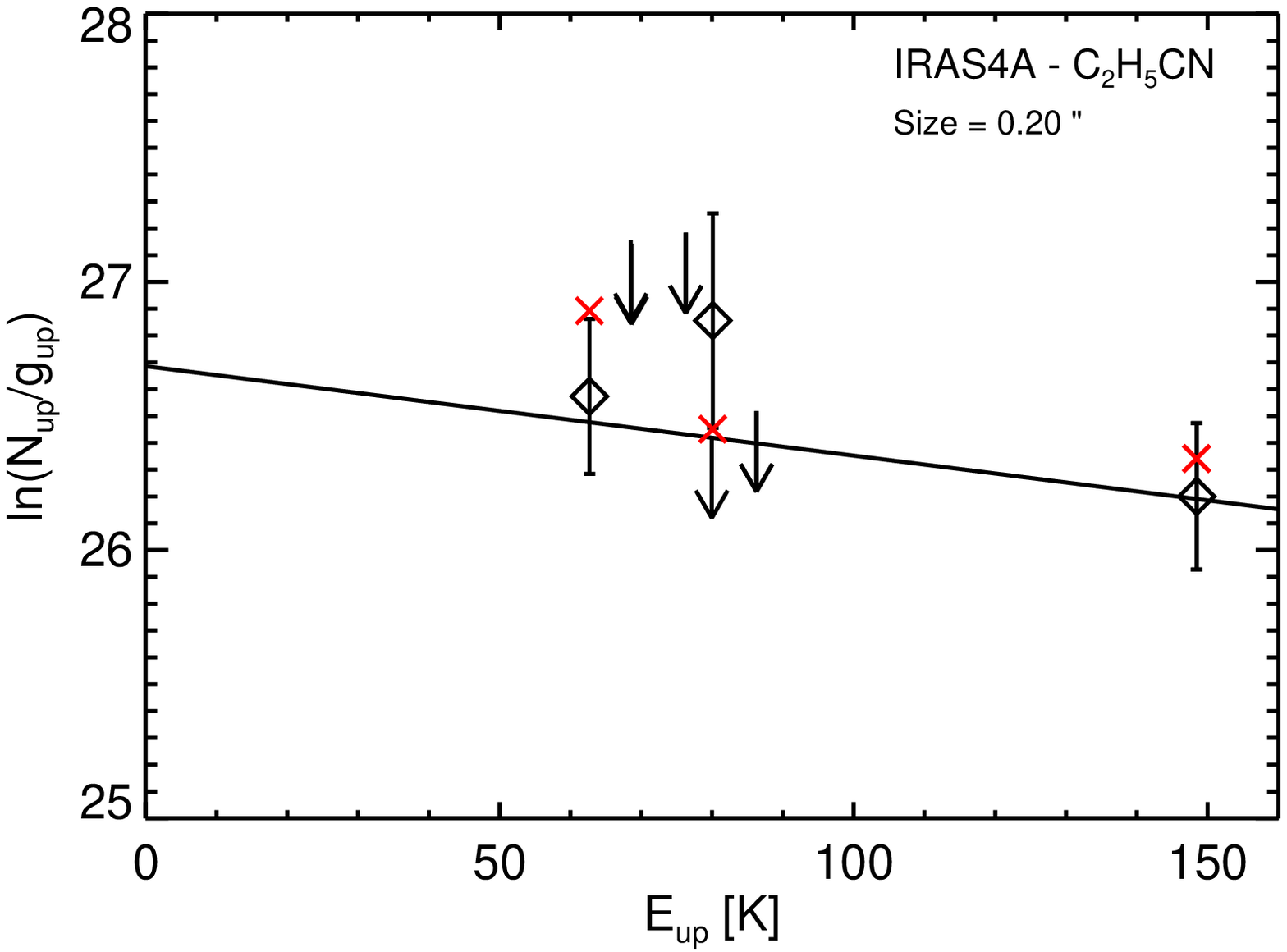} \\
\caption{Rotational and population diagrams of CH$_3$OCH$_3$,
  C$_2$H$_5$OH, HCOCH$_2$OH, C$_2$H$_5$CN for source sizes derived
  from the PD analysis of the methanol population distribution
  (0.36 $\arcsec$ for IRAS2A and 0.20 $\arcsec$ for IRAS4A). 
  Observational data is depicted by the black 
  diamonds. Error bars are derived assuming a calibration
  uncertainty of 20 \% on top of the statistical error.
  Straight lines represent the best fit of the RD analysis to the data. Red
  crosses show the best fit of the PD to the data. }
\label{PD_coms}
\end{figure}


\begin{deluxetable}{l c c c c c c c c c c c c c}
\centering
\rotate
\tabletypesize
\tiny
\tablecaption{COMs abundances observed in high-mass,
  intermediate-mass, and low-mass hot cores.}
\startdata 
\hline
\hline
	&		&		&	\multicolumn{2}{c}{CH$_3$OH}			&		H$_2$CO	&		CH$_2$CO	&		HCOOCH$_3$	&		HCOCH$_2$OH	&		CH$_3$OCH$_3$	&		C$_2$H$_5$OH	&		CH$_3$CN	&		C$_2$H$_5$CN	&	Reference	\\
\hline																																			
Source	&	$L_{\textrm{bol}}$	&	$N$(H$_2$)	&	$T_{\textrm{rot}}$	&	$X_{\textrm{H2}}$	&		$X_{\textrm{meth}}$	&		$X_{\textrm{meth}}$	&		$X_{\textrm{meth}}$	&		$X_{\textrm{meth}}$	&		$X_{\textrm{meth}}$	&		$X_{\textrm{meth}}$	&		$X_{\textrm{meth}}$	&		$X_{\textrm{meth}}$	&		\\
	&	(L$_{\odot}$)	&	(cm$^{-2}$)	&	(K)	&		&			&			&			&			&			&			&			&			&		\\
\hline																																			
\multicolumn{14}{c}{\bf High-mass protostars observed with single-dish telescopes}																																			\\
\hline																																			
IRAS 20126+4104	&	1.3(+4)	&	1.0(+24)	&	300	&	1.1(-7)	&		2.9(-1)	&	$<$	1.1(-2)	&	$<$	1.8(-2)	&	-	&	$<$	9.1(-2)	&	$<$	2.7(-2)	&		5.2(-2)	&	$<$	4.5(-3)	&	1	\\
IRAS 18089-1732	&	3.2(+4)	&	1.0(+24)	&	300	&	2.0(-7)	&		8.6(-2)	&		3.1(-2)	&		1.3(-1)	&	-	&		5.9(-1)	&		1.8(-1)	&		2.1(-2)	&		1.4(-2)	&	1	\\
G31.41+0.31	&	2.6(+5)	&	1.7(+23)	&	200	&	6.0(-6)	&		5.4(-2)	&		2.8(-2)	&		1.4(-1)	&		1.1(-1)	&		4.5(-1)	&		1.3(-1)	&		1.3(-2)	&		1.7(-2)	&	1	\\
AFGL 2591	&	2.0(+4)	&	7.6(+22)	&	147	&	6.2(-7)	&		2.8(-1)	&		2.1(-2)	&	$<$	5.1(-1)	&	-	&	$<$	1.6(-1)	&	$<$	2.1(-2)	&	$<$	7.4(-2)	&	$<$	1.6(-2)	&	2	\\
NGC 7538 IRS1	&	1.3(+5)	&	2.1(+23)	&	156	&	5.7(-7)	&		2.1(-1)	&		5.3(-2)	&		1.2(-1)	&	-	&	$<$	1.3(-1)	&		4.8(-2)	&	$<$	6.8(-2)	&	$<$	7.7(-3)	&	2	\\
G24.78	&	7.9(+5)	&	4.0(+23)	&	211	&	7.0(-7)	&		2.3(-1)	&		5.2(-2)	&		1.1(-1)	&	-	&		4.3(-1)	&		2.5(-2)	&		2.1(-1)	&		1.4(-2)	&	2	\\
G75.78	&	1.9(+5)	&	1.2(+23)	&	113	&	9.2(-7)	&		2.0(-1)	&		6.2(-2)	&		6.5(-2)	&	-	&		2.1(-1)	&	$<$	2.2(-2)	&		1.6(-2)	&	$<$	1.1(-2)	&	2	\\
W33A	&	1.0(+5)	&	2.6(+23)	&	259	&	7.7(-7)	&		2.7(-1)	&		4.9(-2)	&		1.3(-1)	&	-	&		1.4(-1)	&		2.4(-2)	&		1.4(-1)	&	$<$	1.1(-2)	&	2	\\
NGC 6334 IRS1	&	1.7(+5)	&	2.4(+23)	&	178	&	4.0(-6)	&		1.3(-1)	&		2.0(-2)	&		1.2(-1)	&	-	&		6.0(-1)	&		2.0(-2)	&		3.0(-2)	&		5.3(-3)	&	2	\\
W3 (H2O)	&	2.0(+4)	&	1.8(+23)	&	181	&	5.6(-6)	&		1.8(-1)	&		1.5(-2)	&		5.2(-2)	&	-	&		1.5(-1)	&		8.4(-3)	&		7.0(-3)	&		4.5(-3)	&	2	\\
Sgr B2 (M)	&	6.5(+6)	&	3.5(+24)	&	150	&	7.4(-9)	&		1.0(-1)	&		1.9(-2)	&		6.2(-2)	&	-	&	-	&		4.2(-2)	&		6.0(-2)	&	-	&	3	\\
Sgr B2 (N)	&	6.5(+6)	&	8.0(+24)	&	170	&	6.3(-7)	&		4.8(-2)	&		1.4(-2)	&	$<$	3.8(-2)	&	-	&		5.6(-2)	&		3.8(-2)	&		5.1(-2)	&		1.3(-1)	&	4	\\
G327.3-0.6	&	1.0(+5)	&	3.0(+24)	&	118	&	2.0(-5)	&		3.6(-5)	&	-	&		8.0(-2)	&	-	&		5.4(-1)	&		4.1(-3)	&		3.5(-2)	&		2.2(-2)	&	5	\\
Orion KL - HC	&	1.0(+5)	&	3.1(+23)	&	128	&	2.2(-6)	&		5.5(-2)	&	-	&	-	&	-	&		3.1(-2)	&	-	&		1.4(-2)	&		5.0(-3)	&	6	\\
Orion KL - CR	&	1.0(+5)	&	3.9(+23)	&	140	&	1.2(-6)	&		3.7(-2)	&		4.3(-3)	&		2.8(-1)	&	-	&		1.4(-1)	&		1.4(-2)	&		1.1(-2)	&	-	&	6	\\
G34.3+0.15	&	6.3(+5)	&	5.3(+23)	&	336	&	7.0(-8)	&	-	&		1.8(-2)	&		4.3(-1)	&	-	&		2.5(-1)	&		9.5(-2)	&		6.5(-3)	&	-	&	7	\\
G34.3+0.2	&	6.3(+5)	&	1.6(+23)	&	96	&	1.7(-7)	&	-	&	-	&		5.4(-2)	&	-	&		1.4(-1)	&		6.5(-2)	&	-	&		1.0(-2)	&	8	\\
DR21(OH)	&	5.0(+4)	&	2.5(+24)	&	150	&	1.0(-8)	&	-	&	-	&	$<$	4.0(-3)	&	-	&	-	&	$<$	5.2(-3)	&	-	&	$<$	8.4(-4)	&	8	\\
W51	&	1.5(+6)	&	3.3(+23)	&	208	&	3.0(-7)	&	-	&	-	&		1.2(-1)	&	-	&	-	&		3.1(-2)	&	-	&		7.0(-3)	&	8	\\
\hline																																			
\multicolumn{14}{c}{\bf High-mass hot cores observed with interferometers}																																			\\
\hline																																			
Orion KL	&	1.0(+5)	&	4.4(+24)	&	200	&	4.5(-9)	&	-	&	-	&		3.0(-1)	&	-	&		5.0(-1)	&		1.0(-1)	&		1.0(-1)	&		2.5(-1)	&	9	\\
G29.96	&	9.0(+4)	&	3.3(+24)	&	200	&	1.2(-7)	&	-	&	-	&		2.0(-1)	&	-	&		5.0(-1)	&		1.5(-1)	&		2.5(-2)	&		2.5(-2)	&	10	\\
G19.61-0.23	&	1.6(+5)	&	8.4(+23)	&	151	&	6.2(-7)	&	-	&	-	&		4.2(-2)	&	-	&		2.7(-2)	&		1.2(-1)	&		7.9(-2)	&		3.1(-2)	&	11	\\
\hline																																			
\multicolumn{14}{c}{\bf Intermediate-mass hot cores observed with interferometers}																																			\\
\hline																																			
I22198-MM2	&	370	&	2.0(+25)	&	120	&	1.2(-6)	&	-	&	-	&		1.1(-2)	&	-	&	-	&		9.1(-3)	&	-	&	-	&	12	\\
A5142-MM1	&	2300	&	1.0(+25)	&	210	&	2.3(-7)	&	-	&	-	&	$<$	8.7(-2)	&	-	&	-	&		9.1(-2)	&	-	&	-	&	12	\\
A5142-MM2	&	2300	&	2.0(+25)	&	140	&	2.0(-7)	&	-	&	-	&	$<$	5.0(-2)	&	-	&	-	&		2.0(-2)	&	-	&	-	&	12	\\
NGC 7129 FIRS2	&	500	&	2.5(+24)	&	238	&	1.0(-6)	&		1.6(-2)	&	-	&		1.5(-2)	&	-	&		1.2(-2)	&		8.8(-3)	&		5.2(-3)	&		3.5(-4)	&	13	\\
\hline																																			
\multicolumn{14}{c}{\bf Low-mass protostars observed with single-dish telescopes}																																			\\
\hline																																			
IRAS 16293	&	27	&	2.0(+23)	&	84	&	1.0(-7)	&		2.6(-1)	&		1.0(-3)	&		9.0(-2)	&		6.9(-3)	&		4.0(-1)	&	$<$	5.0(-2)	&		9.1(-3)	&	$<$	2.0(-3)	&	14	\\
IRAS2A	&	36	&	2.1(+23)	&	101	&	8.8(-7)	&		2.9(-1)	&	-	&	$<$	8.5(-1)	&	-	&	$<$	5.3(-1)	&	-	&		1.1(-2)	&	$<$	1.3(-1)	&	15	\\
IRAS4A	&	9.1	&	1.6(+24)	&	24	&	1.4(-7)	&		1.4(-1)	&	-	&		5.5(-1)	&	-	&	$<$	2.2(-1)	&	-	&		1.3(-2)	&	$<$	9.2(-3)	&	15	\\
IRAS4B	&	4.4	&	8.1(+22)	&	34	&	6.9(-6)	&		2.0(-1)	&	-	&		1.3(-1)	&	-	&	$<$	1.9(-1)	&	-	&		1.6(-2)	&	$<$	1.2(-1)	&	15	\\
SMM1	&	30	&	1.3(+23)	&	16	&	1.9(-9)	&	-	&	-	&		1.0(-1)	&	-	&		5.3(-2)	&	$<$	3.4(-2)	&	-	&	-	&	16	\\
SMM4	&	1.9	&	1.1(+23)	&	13	&	9.5(-9)	&	-	&	-	&	$<$	1.0(-2)	&	-	&	$<$	8.0(-3)	&	$<$	6.0(-3)	&	-	&	-	&	16	\\
B1-a	&	1.3	&	1.9(+22)	&	15	&	6.4(-10)	&	-	&	-	&		1.0(-1)	&	-	&	$<$	6.7(-2)	&	-	&		1.3(-2)	&	-	&	17	\\
SVS 4-5	&	38	&	5.7(+22)	&	20	&	3.9(-9)	&	-	&	-	&		4.5(-2)	&	-	&		1.0(-1)	&	-	&		7.7(-3)	&	-	&	17	\\
B5 IRS1	&	4.7	&	2.3(+22)	&	17	&	1.0(-9)	&	-	&	-	&	$<$	1.7(-1)	&	-	&	$<$	3.5(-1)	&	-	&		1.7(-2)	&	-	&	17	\\
IRAS 03235	&	1.9	&	1.4(+23)	&	18	&	8.6(-11)	&	-	&	-	&	$<$	1.7(-1)	&	-	&	$<$	4.2(-1)	&	-	&	$<$	2.5(-2)	&	-	&	17	\\
IRAS 04108	&	0.62	&	2.9(+22)	&	9	&	4.1(-10)	&	-	&	-	&	$<$	8.3(-2)	&	-	&	-	&	-	&	-	&	-	&	17	\\
L1489 IRS	&	3.7	&	4.3(+22)	&	8	&	1.2(-10)	&	-	&	-	&	$<$	4.0(-1)	&	-	&	-	&	-	&	-	&	-	&	17	\\
\hline																																			
\multicolumn{14}{c}{\bf Low-mass hot corinos observed with interferometers}																																			\\
\hline																																			
IRAS2A-RD	&	36	&	2.0(+24)	&	179	&	2.5(-7)	&		8.1(-2)	&		1.4(-3)	&		1.9(-2)	&		1.5(-3)	&		1.2(-2)	&		1.5(-2)	&		3.0(-3)	&		2.7(-4)	&	18	\\
IRAS2A-PD	&	36	&	2.0(+24)	&	140	&	1.0(-6)	&	-	&	-	&		1.6(-2)	&		1.4(-3)	&		1.0(-2)	&		1.6(-2)	&		4.0(-3)	&		3.0(-4)	&	18	\\
IRAS4A-RD	&	9.1	&	1.4(+25)	&	300	&	1.7(-8)	&		6.3(-2)	&		2.8(-3)	&		1.5(-2)	&		2.5(-3)	&		8.7(-3)	&		1.2(-2)	&		1.8(-3)	&		4.2(-4)	&	18	\\
IRAS4A-PD	&	9.1	&	1.4(+25)	&	140	&	4.3(-7)	&	-	&	-	&		3.1(-2)	&		3.0(-3)	&		1.0(-2)	&		1.0(-2)	&		3.9(-3)	&		4.0(-4)	&	18	\\
\hline
\enddata
\tablecomments{
1: \citet{Isokoski2013};
2: \citet{Bisschop2007};
3: \citet{Nummelin2000};
4: \citet{Neill2014};
5: \citet{Gibb2000};
6: \citet{Crockett2014};
7: \citet{Macdonald1996};
8: \citet{Ikeda2001};
9: \citet{Beuther2009};
10: \citet{Beuther2007};
11: \citet{Qin2010};
12: \citet{Palau2011};
13: \citet{Fuente2014};
14: \citet{Maret2005, Jaber2014};
15: \citet{Maret2004, Maret2005, Bottinelli2004a, Bottinelli2007};
16: \citet{Oberg2011};
17: \citet{Oberg2014};
18: This work
}
\label{list_hc}
\end{deluxetable}

\end{document}